\documentclass[pdflatex,sn-mathphys-num]{sn-jnl}

\usepackage{graphicx}%
\usepackage{multirow}%
\usepackage{amsmath,amssymb,amsfonts}%
\usepackage{amsthm}%
\usepackage{mathrsfs}%
\usepackage[title]{appendix}%
\usepackage{xcolor}%
\usepackage{textcomp}%
\usepackage{manyfoot}%
\usepackage{booktabs}%
\usepackage{algorithm}%
\usepackage{algorithmicx}%
\usepackage{algpseudocode}%
\usepackage{listings}%
\usepackage[binary-units=true]{siunitx}

\usepackage[symbol]{footmisc}

\usepackage {xr}
\makeatletter
\newcommand*{\addFileDependency}[1]{
  \typeout{(#1)}
  \@addtofilelist{#1}
  \IfFileExists{#1}{}{\typeout{No file #1.}}
}
\makeatother

\newcommand*{\myexternaldocument}[1]{
    \externaldocument{#1}
    \addFileDependency{#1.tex}
    \addFileDependency{#1.aux}
}

\myexternaldocument{sn-supplementary}
\theoremstyle{thmstyleone}%

\theoremstyle{thmstyletwo}%

\theoremstyle{thmstylethree}%

\begin{document}

\title[Article Title]{Mixed-Mode In-Memory Computing: Towards High-Performance Logic Processing In A Memristive Crossbar Array}

\author*[1,2]{\fnm{Nan} \sur{Du}}
\email{nan.du@leibniz-ipht.de}

\author[3]{\fnm{Ilia} \sur{Polian}}

\author[4, †]{\fnm{Christopher} \sur{Bengel} \footnote[0]{† Christopher Bengel works now at HELLA GmbH \& Co. KGaA | Rixbecker Strasse 75, 59552 Lippstadt, Germany. }}

\author[1,2]{\fnm{Kefeng} \sur{Li}}

\author[1,2]{\fnm{Ziang} \sur{Chen}}

\author[1,2]{\fnm{Xianyue} \sur{Zhao}}

\author[1]{\fnm{Uwe} \sur{Hübner}}

\author[3]{\fnm{Li-Wei} \sur{Chen}}

\author[5]{\fnm{Feng} \sur{Liu}}


\author[6]{\fnm{Massimiliano} \sur{Di Ventra}}

\author[5]{\fnm{Stephan} \sur{Menzel}}

\author[1,2]{\fnm{Heidemarie} \sur{Krüger}}

\affil[1]{\orgname{Leibniz Institute of Photonic Technology (IPHT)}, \orgaddress{\street{Albert-Einstein-Str. 9}, \city{Jena}, \postcode{07745}, \country{Germany}}}

\affil[2]{\orgdiv{Institute for Solid State Physics}, \orgname{Friedrich Schiller University Jena}, \orgaddress{\street{Helmholtzweg 3}, \city{Jena}, \postcode{07743}, \country{Germany}}}

\affil[3]{\orgdiv{Institute of Computer Engineering and Computer Architecture}, \orgname{University of Stuttgart}, \orgaddress{\street{Pfaffenwaldring 47}, \city{Stuttgart}, \postcode{70569}, \country{Germany}}}

\affil[4]{\orgdiv{Institute of Materials in Electrical Engineering and Information Technology}, \orgname{RWTH Aachen University}, \orgaddress{\street{Sommerfeldstraße 18}, \city{Aachen}, \postcode{52074}, \country{Germany}}}

\affil[5]{\orgdiv{Peter Grünberg Institut (PGI-7)}, \orgname{Forschungszentrum Jülich}, \orgaddress{\street{Wilhelm-Johnen-Straße}, \city{Jülich}, \postcode{52428}, \country{Germany}}}

\affil[6]{\orgdiv{Department of Physics}, \orgname{University of California, San Diego}, \orgaddress{\street{9500 Gilman Drive}, \city{La Jolla}, \postcode{CA 92093-0319}, \country{USA}}}

\abstract{ In-memory computing is a promising alternative to traditional computer designs, as it helps overcome performance limits caused by the separation of memory and processing units. However, many current approaches struggle with unreliable device behavior, which affects data accuracy and efficiency. In this work, the authors present a new computing method that combines two types of operations—those based on electrical resistance and those based on voltage—within each memory cell. This design improves reliability and avoids the need for expensive current measurements. A new software tool also helps automate the design process, supporting highly parallel operations in dense two-dimensional memory arrays. The approach balances speed and space, making it practical for advanced computing tasks. Demonstrations include a digital adder and a key part of encryption module, showing both strong performance and accuracy. This work offers a new direction for reliable and efficient in-memory computing systems with real-world applications.
}

\keywords{mixed-mode computing, practical in-memory computing, memristive crossbar, automation flow tool, high efficiency, reliability, complex logic processing, voltage-controlled computing mode, memristance-controlled computing mode}

\maketitle

\section*{Introduction}\label{sec1}

The increasing demands of data-intensive computing underscore the necessity for high-performance computing paradigms endowed with robust data processing capabilities. Today's von-Neumann computing architectures, based on mature complementary metal-oxide-semiconductor (CMOS) technology and transistors with latching characteristics for logic gates, face challenges in transistor scaling and the persistent memory wall issue. Despite advances in automation tools for logic circuitry design, these limitations necessitate a paradigm shift \cite{kimovski2023beyond,wright2013beyond}. 
In-memory computing enabled by nanoscale memory technology ~\cite{di2022memcomputing,sebastian2020memory} circumvents the memory-processor bottleneck by executing logic processing within the memory itself. Among enabling nanotechnologies \cite{finocchio2023roadmap}, memristive crossbars stand out as particularly promising candidates due to their analog computing capability \cite{song2024programming,rao2023thousands}, low power cost and high scalability \cite{pi2019memristor,liu2023highly}. The key feature of a memristive element is its resistance change over time based on the current that flows though it (and some other microscopic degrees of freedom), integrating memory effect and latching characteristics in one, distinguishing it fundamentally from its transistor counterparts. In memory applications, data is stored as resistance (memristance) values in each cell, with Low Resistance State (LRS) representing `1' and High Resistance State (HRS) representing `0'. 

For logic processing applications, known memristor-enabled computing paradigms are based on either stateful or nonstateful logic approaches. Stateful logic designs \cite{borghetti2010memristive,kvatinsky2014magic,taherinejad2021sixor,gupta2018felix} capitalize on memristance states stored in individual cells as logic inputs and outputs, offering cost-effectiveness. However, their practical implementation in memristor crossbars faces substantial obstacles due to inherent reliability issues \cite{in2020universal,kim2020stateful}. For instance, notable stateful logic designs \cite{borghetti2010memristive, kvatinsky2014magic} 
typically assess simultaneously multiple cells in a crossbar configuration for computing each logic gate. These designs confront challenges such as writing variations that reduce the programming window between logic `0' and `1' \cite{kopperberg2022endurance}, as well as unintended programming of neighboring cells by programming voltages applied to target cell(s) \cite{schon2023spatio,staudigl2024s}. Alternatively, the nonstateful approaches \cite{papandroulidakis2014boolean,linn2010complementary,xie2017scouting} leverage transient voltages or currents applied to (or sensed from) memristors as logic variables `0' and `1'. Though more resilient to variations, nonstateful designs lack universality (as clarified in Supplementary Information \ref{secA}). They require sense amplifiers controlled by dedicated peripherals to handle cascaded logic functions, leading to high power and latency cost in peripherals and compromising the in-memory computing benefits.
Despite considerable efforts to advance memristive logic designs \cite{taherinejad2021sixor,gupta2018felix}, the inherent trade-off between computing efficiency and accuracy persists in both approaches.

To tackle this issue, we propose an innovative approach: mixed-mode in-memory computing paradigm that maximize the use of memory effect and latching characteristics in each memristive element. Our approach offers effective solutions at the circuit, architecture, and software levels to address the reliability issues pertaining to memristive crossbars, while eliminating high cost current sensing during cascading.

In particular, we see that mapping computations to memristive crossbars differs substantially from classical logic synthesis in CMOS transistors. Simply adapting known tools \cite{8203782}, algorithms \cite{rohani2019semiparallel, fayyazi2020hipe}, or data structures \cite{TSAM:19,CCDS:14,TZW+:20} to memristive crossbars, as done in current solutions, is insufficient to fully exploit the high parallel computing capability of 2D crossbars. Therefore, we develop a dedicated crossbar-oriented automation tool for seamlessly integrating the voltage- and memristance-controlled logic operations. As a proof-of-principle, we have experimentally demonstrated the computation of an $N$-bit carry ripple adder and cryptographic S-Boxes. While striving for an optimal balance between processing latency and area, these demonstrations provide compelling evidence of its capability to solve arbitrary $n$-input functions with enhanced reliability in data processing.

\section*{Results and Discussion}\label{sec_2}
\subsection*{Mixed-mode In-Memory Computing Framework}\label{sec2}
\begin{figure}[!h]
\centering\includegraphics[width=\linewidth]{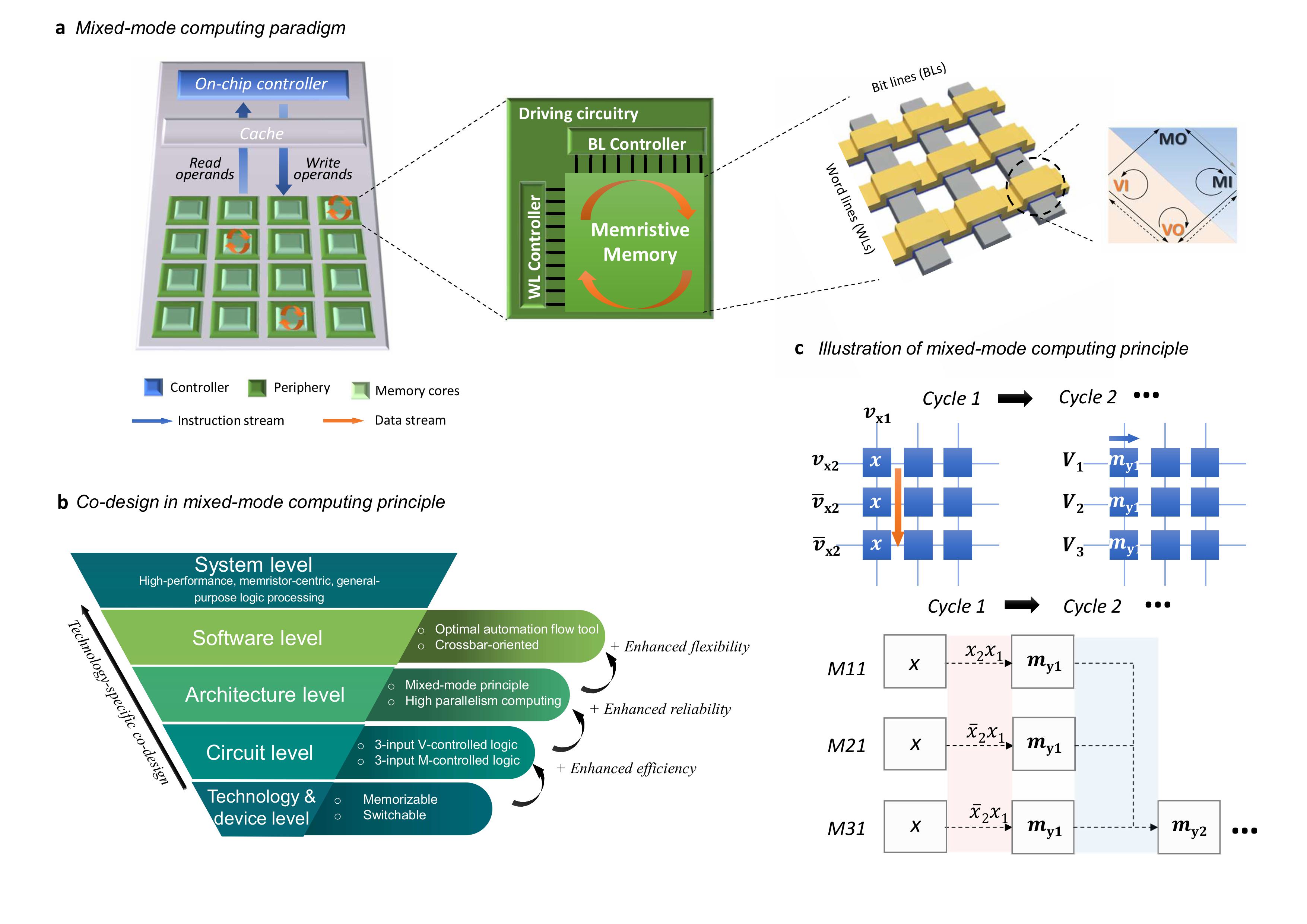}
\caption{Mixed-mode in-memory computing and system-level co-design using memristive crossbars. (a) Illustration of computing structures for in-memory computing with an inset on the right demonstrating working principle of mixed-mode computing, i.e. flexible integration of V- and M-mode logic operations in each memristive cell, including MI (memristance-input), MO (memristance-output), VI (voltage-input), and VO (voltage-output).
(b) Co-design realized in system hierarchy  using memristive crossbars for logic processing. (c) Example of mixed-mode computing using V and M modes in cycle 1 and 2, respectively. The computing is performed on memristive cells M11, M21 and M31 in one bit line in a memristive crossbar. In cycle 1 (V-mode), the voltage values $v_{\text{x1}}$/$v_{\text{x2}}$/$v_{\text{x3}}$ are applied to execute logic encoding based on the combinations of logic input variables $x_{\text{1}}$/$x_{\text{2}}$/$x_{\text{3}}$. In Cycle 2 (M-mode), fixed voltages $V_{\text{1}}$/$V_{\text{2}}$/$V_{\text{3}}$ are applied, and outputs are stored as resistance states  $m_{\text{y1}}$ and $m_{\text{y2}}$ in corresponding memristive cells. 
}
\label{fig_1}
\end{figure}

The key of innovative mixed-mode in-memory computing paradigm is to  utilize each memristive cell for either M-mode, where information is represented as resistance or memristance stored in the cell, or V-mode, where voltage applied to or current sensed from the same cell is used for this purpose, as illustrated in Figure~\ref{fig_1}a. This allows firstly arbitrary combinations of stateful and nonstateful operations on user-defined subsets of the crossbar, maximizing their benefits while keeping their expected drawbacks in check. We propose here effective solutions at circuit-, architecture-, and software-level (Figure~\ref{fig_1}b), for solving arbitrary complex logic functions with high performance while mitigating the intrinsic reliability issues in nanoscale memory devices. 

At circuit- and architecture-level, we determine distinct four logic kernels serving as building blocks of mixed-mode computation: memristance-input (MI), memristance-output (MO), voltage-input (VI), and voltage-output (VO). MI or VI kernels employ memristance M or, respectively, voltage V as the logic input variable, while in MO or VO kernels, respectively, M or V serves as the logic output variable. Supplementary Information \ref{secA} highlights that all existing representative memristive logic designs in the literature can be systematically classified according to newly determined logic kernels. Remarkably, existing memristive logic designs in the same logic kernels show analogous benefits and constraints, affirming the validity of employing these distinct logic kernels for systematic categorization of memristive logic designs.

MI and VI kernels are representing the latching feature inherent in memristors, while MO and VO representing memory effect, enabling logic operations without changing their resistance states. 
Under the mixed-mode principle, a computation is mapped to a series of logic operations in MI, MO, VI, and VO kernels, which can be flexibly interchanged and can use combinations of M- and V-modes at their inputs and outputs. 
Moreover, each cell in the array can be utilized for implementing single or multiple logic operations in each kernel, with MO facilitating low-cost cascading and storing the output memristance in the circuitry. For instance, in Figure \ref{fig_1}c, cells M11--M31 in a BL in crossbar undergo one cycle of logic operations in VI, followed by logic operations in MI during the subsequent cycle.

At software- and system-level, in order to maximize the considerable computing efficiency achievable through parallel operations across multiple memristive cells in a two-dimensional grid, we developed an approach called M$^3$S (mixed-mode mapping and synthesis) that automatically synthesizes M-and V-mode operations to the cells for solving arbitrary logic function and determines the corresponding control signals on wordlines (WLs) and bitlines (BLs). 
While its algorithmic principle is generic, the tool used for demonstrations in this work balances between benefits and costs of the VO kernel (Supplementary Information \ref{secB}) by permitting only the initial values as logic input variables and restricting readout operations to the beginning of computation. Despite minimizing the power and latency costs in peripherals, it also mitigates the necessity for data copying or transmission within the crossbar, and further allows logic processing at arbitrary positions while ensuring necessary logic inputs for the VI kernel at any
computing cycle.
 
\subsection*{Highly Efficient 3-input Logic Operations}\label{sec3} 

\begin{figure}[!h]
\centering\includegraphics[width=0.98 \linewidth]{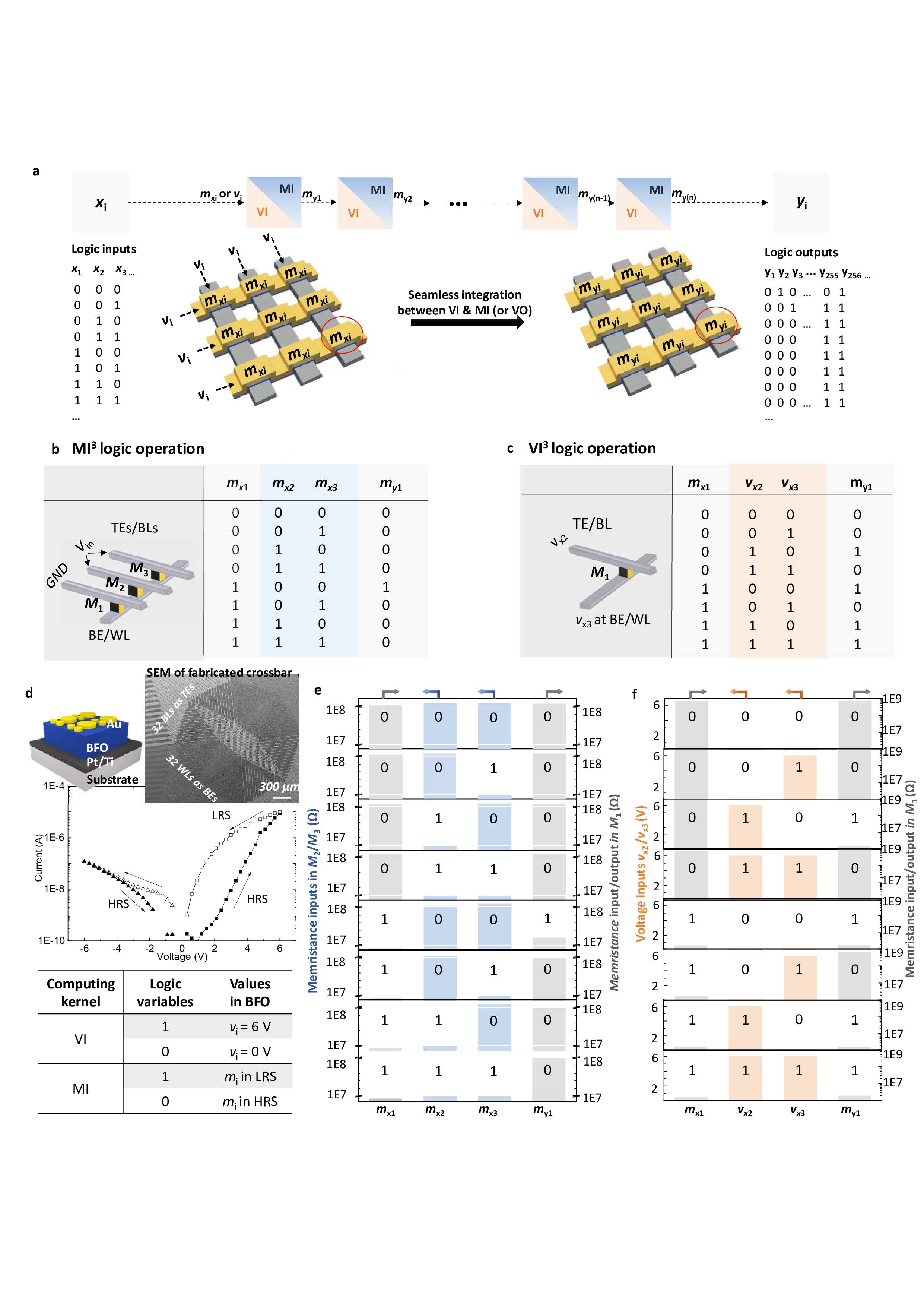}
\caption{Experimental demonstration of 3-input MI$^{3}$ and VI$^{3}$ operations by exploiting BiFeO$_{\text{3}}$ memristive cells. (a) Demonstration of mixed-mode computing principle, transferring logic inputs $x_{\text{i}}$ into $y_{\text{i}}$ through seamless integration between VI and MI kernels (VO kernel is strictly excluded during cascading due to high cost as depicted in Supplementary Information \ref{secB}). Inset below illustrates the cell states in self-rectifying passive memristive crossbar structures before and after processing mixed-mode computing. Illustration of truth tables and circuit designs for (b) MI$^{3}$ and (c) VI$^{3}$ logic operations. GND indicates the grounded electrode. (d) Schematics and current-voltage ($I$-$V$) characteristics of BiFeO$_{\text{3}}$ memristive devices applied in this work with insets showing the SEM of fabricated passive crossbar and the definition of logic variables. The accompanying table summarizes the applied bias voltages $v_{\text{i}}$ for logic `1' and `0' in the VI computing kernel, and defines the corresponding memristance states $m_{\text{i}}$ in the MI kernel, where logic `1' and `0' are represented by low-resistance state (LRS) and high-resistance state (HRS), respectively. Experimental demonstration of 3-input (e) MI$^{3}$ ($V_{in}$ = 5.3 V) and (f) VI$^{3}$ (input logic `1' as $v_{W}$ = 6 V)  for each input combination in the truth table. The memristance logic inputs $m_{\text{x2}}$/$m_{\text{x3}}$ (marked in blue) in MI$^{3}$ and the voltage logic inputs $v_{\text{x2}}$/$v_{\text{x3}}$ (marked in red) in VI$^{3}$ are demonstrated on left y axis, while the memristance input $m_{\text{x1}}$ and output $m_{\text{y1}}$  states (marked in gray) are shown on the right y-axis. State verification of these results is conducted through memory validation MO by applying a readout step at the end with a bias of 2 V to the top electrode. }
\label{fig_2}
\end{figure}

For facilitating the mixed-mode logic cascading introduced in last section, we propose 3-input logic operations in each kernel, i.e. MI$^{3}$, VI$^{3}$, and VO$^{3}$ operations, with MO as low cost logic cascading among them, as shown in Figure \ref{fig_2}a. All 3-input logic designs execute distinct logic functions within one single cycle. We determine  $\text{M}_{\text{1}}$ as input and output cell. the memristance state $m_{\text{x1}}$ of $\text{M}_{\text{1}}$ prior to the logic operation and the memristance state $m_{\text{y1}}$ subsequent to it serves as one of the logic inputs and the logic output, respectively.

The MI$^{3}$ operation from MI kernel exploits three parallel-connected memristive cells in one word line (WL) or bit line (BL) with memristance logic inputs $m_{\text{x1-x3}}$ that are stored in each device as a logic input variable. 
In contrast, the 3-input VI$^{3}$ operation from VI operates exploiting deterministic writing operations on a single cell, utilizing the initial state $m_{\text{x1}}$ and the voltages applied to its device terminals $v_{\text{x2}}$/$v_{\text{x3}}$ as logic input variables. 
MI$^{3}$ and VI$^{3}$ operations implement the functions

\begin{align} \label{eq1}
(m_{\text{y1}}) &= \text{MI}^{3}(m_{\text{x1}}, m_{\text{x2}}, m_{\text{x3}}),  \\ 
(m_{\text{y1}}) &= \text{VI}^{3}(m_{\text{x1}}, v_{\text{x2}}, v_{\text{x3}}). 
\end{align}

Figure \ref{fig_2}b and Figure \ref{fig_2}c illustrate the corresponding truth tables, respectively. The logic functions are implemented on Au/BiFeO$_{\text{3}}$/Pt/Ti memristors (Figure \ref{fig_2}d). The output $m_{\text{y1}}$ can subsequently function as logic input for both cascaded VI$^3$ and MI$^3$, thereby greatly facilitating logic cascading during automation design. 
Intuitively, the 3-input MI$^{3}$ and VI$^{3}$ operations can be considered as extended 2-input memristor-aided logic (MAGIC) \cite{kvatinsky2014magic} and complementary resistive switching (CRS, without readout) logic designs \cite{linn2010complementary}, which encounter inevitable high error rates and universality issues, respectively, as detailed in Supplementary Information \ref{secA}. 

The mixed-mode computing framework offers solutions enabling the practical logic processing by exploiting MI$^{3}$ and VI$^{3}$ operations. Traditional MI kernels alone suffer from longer cycles and higher error rates due to the stochastic variability of memristive technologies \cite{in2020universal}, while VI kernels alone are not universal and require readout VO to realize arbitrary functions (Supplementary Information \ref{secA}). These limitations are overcome with the proposed mixed-mode approach, which integrates VI$^{3}$ with MI$^{3}$ operations to achieve universality without the need for VO, resulting in shorter processing cycles and enhanced resilience by offloading part of the functionality to the more reliable VI$^{3}$ operations. Additionally, the mixed-mode computing principle addresses common issues in stateful logic designs such as ``input drift'' and ``partial switching'' (Supplementary Information \ref{secC}), enabling practically cascadable MI$^{3}$ operations. By setting the programming bias to $V_\text{in}$ = 5.3 V, we ensure correct transitions from LRS to HRS in $M_\text{1}$ with input combinations `101', `110', and `111', while preventing changes in memristance values in the input cells $M_\text{2-3}$. As expected, this $V_\text{in}$ results in a compromised LRS in $M_\text{1}$ for the '100' input combination (27.2 M$\si{\ohm}$ compared to the initialized 1.1 M$\si{\ohm}$). As next, the M$^3$S automation tool prioritizes using the output cell $M_\text{1}$ as an input cell in cascaded MI$^{3}$ (or VI$^{3}$) operations, allowing the compromised LRS in $M_\text{1}$ to be re-switched back to logic `1' through a positive $V_\text{in}$ applied to the top electrode of $M_\text{1}$ while correctly computing the cascaded MI$^{3}$ output. These methods enable robust and deeply cascadable MI$^{3}$ operations in practical implementations.

Figure \ref{fig_2}e and Figure \ref{fig_2}f demonstrate the experimental results on MI$^{3}$ and VI$^{3}$ operations recorded from the fabricated BiFeO$_{\text{3}}$-based self-rectifying crossbar array (extended experimental results in Supplementary Information \ref{secC}). 
The Experimental Section depicts the fabrication and characterization details of the self-rectifying crossbar array based on BiFeO$_{\text{3}}$ memristors, with Au top electrodes (bottom electrodes) interconnecting as BLs (WLs). 
It is important to emphasize that the proposed mixed-mode approach is compatible with various nanotechnologies, including passive crossbar arrays and 1-transistor-1-resistor (1T1R) arrays (Supplementary Information D).
Supplementary Information \ref{secB} presents the designed 3-input VO$^{3}$ operation using different types of BiFeO$_{\text{3}}$ memristors in a comparative manner, effectively illustrating the technology dependency in the VO kernel. With MI operations at our disposal, VO is not necessary for universality.

\subsection*{M$^3$S: Mixed-Mode Mapping and Synthesis Tool}\label{sec4}

Mixed-mode in-memory computing, with its V-mode and M-mode operations running simultaneously on a memristive crossbar, is so distinct from a conventional gate-based CMOS circuit that a mere adaptation of an existing synthesis tool is insufficient. The key to achieve an optimal solution with minimized requirements for cell and cycle numbers in in-memory computing, is to align the physical data location (mapping) following each gate operation in every computing cycle (synthesis) for enabling highly parallel computing in a two-dimensional memristive crossbar given its sequential computing nature.
We developed the M$^3$S tool, which concurrently addresses both mapping and synthesis tasks by formulating them as a Boolean satisfiability formula in conjunctive normal form , such that the solution (satisfying assignment) of this formula gives, for each crossbar location and each cycle, the V-mode or M-mode operations executed.

While M$^3$S tool is capable of generating conjunctive normal forms that flexibly combine arbitrary V-mode and M-mode operations, the proof-of-principle demonstrations in this study utilize V-mode phase followed by M-mode phase, satisfying the computational requirements for the desired logic functions. Taking VI$^3$ operation in V-mode as an example (Figure \ref{fig_2}c), we assume that the peripherals can apply input voltages $v_{\text{x2-x3}}$ according to functions from the list of literals $\{l_1 = \text{const-0}, l_2 = \text{const-1}, l_3 = x_1, l_4 = \overline x_1, l_3 = x_2, l_4 = \overline x_2, \ldots \}$. 
Therefore, the Boolean satisfiability formula determines for each crossbar location M$_{ab}$ $(a, b)$ and each cycle $c$ 
the variables $g^\text{TE}_{a,b,j,c}$ and $g^\text{BE}_{a,b,k,c}$ for voltage inputs $v_{\text{x2}}$ = $l_j$ to top electrode and $v_{\text{x3}}$ = $l_k$ to bottom electrode while executing each VI$^3$. 
For instance, the VI$^3$ operation on the crossbar cell M$_{78}$ in cycle 2 can be transformed into conjunctive normal form as shown in Eq.~\ref{eq:cnf}:
\begin{equation}
    \bigwedge_{\begin{array}{c}\scriptscriptstyle 1 \leq j,k \leq 2n + 2\\[-4pt]\scriptscriptstyle 1 \leq q \leq 2^n\end{array}} \left( (g_{7,8,j,2}^\text{TE} \wedge g_{7,8,k,2}^\text{BE}) \to
\left(v_{7,8,2,q} \equiv \text{VI}^3(v_{7,8,1,q}, l_{j,q}, l_{k,q}) \right)\right).\label{eq:cnf}
\end{equation}
 
Here, 
the expression $g_{7,8,j,2}^\text{TE} \wedge g_{7,8,k,2}^\text{BE}$ on the left side of the implication ($\to$'') defines the mapping constraints, including cell location $(a, b)$, cycle number $c$, and input bias-related literals $j/k$. If the solution to the Boolean satisfiability formula results in $g_{7,8,1,2}^\text{TE} = 1$ and $g_{7,8,6,2}^\text{BE} = 1$, it indicates that during cycle 2, the top electrode and bottom electrode of M${78}$ are driven by $l_1 = \text{const-0}$ and $l_6 = \overline{x_2}$. 
Conversely, the right side of the implication ($\to$'') with $v_{7,8,2,q} \equiv \text{VI}^3(v_{7,8,1,q}, l_{j,q}, l_{k,q})$ specifies the conditions for the VI$^3$ operation in cycle 2. This encompasses the synthesis conditions, such as the definitions of the logic functions VI$^3$ or MI$^3$ and the operated $q$-th entry. Hence, based on the memristance input variable $v_{7,8,1,q}$ in M${78}$ during cycle 1, the output variable $v_{7,8,2,q}$ in cycle 2 is set to 1 if the $q$-th entry of the VI$^3$ truth table is 1, and 0 otherwise. Detailed descriptions of the M$^3$S design in V- and M-mode can be found in Supplementary Information F.
The parallel computing strategy, which cascades across multiple cells in the 2D crossbar array, is ensured by transforming all VI$^3$ and MI$^3$ operations into conjunctive normal forms for each cycle.
Additional constraints ensure that all cells sharing the same WL (bottom electrode) and BL (top electrode) as M${ab}$ adhere to the same input-related literals $l{j}$. M$^3$S computes all coordinates simultaneously, achieving an optimal solution with minimized crossbar size and cycle count while adhering to the robust logic cascading conditions outlined previously.

Unlike the previous work \cite{ben2019simpler}, which relies on classical synthesis tools like ABC and CPLEX, we build directly from Boolean functions, and employ at least two modes (MI and VI) rather than the single MAGIC logic (MI mode). Additionally, our approach integrates synthesis and mapping from the outset, simultaneously generating a gate netlist while considering the physical locations of memory cells and the parallelism inherent in crossbar architectures. This co-design methodology ensures that synthesis directly exploits the crossbar’s computing potential, leading to superior efficiency and performance and offering optimality guarantees. 

\subsection*{Experimental Proof-of-principle Demonstrations}\label{sec5}

We demonstrate the advantages of using the mixed-mode computing paradigm on two examples: \(N\)-bit carry-ripple adders and a small-scale (4-bit) version of the S-Box from the Advanced Encryption Algorithm (AES) \cite{cid2005small}. An \(N\)-bit carry-ripple adder is an example of modular design that consists of identical cells (full adders); there are numerous prior implementations of adders that can serve for comparison. An S-Box, which provides the necessary nonlinearity during encryption, is an algebraically complex construction that is too large to be optimized in a handcrafted manner. Note that security aspects of the S-Box (e.g., its vulnerability to side-channel attacks) are not in scope of this work. 

\begin{figure}[!h]
\centering\includegraphics[width=\linewidth]{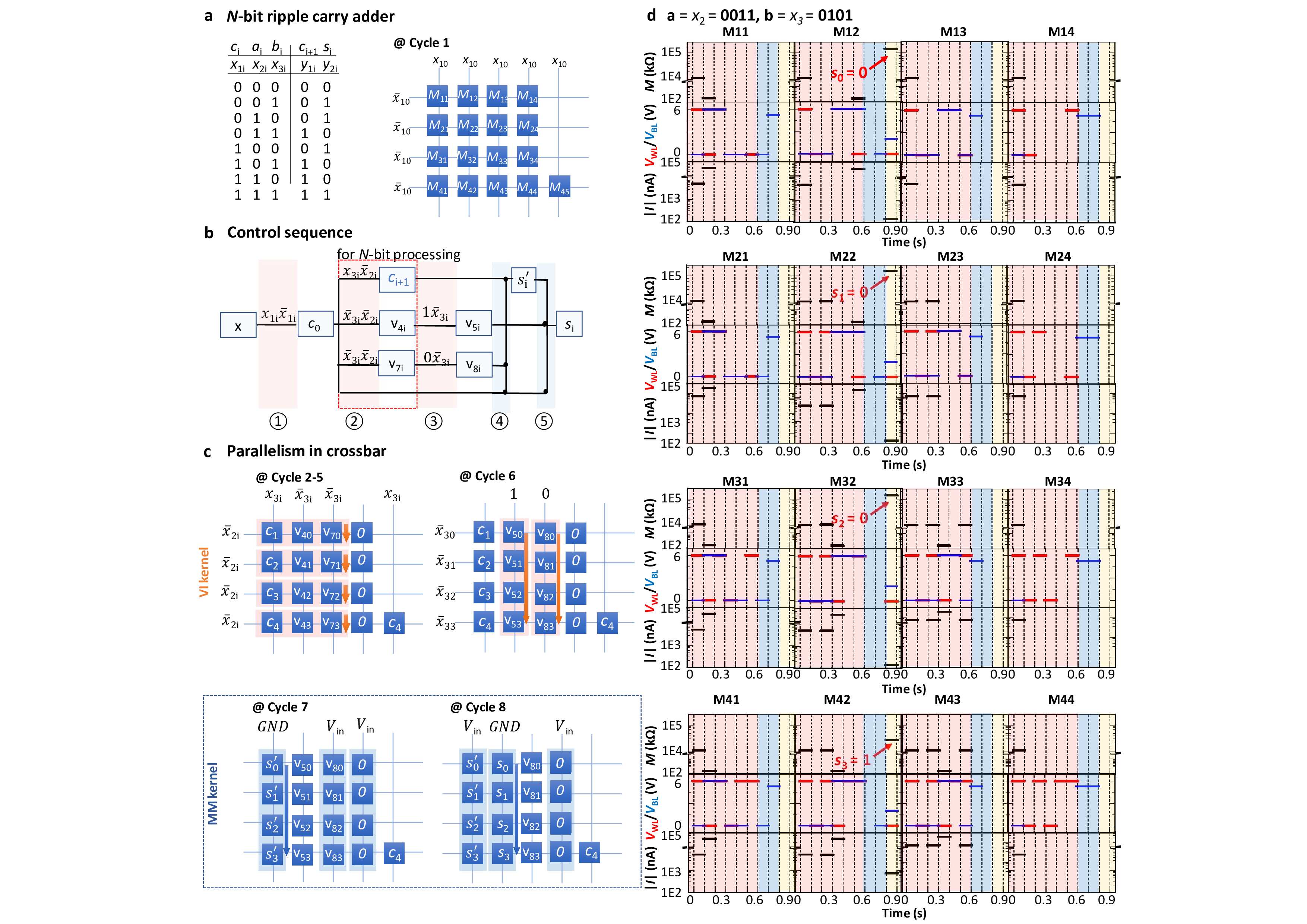}
\caption{Experimental implementation of \(N\)-bit carry-ripple adder by exploiting automation tool M$^3$S. (a) Demonstration of the truth table of \(N\)-bit carry-ripple adder and the exploited crossbar structure. (b) Control sequence of \(N\)-bit carry-ripple adder using BiFeO$_{\text{3}}$ memristive crossbar synthesized exploiting M$^3$S. Each VI$^{3}$ operation is noted as $v$ in the diagram. (c) Illustration of logic operations for 4-bit carry-ripple adder with arbitrary inputs, which requires in total 8 cycles, i.e., 6 of VI$^{3}$ and 2 of MI$^{3}$ operations. (d) Experimental results of 4-bit carry-ripple adder, including memristance of each cell $M_i$, voltage on WL to shared bottom electrode $V_\text{WL}$, voltage on BL to top electrode $V_\text{BL}$, and the absolute values of current across each cell $|I_i|$ of 4-bit carry-ripple adder with inputs $a = 0011$, $b = 0101$. 
}
\label{fig_3}
\end{figure}

Utilizing the automation tool M$^3$S, a control sequence has been generated for implementing the full adder \(N\)-bit carry-ripple adder as depicted in Figure \ref{fig_3}a, utilizing BiFeO$_{\text{3}}$-based crossbar architectures. The control sequence illustrated in Figure \ref{fig_3}b maps arbitrary combinations of logic inputs $c_\text{i}$, $a_\text{i}$, and $b_\text{i}$ (represented as $x_{\text{1-3}_\text{i}}$, respectively) onto the desired logic outputs $c_\text{i+1}$ and $s_\text{i}$ (represented as $y_{\text{1-2}_\text{i}}$) through VI$^{3}$ and MI$^{3}$ operations.     
According to carry-ripple adder’s three input $x_{\text{1-3}_\text{i}}$, one of the eight values (literals)
$x_{\text{1}_\text{i}}$, $\overline{x}_{\text{1}_\text{i}}$, $x_{\text{2}_\text{i}}$, $\overline{x}_{\text{2}_\text{i}}$, $x_{\text{3}_\text{i}}$, $\overline{x}_{\text{3}_\text{i}}$, $\text{const-0}$ and $\text{const-1}$ can be applied to either of the memristive cell’s two electrodes, with the first connected to the top electrode and the second to the bottom electrode of each cell, which correspond to BLs and WLs in a crossbar array (Fig. \ref{fig_3}b). 
Starting with the unknown state $x$, each VI$^{3}$ operation computes $m_{\text{y}_\text{i}}$ and stores it directly as a memristive state, which can serve as input to subsequent VI$^{3}$ and MI$^{3}$ operations. 
The control sequence depicted in Figure \ref{fig_3}b contains 
4 rows, each representing a cell required for computing a $1$-bit carry-ripple adder, with 1 additional cell dedicated to storing the output-carry bit \(c_\text{i+1}\)). Furthermore, the 5 numbered columns signify the requisite five cycles for executing the $1$-bit carry-ripple adder: 3 cycles involve VI$^{3}$ operations concurrently across all cells (highlighted in red), while 2 subsequent cycles involve MI$^{3}$ operations (highlighted in blue). Extending such design to an $N$-bit adder reveals that only the variation-resilient VI$^{3}$ operation for computing the output carry bit \(c_\text{i+1}\) in cycle 2 necessitates operations for each bit in $a_{\text{i}}$ and $b_{\text{i}}$, resulting in a total of \(N + 4\) cycles and \(4N + 1\) cells. 
As aforementioned, experimental processing typically incurs high error rates with the MI kernel, while the VI kernel remains relatively error-free. It is notable that regardless of $N$'s size, an $N$-bit adder utilizing this design contains only 2 cycles of parallel MI$^{3}$ operations, implying consistent error rates across varying $N$. Due to space constraints, herewith we demonstrate the experimental results of a $4$-bit carry-ripple adder.

\begin{figure}[!h]
\centering\includegraphics[width=\linewidth]{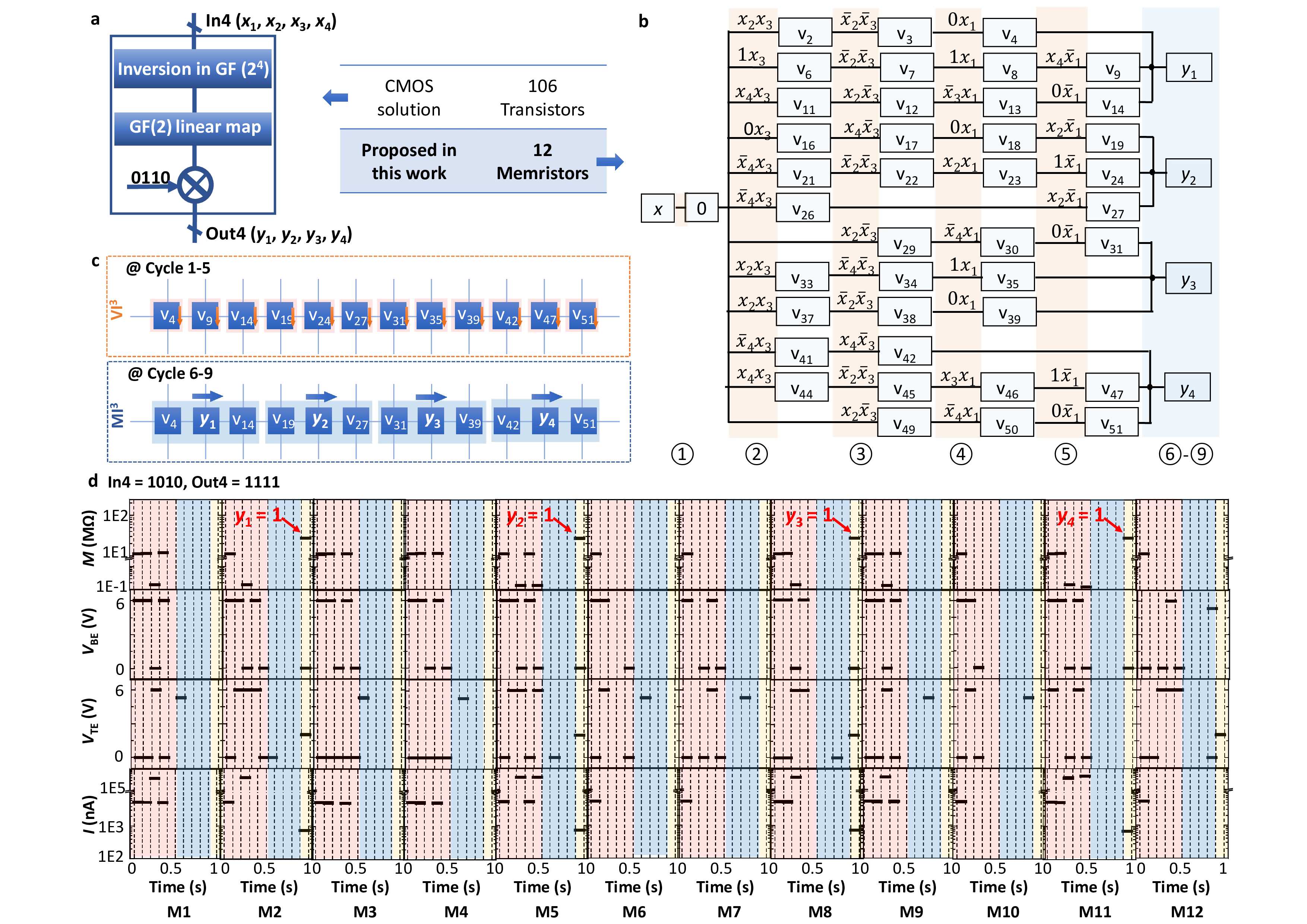}
\caption{Demonstration and implementation of 4-bit S-Box based on automatic synthesis algorithm. (a) Block diagram and comparison of cell numbers between proposed solution and CMOS implementation from Ref. \cite{neugebauer2017building}. GF denotes Galois Field. (b) Control sequence provided by M$^3$S 
using BiFeO$_{\text{3}}$ based memristive crossbar. Each VI$^{3}$ operation is noted as v in the diagram. The red and blue shadowed cycles are VI and MI cycles, respectively. (c) Demonstration of sequential operations with arbitrary inputs. (d) Experimental results with input $A_\text{hex}$ ($x_\text{1-4}$ = 1010) and output $F_\text{hex}$ ($y_\text{1-4}$ = 1111). The results of the memristance of each cell $M_\text{i}$, the voltage on WL to shared bottom electrode $V_\text{WL}$, the voltage on BL to top electrode $V_\text{BLi}$, and the absolute values of the current $|I_\text{i}|$ across each cell during operation are demonstrated. 
}
\label{FigS_Sbox}
\end{figure}

Table S4 in the Supplementary Information I shows that all representative $N$-bit adder designs \cite{lehtonen2009stateful,kvatinsky2013memristor,liu2015signal,talati2016logic,rohani2017improved,thangkhiew2017efficient,mandal2019design,cheng2019functional,rohani2019semiparallel,siemon2019stateful,xu2020stateful,ali2021hybrid,taherinejad2021sixor,fu2022high,kaushik2023imply} are based on either MI or VI kernels. Our mixed-mode approach, being the first to exploit both kernels without current sensing during cascading, showcases an optimal balance between cell and cycle numbers. The only design with a better Area-Delay-Product lacks crossbar compatibility or necessitates substantial structural modifications for practical implementation. We highlight that our adder has been experimentally demonstrated in full, while many previous papers have characterize individual gates experimentally and then plug the results of such measurements into a simulation. Moreover, while our design eliminates current sensing during cascading, a separate set of results for adders incorporating the VO$^3$ operation is found in Supplementary Information G, along with comparable adders from literature \cite{siemon2015complementary,pinto2021robust,reuben2021accelerated,wang2018efficient,siemon2019sklansky,brackmann2024improved}. Supplementary Information H summarizes the energy costs of adders with and without readout. 

The block diagram of the $4$-bit S-Box and its truth table are shown in Figure \ref{FigS_Sbox}a. Note that its input $x_1$-$x_4$ and output $y_1$-$y_4$ are encoded using 4 bits. 
Its control sequence found by M$^3$S and shown in Figure \ref{FigS_Sbox}b,
has the minimum possible number of 4 MI$^{3}$ operations, equal to 4 outputs and 12 cells (the 4 MI cycles are shown in parallel in Figure \ref{FigS_Sbox}b to save space). Figure \ref{FigS_Sbox}c demonstrates the VI$^{3}$ and MI$^{3}$ operations of 4-bit S-Box with arbitrary inputs, where the bitwise parallel operations are performed with shared bottom electrodes in one BL in cycles 1-5 in individual cells, and the MI$^{3}$ operations are executed sequentially in cells M1-M3, M4-M6, M7-M9, and M10-M12, respectively. Compared with 4-bit S-Box exploiting 12 memristive cells by using mixed-mode principle, the prior CMOS implementation in \cite{neugebauer2017building} obtained by Altera synthesis software had roughly 106 transistors (8 two-input gates, 8 three-input gates, 2 four-input gates, 1 five-input gate), and the CMOS circuit using as the basis for the memristive implementation in \cite{chen2022side} had 32 two-input gates or roughly 128 transistors. The experimental results of $4$-bit Sbox on a passive crossbar based on BiFeO$_{\text{3}}$ memristive crossbar with inputs 1010 and 1111 by using the control sequence are presented in Figure \ref{FigS_Sbox}b.

This analysis does not consider the effort for peripherals that supply the top electrode and bottom electrode values. In comparison to peripherals required for other memristive implementations, the mixed-mode VI-MI operation incurs the same effort for reading but might add complexity for input-dependent writing biases. 
While we do not explicitly examine the endurance or retention properties of the memristors in this work, we are encouraging the synthesis procedure to use the variation-resilient VI kernel as much as possible, while resorting to the MI kernel only when this is required for universality. This consideration supports the flexibility of our synthesis approach: once M-mode operations will get more reliable in the future, the designer can simply rerun M$^3$S with more clock cycles allocated to them.

\section*{Conclusions}\label{sec6}

In this work, we propose a mixed-mode in-memory computing paradigm by seamlessly integrating V-mode and M-mode operations within each physical memristive cell. This approach combines the strengths of diverse gate processing (V-mode) and universality (M-mode) while mitigating the impact of device variations inherent in nanodevices. According to our review of current research, this approach represents the first practical implementation enabling complex functions in memristive crossbars, while eliminating the need for high-cost current sensing in peripherals during cascading. 

Moreover, the first-of-its-kind memristor crossbar-oriented mapping and synthesis (M$^3$S) automation tool is developed, enabling flexible combinations of V-mode and M-mode operations for implementing arbitrary logic functions by carrying out synthesis and mapping tasks in one. M$^3$S achieves an optimal balance between latency and area, reducing the typical trade-offs of conventional in-memory computing and enhancing power efficiency. Our experimental demonstrations include a best-in-class $N$-bit carry ripple adder and the first implementation of complex S-Box circuitry with 12 memristive cells. This showcases not only a significant reduction in device count—by an order of magnitude compared to conventional CMOS technology—but also highlights the enhanced robustness and potential of the mixed-mode computing paradigm.

The proposed mixed-mode computing is adaptable and transferable across various nanoscale memory technologies that feature nonvolatile and latching characteristics. In this work, while utilizing passive 1R crossbar array for its potential high density, the future work can extend to the 1T1R crossbar array which allows additional inputs to transistors to serve as logic inputs, thus facilitating advanced logic operations in V and M kernels within the mixed-mode computing framework. Furthermore, the inherent analog computing capabilities of memristive devices enabling the design of ternary and multi-nary logic operations within the mixed-mode computing paradigm not only expands the computational horizons into mixed-nary computing but also heralds a new era of unparalleled efficiency in computing paradigms, promising unmatched computational power and efficacy.

\section*{Methods}
\subsection*{Passive memristive crossbar fabrication}
To fabricate the passive memristive crossbar array based on BiFeO$_3$ thin film series, i.e. BiFeO$_{\text{3}}$, BiFeTiO$_{\text{3}}$, and BiFeTiO$_{\text{3}}$/BiFeO$_{\text{3}}$ thin films, the corresponding polycrystalline BiFeO$_{\text{3}}$ film series are deposited by pulsed laser deposition at 650 °C in oxygen ambient, respectively. Upon the Pt/Ti-bottom electrode patterned by photolithography and ion beam etching, for fabricating the BiFeTiO$_{\text{3}}$ or BiFeO$_{\text{3}}$ memristive crossbar arrays, the Ti doped BiFeO$_{\text{3}}$ thin film or undoped BiFeO$_{\text{3}}$ thin film are deposited on structured Pt (100 nm)/Sapphire and Pt (100 nm)/Ti (50 nm)/Sapphire substrates, respectively. Both BiFeTiO$_3$ and BiFeO$_3$ thin films possess rhombohedrally distorted perovskite structure (R3c space group) and a nominal thickness of 550 nm. In BiFeO$_{\text{3}}$ thin film, the ambient hitting during deposition provokes the substitutional incorporation of Ti donors into the BiFeO$_{\text{3}}$ lattice, which constructs a rectifying/non-rectifying contact with flexible barrier height near to top/bottom electrode region, whereas, in BiFeTiO$_{\text{3}}$ thin film the Ti content of nominal 1 at\% promotes the modulation of the flexible barrier in the top electrode region. The subsequent deposition of BiFeTiO$_{\text{3}}$ (100 nm) and BiFeO$_{\text{3}}$ films (500 nm) on structured Pt/Sapphire substrate sets up the BiFeTiO$_{\text{3}}$/BiFeO$_{\text{3}}$ bilayer structure in BiFeTiO$_{\text{3}}$/BiFeO$_{\text{3}}$ memristive crossbar array with the flexible Schottky-like barriers formed at the top and bottom interfaces for the bilayer structure. The Au top electrode with thickness of 180 nm was evaporated and patterned by photolithography followed by a lift-off. As an example, the SEM image of 32×32 BiFeO$_{\text{3}}$ based crossbar array (with a junction area of each cell as 20×20 µm\textsuperscript{2} and a pitch of 25 µm) is demonstrated in Fig. \ref{fig_2}a.\\

\subsection*{Electrical characterization}
All the experimental electrical measurement illustrated in this work were recorded using a probe station and a Keithley source meter 2400, which is controlled by a home-made PCB board for selecting and applying biases to WLs and BLs through LabVIEW program.

\section*{Data Availability Statement}
The data that support the plots within this paper and other findings of this study are available from the corresponding author upon reasonable request.

\bibliography{sn-bibliography}


\begin{thebibliography}{50}
\ifx \bisbn   \undefined \def \bisbn  #1{ISBN #1}\fi
\ifx \binits  \undefined \def \binits#1{#1}\fi
\ifx \bauthor  \undefined \def \bauthor#1{#1}\fi
\ifx \batitle  \undefined \def \batitle#1{#1}\fi
\ifx \bjtitle  \undefined \def \bjtitle#1{#1}\fi
\ifx \bvolume  \undefined \def \bvolume#1{\textbf{#1}}\fi
\ifx \byear  \undefined \def \byear#1{#1}\fi
\ifx \bissue  \undefined \def \bissue#1{#1}\fi
\ifx \bfpage  \undefined \def \bfpage#1{#1}\fi
\ifx \blpage  \undefined \def \blpage #1{#1}\fi
\ifx \burl  \undefined \def \burl#1{\textsf{#1}}\fi
\ifx \doiurl  \undefined \def \doiurl#1{\url{https://doi.org/#1}}\fi
\ifx \betal  \undefined \def \betal{\textit{et al.}}\fi
\ifx \binstitute  \undefined \def \binstitute#1{#1}\fi
\ifx \binstitutionaled  \undefined \def \binstitutionaled#1{#1}\fi
\ifx \bctitle  \undefined \def \bctitle#1{#1}\fi
\ifx \beditor  \undefined \def \beditor#1{#1}\fi
\ifx \bpublisher  \undefined \def \bpublisher#1{#1}\fi
\ifx \bbtitle  \undefined \def \bbtitle#1{#1}\fi
\ifx \bedition  \undefined \def \bedition#1{#1}\fi
\ifx \bseriesno  \undefined \def \bseriesno#1{#1}\fi
\ifx \blocation  \undefined \def \blocation#1{#1}\fi
\ifx \bsertitle  \undefined \def \bsertitle#1{#1}\fi
\ifx \bsnm \undefined \def \bsnm#1{#1}\fi
\ifx \bsuffix \undefined \def \bsuffix#1{#1}\fi
\ifx \bparticle \undefined \def \bparticle#1{#1}\fi
\ifx \barticle \undefined \def \barticle#1{#1}\fi
\bibcommenthead
\ifx \bconfdate \undefined \def \bconfdate #1{#1}\fi
\ifx \botherref \undefined \def \botherref #1{#1}\fi
\ifx \url \undefined \def \url#1{\textsf{#1}}\fi
\ifx \bchapter \undefined \def \bchapter#1{#1}\fi
\ifx \bbook \undefined \def \bbook#1{#1}\fi
\ifx \bcomment \undefined \def \bcomment#1{#1}\fi
\ifx \oauthor \undefined \def \oauthor#1{#1}\fi
\ifx \citeauthoryear \undefined \def \citeauthoryear#1{#1}\fi
\ifx \endbibitem  \undefined \def \endbibitem {}\fi
\ifx \bconflocation  \undefined \def \bconflocation#1{#1}\fi
\ifx \arxivurl  \undefined \def \arxivurl#1{\textsf{#1}}\fi
\csname PreBibitemsHook\endcsname

\bibitem[\protect\citeauthoryear{Kimovski et~al.}{2023}]{kimovski2023beyond}
\begin{botherref}
\oauthor{\bsnm{Kimovski}, \binits{D.}},
\oauthor{\bsnm{Saurabh}, \binits{N.}},
\oauthor{\bsnm{Jansen}, \binits{M.}},
\oauthor{\bsnm{Aral}, \binits{A.}},
\oauthor{\bsnm{Al-Dulaimy}, \binits{A.}},
\oauthor{\bsnm{Bondi}, \binits{A.B.}},
\oauthor{\bsnm{Galletta}, \binits{A.}},
\oauthor{\bsnm{Papadopoulos}, \binits{A.V.}},
\oauthor{\bsnm{Iosup}, \binits{A.}},
\oauthor{\bsnm{Prodan}, \binits{R.}}:
Beyond von neumann in the computing continuum: Architectures, applications, and future directions.
IEEE Internet Computing
(2023)
\end{botherref}
\endbibitem

\bibitem[\protect\citeauthoryear{Wright et~al.}{2013}]{wright2013beyond}
\begin{barticle}
\bauthor{\bsnm{Wright}, \binits{C.D.}},
\bauthor{\bsnm{Hosseini}, \binits{P.}},
\bauthor{\bsnm{Diosdado}, \binits{J.A.V.}}:
\batitle{Beyond von-neumann computing with nanoscale phase-change memory devices}.
\bjtitle{Advanced Functional Materials}
\bvolume{23}(\bissue{18}),
\bfpage{2248}--\blpage{2254}
(\byear{2013})
\end{barticle}
\endbibitem

\bibitem[\protect\citeauthoryear{Di~Ventra}{2022}]{di2022memcomputing}
\begin{bbook}
\bauthor{\bsnm{Di~Ventra}, \binits{M.}}:
\bbtitle{MemComputing: Fundamentals and Applications}.
\bpublisher{Oxford University Press, Oxford},
\blocation{Oxford, UK}
(\byear{2022})
\end{bbook}
\endbibitem

\bibitem[\protect\citeauthoryear{Sebastian et~al.}{2020}]{sebastian2020memory}
\begin{barticle}
\bauthor{\bsnm{Sebastian}, \binits{A.}},
\bauthor{\bsnm{Le~Gallo}, \binits{M.}},
\bauthor{\bsnm{Khaddam-Aljameh}, \binits{R.}},
\bauthor{\bsnm{Eleftheriou}, \binits{E.}}:
\batitle{Memory devices and applications for in-memory computing}.
\bjtitle{Nature nanotechnology}
\bvolume{15}(\bissue{7}),
\bfpage{529}--\blpage{544}
(\byear{2020})
\end{barticle}
\endbibitem

\bibitem[\protect\citeauthoryear{Finocchio et~al.}{2023}]{finocchio2023roadmap}
\begin{botherref}
\oauthor{\bsnm{Finocchio}, \binits{G.}},
\oauthor{\bsnm{Incorvia}, \binits{J.A.C.}},
\oauthor{\bsnm{Friedman}, \binits{J.S.}},
\oauthor{\bsnm{Yang}, \binits{Q.}},
\oauthor{\bsnm{Giordano}, \binits{A.}},
\oauthor{\bsnm{Grollier}, \binits{J.}},
\oauthor{\bsnm{Yang}, \binits{H.}},
\oauthor{\bsnm{Ciubotaru}, \binits{F.}},
\oauthor{\bsnm{Chumak}, \binits{A.}},
\oauthor{\bsnm{Naeemi}, \binits{A.}}, et al.:
Roadmap for unconventional computing with nanotechnology.
Nano Futures
(2023)
\end{botherref}
\endbibitem

\bibitem[\protect\citeauthoryear{Song et~al.}{2024}]{song2024programming}
\begin{barticle}
\bauthor{\bsnm{Song}, \binits{W.}},
\bauthor{\bsnm{Rao}, \binits{M.}},
\bauthor{\bsnm{Li}, \binits{Y.}},
\bauthor{\bsnm{Li}, \binits{C.}},
\bauthor{\bsnm{Zhuo}, \binits{Y.}},
\bauthor{\bsnm{Cai}, \binits{F.}},
\bauthor{\bsnm{Wu}, \binits{M.}},
\bauthor{\bsnm{Yin}, \binits{W.}},
\bauthor{\bsnm{Li}, \binits{Z.}},
\bauthor{\bsnm{Wei}, \binits{Q.}}, \betal:
\batitle{Programming memristor arrays with arbitrarily high precision for analog computing}.
\bjtitle{Science}
\bvolume{383}(\bissue{6685}),
\bfpage{903}--\blpage{910}
(\byear{2024})
\end{barticle}
\endbibitem

\bibitem[\protect\citeauthoryear{Rao et~al.}{2023}]{rao2023thousands}
\begin{barticle}
\bauthor{\bsnm{Rao}, \binits{M.}},
\bauthor{\bsnm{Tang}, \binits{H.}},
\bauthor{\bsnm{Wu}, \binits{J.}},
\bauthor{\bsnm{Song}, \binits{W.}},
\bauthor{\bsnm{Zhang}, \binits{M.}},
\bauthor{\bsnm{Yin}, \binits{W.}},
\bauthor{\bsnm{Zhuo}, \binits{Y.}},
\bauthor{\bsnm{Kiani}, \binits{F.}},
\bauthor{\bsnm{Chen}, \binits{B.}},
\bauthor{\bsnm{Jiang}, \binits{X.}}, \betal:
\batitle{Thousands of conductance levels in memristors integrated on cmos}.
\bjtitle{Nature}
\bvolume{615}(\bissue{7954}),
\bfpage{823}--\blpage{829}
(\byear{2023})
\end{barticle}
\endbibitem

\bibitem[\protect\citeauthoryear{Pi et~al.}{2019}]{pi2019memristor}
\begin{barticle}
\bauthor{\bsnm{Pi}, \binits{S.}},
\bauthor{\bsnm{Li}, \binits{C.}},
\bauthor{\bsnm{Jiang}, \binits{H.}},
\bauthor{\bsnm{Xia}, \binits{W.}},
\bauthor{\bsnm{Xin}, \binits{H.}},
\bauthor{\bsnm{Yang}, \binits{J.J.}},
\bauthor{\bsnm{Xia}, \binits{Q.}}:
\batitle{Memristor crossbar arrays with 6-nm half-pitch and 2-nm critical dimension}.
\bjtitle{Nature nanotechnology}
\bvolume{14}(\bissue{1}),
\bfpage{35}--\blpage{39}
(\byear{2019})
\end{barticle}
\endbibitem

\bibitem[\protect\citeauthoryear{Liu et~al.}{2023}]{liu2023highly}
\begin{barticle}
\bauthor{\bsnm{Liu}, \binits{L.}},
\bauthor{\bsnm{Geng}, \binits{B.}},
\bauthor{\bsnm{Ji}, \binits{W.}},
\bauthor{\bsnm{Wu}, \binits{L.}},
\bauthor{\bsnm{Lei}, \binits{S.}},
\bauthor{\bsnm{Hu}, \binits{W.}}:
\batitle{A highly crystalline single layer 2d polymer for low variability and excellent scalability molecular memristors}.
\bjtitle{Advanced Materials}
\bvolume{35}(\bissue{6}),
\bfpage{2208377}
(\byear{2023})
\end{barticle}
\endbibitem

\bibitem[\protect\citeauthoryear{Borghetti et~al.}{2010}]{borghetti2010memristive}
\begin{barticle}
\bauthor{\bsnm{Borghetti}, \binits{J.}},
\bauthor{\bsnm{Snider}, \binits{G.S.}},
\bauthor{\bsnm{Kuekes}, \binits{P.J.}},
\bauthor{\bsnm{Yang}, \binits{J.J.}},
\bauthor{\bsnm{Stewart}, \binits{D.R.}},
\bauthor{\bsnm{Williams}, \binits{R.S.}}:
\batitle{‘memristive’switches enable ‘stateful’logic operations via material implication}.
\bjtitle{Nature}
\bvolume{464}(\bissue{7290}),
\bfpage{873}--\blpage{876}
(\byear{2010})
\end{barticle}
\endbibitem

\bibitem[\protect\citeauthoryear{Kvatinsky et~al.}{2014}]{kvatinsky2014magic}
\begin{barticle}
\bauthor{\bsnm{Kvatinsky}, \binits{S.}},
\bauthor{\bsnm{Belousov}, \binits{D.}},
\bauthor{\bsnm{Liman}, \binits{S.}},
\bauthor{\bsnm{Satat}, \binits{G.}},
\bauthor{\bsnm{Wald}, \binits{N.}},
\bauthor{\bsnm{Friedman}, \binits{E.G.}},
\bauthor{\bsnm{Kolodny}, \binits{A.}},
\bauthor{\bsnm{Weiser}, \binits{U.C.}}:
\batitle{Magic—memristor-aided logic}.
\bjtitle{IEEE Transactions on Circuits and Systems II: Express Briefs}
\bvolume{61}(\bissue{11}),
\bfpage{895}--\blpage{899}
(\byear{2014})
\end{barticle}
\endbibitem

\bibitem[\protect\citeauthoryear{TaheriNejad}{2021}]{taherinejad2021sixor}
\begin{barticle}
\bauthor{\bsnm{TaheriNejad}, \binits{N.}}:
\batitle{Sixor: Single-cycle in-memristor xor}.
\bjtitle{IEEE Transactions on Very Large Scale Integration (VLSI) Systems}
\bvolume{29}(\bissue{5}),
\bfpage{925}--\blpage{935}
(\byear{2021})
\end{barticle}
\endbibitem

\bibitem[\protect\citeauthoryear{Gupta et~al.}{2018}]{gupta2018felix}
\begin{bchapter}
\bauthor{\bsnm{Gupta}, \binits{S.}},
\bauthor{\bsnm{Imani}, \binits{M.}},
\bauthor{\bsnm{Rosing}, \binits{T.}}:
\bctitle{Felix: Fast and energy-efficient logic in memory}.
In: \bbtitle{2018 IEEE/ACM International Conference on Computer-Aided Design (ICCAD)},
pp. \bfpage{1}--\blpage{7}
(\byear{2018}).
\bcomment{IEEE}
\end{bchapter}
\endbibitem

\bibitem[\protect\citeauthoryear{In et~al.}{2020}]{in2020universal}
\begin{barticle}
\bauthor{\bsnm{In}, \binits{J.H.}},
\bauthor{\bsnm{Kim}, \binits{Y.S.}},
\bauthor{\bsnm{Song}, \binits{H.}},
\bauthor{\bsnm{Kim}, \binits{G.M.}},
\bauthor{\bsnm{An}, \binits{J.}},
\bauthor{\bsnm{Jeon}, \binits{J.B.}},
\bauthor{\bsnm{Kim}, \binits{K.M.}}:
\batitle{A universal error correction method for memristive stateful logic devices for practical near-memory computing}.
\bjtitle{Advanced Intelligent Systems}
\bvolume{2}(\bissue{9}),
\bfpage{2000081}
(\byear{2020})
\end{barticle}
\endbibitem

\bibitem[\protect\citeauthoryear{Kim et~al.}{2020}]{kim2020stateful}
\begin{barticle}
\bauthor{\bsnm{Kim}, \binits{Y.S.}},
\bauthor{\bsnm{Son}, \binits{M.W.}},
\bauthor{\bsnm{Song}, \binits{H.}},
\bauthor{\bsnm{Park}, \binits{J.}},
\bauthor{\bsnm{An}, \binits{J.}},
\bauthor{\bsnm{Jeon}, \binits{J.B.}},
\bauthor{\bsnm{Kim}, \binits{G.Y.}},
\bauthor{\bsnm{Son}, \binits{S.}},
\bauthor{\bsnm{Kim}, \binits{K.M.}}:
\batitle{Stateful in-memory logic system and its practical implementation in a taox-based bipolar-type memristive crossbar array}.
\bjtitle{Advanced Intelligent Systems}
\bvolume{2}(\bissue{3}),
\bfpage{1900156}
(\byear{2020})
\end{barticle}
\endbibitem

\bibitem[\protect\citeauthoryear{Kopperberg et~al.}{2022}]{kopperberg2022endurance}
\begin{barticle}
\bauthor{\bsnm{Kopperberg}, \binits{N.}},
\bauthor{\bsnm{Wiefels}, \binits{S.}},
\bauthor{\bsnm{Hofmann}, \binits{K.}},
\bauthor{\bsnm{Otterstedt}, \binits{J.}},
\bauthor{\bsnm{Wouters}, \binits{D.J.}},
\bauthor{\bsnm{Waser}, \binits{R.}},
\bauthor{\bsnm{Menzel}, \binits{S.}}:
\batitle{Endurance of 2 mbit based beol integrated reram}.
\bjtitle{IEEE Access}
\bvolume{10},
\bfpage{122696}--\blpage{122705}
(\byear{2022})
\end{barticle}
\endbibitem

\bibitem[\protect\citeauthoryear{Sch{\"o}n and Menzel}{2023}]{schon2023spatio}
\begin{barticle}
\bauthor{\bsnm{Sch{\"o}n}, \binits{D.}},
\bauthor{\bsnm{Menzel}, \binits{S.}}:
\batitle{Spatio-temporal correlations in memristive crossbar arrays due to thermal effects}.
\bjtitle{Advanced functional materials}
\bvolume{33}(\bissue{22}),
\bfpage{2213943}
(\byear{2023})
\end{barticle}
\endbibitem

\bibitem[\protect\citeauthoryear{Staudigl et~al.}{2024}]{staudigl2024s}
\begin{botherref}
\oauthor{\bsnm{Staudigl}, \binits{F.}},
\oauthor{\bsnm{Al~Indari}, \binits{H.}},
\oauthor{\bsnm{Sch{\"o}n}, \binits{D.}},
\oauthor{\bsnm{Chen}, \binits{H.-Y.}},
\oauthor{\bsnm{Sisejkovic}, \binits{D.}},
\oauthor{\bsnm{Joseph}, \binits{J.M.}},
\oauthor{\bsnm{Rana}, \binits{V.}},
\oauthor{\bsnm{Menzel}, \binits{S.}},
\oauthor{\bsnm{Hagelauer}, \binits{A.}},
\oauthor{\bsnm{Leupers}, \binits{R.}}:
It's getting hot in here: Hardware security implications of thermal crosstalk on rerams.
IEEE Transactions on Reliability
(2024)
\end{botherref}
\endbibitem

\bibitem[\protect\citeauthoryear{Papandroulidakis et~al.}{2014}]{papandroulidakis2014boolean}
\begin{barticle}
\bauthor{\bsnm{Papandroulidakis}, \binits{G.}},
\bauthor{\bsnm{Vourkas}, \binits{I.}},
\bauthor{\bsnm{Vasileiadis}, \binits{N.}},
\bauthor{\bsnm{Sirakoulis}, \binits{G.C.}}:
\batitle{Boolean logic operations and computing circuits based on memristors}.
\bjtitle{IEEE Transactions on Circuits and Systems II: Express Briefs}
\bvolume{61}(\bissue{12}),
\bfpage{972}--\blpage{976}
(\byear{2014})
\end{barticle}
\endbibitem

\bibitem[\protect\citeauthoryear{Linn et~al.}{2010}]{linn2010complementary}
\begin{barticle}
\bauthor{\bsnm{Linn}, \binits{E.}},
\bauthor{\bsnm{Rosezin}, \binits{R.}},
\bauthor{\bsnm{K{\"u}geler}, \binits{C.}},
\bauthor{\bsnm{Waser}, \binits{R.}}:
\batitle{Complementary resistive switches for passive nanocrossbar memories}.
\bjtitle{Nature materials}
\bvolume{9}(\bissue{5}),
\bfpage{403}--\blpage{406}
(\byear{2010})
\end{barticle}
\endbibitem

\bibitem[\protect\citeauthoryear{Xie et~al.}{2017}]{xie2017scouting}
\begin{bchapter}
\bauthor{\bsnm{Xie}, \binits{L.}},
\bauthor{\bsnm{Du~Nguyen}, \binits{H.A.}},
\bauthor{\bsnm{Yu}, \binits{J.}},
\bauthor{\bsnm{Kaichouhi}, \binits{A.}},
\bauthor{\bsnm{Taouil}, \binits{M.}},
\bauthor{\bsnm{AlFailakawi}, \binits{M.}},
\bauthor{\bsnm{Hamdioui}, \binits{S.}}:
\bctitle{Scouting logic: A novel memristor-based logic design for resistive computing}.
In: \bbtitle{2017 IEEE Computer Society Annual Symposium on VLSI (ISVLSI)},
pp. \bfpage{176}--\blpage{181}
(\byear{2017}).
\bcomment{IEEE}
\end{bchapter}
\endbibitem

\bibitem[\protect\citeauthoryear{Ben~Hur et~al.}{2017}]{8203782}
\begin{bchapter}
\bauthor{\bsnm{Ben~Hur}, \binits{R.}},
\bauthor{\bsnm{Wald}, \binits{N.}},
\bauthor{\bsnm{Talati}, \binits{N.}},
\bauthor{\bsnm{Kvatinsky}, \binits{S.}}:
\bctitle{Simple magic: Synthesis and in-memory mapping of logic execution for memristor-aided logic}.
In: \bbtitle{2017 IEEE/ACM International Conference on Computer-Aided Design (ICCAD)},
pp. \bfpage{225}--\blpage{232}
(\byear{2017}).
\doiurl{10.1109/ICCAD.2017.8203782}
\end{bchapter}
\endbibitem

\bibitem[\protect\citeauthoryear{Rohani et~al.}{2019}]{rohani2019semiparallel}
\begin{barticle}
\bauthor{\bsnm{Rohani}, \binits{S.G.}},
\bauthor{\bsnm{Taherinejad}, \binits{N.}},
\bauthor{\bsnm{Radakovits}, \binits{D.}}:
\batitle{A semiparallel full-adder in imply logic}.
\bjtitle{IEEE Transactions on Very Large Scale Integration (VLSI) Systems}
\bvolume{28}(\bissue{1}),
\bfpage{297}--\blpage{301}
(\byear{2019})
\end{barticle}
\endbibitem

\bibitem[\protect\citeauthoryear{Fayyazi et~al.}{2020}]{fayyazi2020hipe}
\begin{bchapter}
\bauthor{\bsnm{Fayyazi}, \binits{A.}},
\bauthor{\bsnm{Esmaili}, \binits{A.}},
\bauthor{\bsnm{Pedram}, \binits{M.}}:
\bctitle{Hipe-magic: a technology-aware synthesis and mapping flow for highly parallel execution of memristor-aided logic}.
In: \bbtitle{Proceedings of the ACM/IEEE International Symposium on Low Power Electronics and Design},
pp. \bfpage{235}--\blpage{240}
(\byear{2020})
\end{bchapter}
\endbibitem

\bibitem[\protect\citeauthoryear{Testa et~al.}{2019}]{TSAM:19}
\begin{barticle}
\bauthor{\bsnm{Testa}, \binits{E.}},
\bauthor{\bsnm{Soeken}, \binits{M.}},
\bauthor{\bsnm{Amar{\`{u}}}, \binits{L.G.}},
\bauthor{\bsnm{Micheli}, \binits{G.D.}}:
\batitle{Logic synthesis for established and emerging computing}.
\bjtitle{Proc. {IEEE}}
\bvolume{107}(\bissue{1}),
\bfpage{165}--\blpage{184}
(\byear{2019})
\doiurl{10.1109/JPROC.2018.2869760}
\end{barticle}
\endbibitem

\bibitem[\protect\citeauthoryear{Chakraborti et~al.}{2014}]{CCDS:14}
\begin{bchapter}
\bauthor{\bsnm{Chakraborti}, \binits{S.}}, \betal:
\bctitle{{BDD} based synthesis of {Boolean} functions using memristors}.
In: \bbtitle{Int'l Design \& Test Symp. (IDT)},
pp. \bfpage{136}--\blpage{141}
(\byear{2014}).
\doiurl{10.1109/IDT.2014.7038601}
\end{bchapter}
\endbibitem

\bibitem[\protect\citeauthoryear{Thangkhiew et~al.}{2020}]{TZW+:20}
\begin{barticle}
\bauthor{\bsnm{Thangkhiew}, \binits{P.L.}},
\bauthor{\bsnm{Zulehner}, \binits{A.}},
\bauthor{\bsnm{Wille}, \binits{R.}},
\bauthor{\bsnm{Datta}, \binits{K.}},
\bauthor{\bsnm{Sengupta}, \binits{I.}}:
\batitle{An efficient memristor crossbar architecture for mapping boolean functions using binary decision diagrams {(BDD)}}.
\bjtitle{Integr.}
\bvolume{71},
\bfpage{125}--\blpage{133}
(\byear{2020})
\end{barticle}
\endbibitem

\bibitem[\protect\citeauthoryear{Ben-Hur et~al.}{2019}]{ben2019simpler}
\begin{barticle}
\bauthor{\bsnm{Ben-Hur}, \binits{R.}},
\bauthor{\bsnm{Ronen}, \binits{R.}},
\bauthor{\bsnm{Haj-Ali}, \binits{A.}},
\bauthor{\bsnm{Bhattacharjee}, \binits{D.}},
\bauthor{\bsnm{Eliahu}, \binits{A.}},
\bauthor{\bsnm{Peled}, \binits{N.}},
\bauthor{\bsnm{Kvatinsky}, \binits{S.}}:
\batitle{Simpler magic: Synthesis and mapping of in-memory logic executed in a single row to improve throughput}.
\bjtitle{IEEE Transactions on Computer-Aided Design of Integrated Circuits and Systems}
\bvolume{39}(\bissue{10}),
\bfpage{2434}--\blpage{2447}
(\byear{2019})
\end{barticle}
\endbibitem

\bibitem[\protect\citeauthoryear{Cid et~al.}{2005}]{cid2005small}
\begin{bchapter}
\bauthor{\bsnm{Cid}, \binits{C.}},
\bauthor{\bsnm{Murphy}, \binits{S.}},
\bauthor{\bsnm{Robshaw}, \binits{M.J.}}:
\bctitle{Small scale variants of the aes}.
In: \bbtitle{International Workshop on Fast Software Encryption},
pp. \bfpage{145}--\blpage{162}
(\byear{2005}).
\bcomment{Springer}
\end{bchapter}
\endbibitem

\bibitem[\protect\citeauthoryear{Neugebauer et~al.}{2017}]{neugebauer2017building}
\begin{bchapter}
\bauthor{\bsnm{Neugebauer}, \binits{F.}},
\bauthor{\bsnm{Polian}, \binits{I.}},
\bauthor{\bsnm{Hayes}, \binits{J.P.}}:
\bctitle{Building a better random number generator for stochastic computing}.
In: \bbtitle{2017 Euromicro Conference on Digital System Design (DSD)},
pp. \bfpage{1}--\blpage{8}
(\byear{2017}).
\bcomment{IEEE}
\end{bchapter}
\endbibitem

\bibitem[\protect\citeauthoryear{Lehtonen and Laiho}{2009}]{lehtonen2009stateful}
\begin{bchapter}
\bauthor{\bsnm{Lehtonen}, \binits{E.}},
\bauthor{\bsnm{Laiho}, \binits{M.}}:
\bctitle{Stateful implication logic with memristors}.
In: \bbtitle{2009 IEEE/ACM International Symposium on Nanoscale Architectures},
pp. \bfpage{33}--\blpage{36}
(\byear{2009}).
\bcomment{IEEE}
\end{bchapter}
\endbibitem

\bibitem[\protect\citeauthoryear{Kvatinsky et~al.}{2013}]{kvatinsky2013memristor}
\begin{barticle}
\bauthor{\bsnm{Kvatinsky}, \binits{S.}},
\bauthor{\bsnm{Satat}, \binits{G.}},
\bauthor{\bsnm{Wald}, \binits{N.}},
\bauthor{\bsnm{Friedman}, \binits{E.G.}},
\bauthor{\bsnm{Kolodny}, \binits{A.}},
\bauthor{\bsnm{Weiser}, \binits{U.C.}}:
\batitle{Memristor-based material implication (imply) logic: Design principles and methodologies}.
\bjtitle{IEEE Transactions on Very Large Scale Integration (VLSI) Systems}
\bvolume{22}(\bissue{10}),
\bfpage{2054}--\blpage{2066}
(\byear{2013})
\end{barticle}
\endbibitem

\bibitem[\protect\citeauthoryear{Liu et~al.}{2015}]{liu2015signal}
\begin{barticle}
\bauthor{\bsnm{Liu}, \binits{B.}},
\bauthor{\bsnm{Wang}, \binits{Y.}},
\bauthor{\bsnm{You}, \binits{Z.}},
\bauthor{\bsnm{Han}, \binits{Y.}},
\bauthor{\bsnm{Li}, \binits{X.}}:
\batitle{A signal degradation reduction method for memristor ratioed logic (mrl) gates}.
\bjtitle{IEICE Electronics Express}
\bvolume{12}(\bissue{8}),
\bfpage{20150062}--\blpage{20150062}
(\byear{2015})
\end{barticle}
\endbibitem

\bibitem[\protect\citeauthoryear{Talati et~al.}{2016}]{talati2016logic}
\begin{barticle}
\bauthor{\bsnm{Talati}, \binits{N.}},
\bauthor{\bsnm{Gupta}, \binits{S.}},
\bauthor{\bsnm{Mane}, \binits{P.}},
\bauthor{\bsnm{Kvatinsky}, \binits{S.}}:
\batitle{Logic design within memristive memories using memristor-aided logic (magic)}.
\bjtitle{IEEE Transactions on Nanotechnology}
\bvolume{15}(\bissue{4}),
\bfpage{635}--\blpage{650}
(\byear{2016})
\end{barticle}
\endbibitem

\bibitem[\protect\citeauthoryear{Rohani and TaheriNejad}{2017}]{rohani2017improved}
\begin{bchapter}
\bauthor{\bsnm{Rohani}, \binits{S.G.}},
\bauthor{\bsnm{TaheriNejad}, \binits{N.}}:
\bctitle{An improved algorithm for imply logic based memristive full-adder}.
In: \bbtitle{2017 IEEE 30th Canadian Conference on Electrical and Computer Engineering (CCECE)},
pp. \bfpage{1}--\blpage{4}
(\byear{2017}).
\bcomment{IEEE}
\end{bchapter}
\endbibitem

\bibitem[\protect\citeauthoryear{Thangkhiew et~al.}{2017}]{thangkhiew2017efficient}
\begin{bchapter}
\bauthor{\bsnm{Thangkhiew}, \binits{P.}},
\bauthor{\bsnm{Gharpinde}, \binits{R.}},
\bauthor{\bsnm{Yadav}, \binits{D.N.}},
\bauthor{\bsnm{Datta}, \binits{K.}},
\bauthor{\bsnm{Sengupta}, \binits{I.}}:
\bctitle{Efficient implementation of adder circuits in memristive crossbar array}.
In: \bbtitle{TENCON 2017-2017 IEEE Region 10 Conference},
pp. \bfpage{207}--\blpage{212}
(\byear{2017}).
\bcomment{IEEE}
\end{bchapter}
\endbibitem

\bibitem[\protect\citeauthoryear{Mandal et~al.}{2019}]{mandal2019design}
\begin{bchapter}
\bauthor{\bsnm{Mandal}, \binits{S.}},
\bauthor{\bsnm{Sinha}, \binits{J.}},
\bauthor{\bsnm{Chakraborty}, \binits{A.}}:
\bctitle{Design of memristor--cmos based logic gates and logic circuits}.
In: \bbtitle{2019 2nd International Conference on Innovations in Electronics, Signal Processing and Communication (IESc)},
pp. \bfpage{215}--\blpage{220}
(\byear{2019}).
\bcomment{IEEE}
\end{bchapter}
\endbibitem

\bibitem[\protect\citeauthoryear{Cheng et~al.}{2019}]{cheng2019functional}
\begin{barticle}
\bauthor{\bsnm{Cheng}, \binits{L.}},
\bauthor{\bsnm{Li}, \binits{Y.}},
\bauthor{\bsnm{Yin}, \binits{K.-S.}},
\bauthor{\bsnm{Hu}, \binits{S.-Y.}},
\bauthor{\bsnm{Su}, \binits{Y.-T.}},
\bauthor{\bsnm{Jin}, \binits{M.-M.}},
\bauthor{\bsnm{Wang}, \binits{Z.-R.}},
\bauthor{\bsnm{Chang}, \binits{T.-C.}},
\bauthor{\bsnm{Miao}, \binits{X.-S.}}:
\batitle{Functional demonstration of a memristive arithmetic logic unit (memalu) for in-memory computing}.
\bjtitle{Advanced Functional Materials}
\bvolume{29}(\bissue{49}),
\bfpage{1905660}
(\byear{2019})
\end{barticle}
\endbibitem

\bibitem[\protect\citeauthoryear{Siemon et~al.}{2019}]{siemon2019stateful}
\begin{barticle}
\bauthor{\bsnm{Siemon}, \binits{A.}},
\bauthor{\bsnm{Drabinski}, \binits{R.}},
\bauthor{\bsnm{Schultis}, \binits{M.}},
\bauthor{\bsnm{Hu}, \binits{X.}},
\bauthor{\bsnm{Linn}, \binits{E.}},
\bauthor{\bsnm{Heittmann}, \binits{A.}},
\bauthor{\bsnm{Waser}, \binits{R.}},
\bauthor{\bsnm{Querlioz}, \binits{D.}},
\bauthor{\bsnm{Menzel}, \binits{S.}},
\bauthor{\bsnm{Friedman}, \binits{J.}}:
\batitle{Stateful three-input logic with memristive switches}.
\bjtitle{Scientific reports}
\bvolume{9}(\bissue{1}),
\bfpage{14618}
(\byear{2019})
\end{barticle}
\endbibitem

\bibitem[\protect\citeauthoryear{Xu et~al.}{2020}]{xu2020stateful}
\begin{barticle}
\bauthor{\bsnm{Xu}, \binits{N.}},
\bauthor{\bsnm{Park}, \binits{T.G.}},
\bauthor{\bsnm{Kim}, \binits{H.J.}},
\bauthor{\bsnm{Shao}, \binits{X.}},
\bauthor{\bsnm{Yoon}, \binits{K.J.}},
\bauthor{\bsnm{Park}, \binits{T.H.}},
\bauthor{\bsnm{Fang}, \binits{L.}},
\bauthor{\bsnm{Kim}, \binits{K.M.}},
\bauthor{\bsnm{Hwang}, \binits{C.S.}}:
\batitle{A stateful logic family based on a new logic primitive circuit composed of two antiparallel bipolar memristors}.
\bjtitle{Advanced Intelligent Systems}
\bvolume{2}(\bissue{1}),
\bfpage{1900082}
(\byear{2020})
\end{barticle}
\endbibitem

\bibitem[\protect\citeauthoryear{Ali et~al.}{2021}]{ali2021hybrid}
\begin{barticle}
\bauthor{\bsnm{Ali}, \binits{K.A.}},
\bauthor{\bsnm{Rizk}, \binits{M.}},
\bauthor{\bsnm{Baghdadi}, \binits{A.}},
\bauthor{\bsnm{Diguet}, \binits{J.-P.}},
\bauthor{\bsnm{Jomaah}, \binits{J.}}:
\batitle{Hybrid memristor--cmos implementation of combinational logic based on x-mrl}.
\bjtitle{Electronics}
\bvolume{10}(\bissue{9}),
\bfpage{1018}
(\byear{2021})
\end{barticle}
\endbibitem

\bibitem[\protect\citeauthoryear{Fu et~al.}{2022}]{fu2022high}
\begin{barticle}
\bauthor{\bsnm{Fu}, \binits{X.}},
\bauthor{\bsnm{Li}, \binits{Q.}},
\bauthor{\bsnm{Wang}, \binits{W.}},
\bauthor{\bsnm{Xu}, \binits{H.}},
\bauthor{\bsnm{Wang}, \binits{Y.}},
\bauthor{\bsnm{Wang}, \binits{W.}},
\bauthor{\bsnm{Yu}, \binits{H.}},
\bauthor{\bsnm{Li}, \binits{Z.}}:
\batitle{High-speed memristor-based ripple carry adders in 1t1r array structure}.
\bjtitle{IEEE Transactions on Circuits and Systems II: Express Briefs}
\bvolume{69}(\bissue{9}),
\bfpage{3889}--\blpage{3893}
(\byear{2022})
\end{barticle}
\endbibitem

\bibitem[\protect\citeauthoryear{Kaushik and Srinivasu}{2023}]{kaushik2023imply}
\begin{botherref}
\oauthor{\bsnm{Kaushik}, \binits{N.}},
\oauthor{\bsnm{Srinivasu}, \binits{B.}}:
Imply-based high-speed conditional carry and carry select adders for in-memory computing.
IEEE Transactions on Nanotechnology
(2023)
\end{botherref}
\endbibitem

\bibitem[\protect\citeauthoryear{Siemon et~al.}{2015}]{siemon2015complementary}
\begin{barticle}
\bauthor{\bsnm{Siemon}, \binits{A.}},
\bauthor{\bsnm{Menzel}, \binits{S.}},
\bauthor{\bsnm{Waser}, \binits{R.}},
\bauthor{\bsnm{Linn}, \binits{E.}}:
\batitle{A complementary resistive switch-based crossbar array adder}.
\bjtitle{IEEE journal on emerging and selected topics in circuits and systems}
\bvolume{5}(\bissue{1}),
\bfpage{64}--\blpage{74}
(\byear{2015})
\end{barticle}
\endbibitem

\bibitem[\protect\citeauthoryear{Pinto and Vourkas}{2021}]{pinto2021robust}
\begin{barticle}
\bauthor{\bsnm{Pinto}, \binits{F.}},
\bauthor{\bsnm{Vourkas}, \binits{I.}}:
\batitle{Robust circuit and system design for general-purpose computational resistive memories}.
\bjtitle{Electronics}
\bvolume{10}(\bissue{9}),
\bfpage{1074}
(\byear{2021})
\end{barticle}
\endbibitem

\bibitem[\protect\citeauthoryear{Reuben and Pechmann}{2021}]{reuben2021accelerated}
\begin{barticle}
\bauthor{\bsnm{Reuben}, \binits{J.}},
\bauthor{\bsnm{Pechmann}, \binits{S.}}:
\batitle{Accelerated addition in resistive ram array using parallel-friendly majority gates}.
\bjtitle{IEEE Transactions on Very Large Scale Integration (VLSI) Systems}
\bvolume{29}(\bissue{6}),
\bfpage{1108}--\blpage{1121}
(\byear{2021})
\end{barticle}
\endbibitem

\bibitem[\protect\citeauthoryear{Wang et~al.}{2018}]{wang2018efficient}
\begin{barticle}
\bauthor{\bsnm{Wang}, \binits{Z.-R.}},
\bauthor{\bsnm{Li}, \binits{Y.}},
\bauthor{\bsnm{Su}, \binits{Y.-T.}},
\bauthor{\bsnm{Zhou}, \binits{Y.-X.}},
\bauthor{\bsnm{Cheng}, \binits{L.}},
\bauthor{\bsnm{Chang}, \binits{T.-C.}},
\bauthor{\bsnm{Xue}, \binits{K.-H.}},
\bauthor{\bsnm{Sze}, \binits{S.M.}},
\bauthor{\bsnm{Miao}, \binits{X.-S.}}:
\batitle{Efficient implementation of boolean and full-adder functions with 1t1r rrams for beyond von neumann in-memory computing}.
\bjtitle{IEEE Transactions on Electron Devices}
\bvolume{65}(\bissue{10}),
\bfpage{4659}--\blpage{4666}
(\byear{2018})
\end{barticle}
\endbibitem

\bibitem[\protect\citeauthoryear{Siemon et~al.}{2019}]{siemon2019sklansky}
\begin{barticle}
\bauthor{\bsnm{Siemon}, \binits{A.}},
\bauthor{\bsnm{Menzel}, \binits{S.}},
\bauthor{\bsnm{Bhattacharjee}, \binits{D.}},
\bauthor{\bsnm{Waser}, \binits{R.}},
\bauthor{\bsnm{Chattopadhyay}, \binits{A.}},
\bauthor{\bsnm{Linn}, \binits{E.}}:
\batitle{Sklansky tree adder realization in 1s1r resistive switching memory architecture}.
\bjtitle{The European Physical Journal Special Topics}
\bvolume{228},
\bfpage{2269}--\blpage{2285}
(\byear{2019})
\end{barticle}
\endbibitem

\bibitem[\protect\citeauthoryear{Brackmann et~al.}{2024}]{brackmann2024improved}
\begin{barticle}
\bauthor{\bsnm{Brackmann}, \binits{L.}},
\bauthor{\bsnm{Ziegler}, \binits{T.}},
\bauthor{\bsnm{Jafari}, \binits{A.}},
\bauthor{\bsnm{Wouters}, \binits{D.J.}},
\bauthor{\bsnm{Tahoori}, \binits{M.B.}},
\bauthor{\bsnm{Menzel}, \binits{S.}}:
\batitle{Improved arithmetic performance by combining stateful and non-stateful logic in resistive random access memory 1t--1r crossbars}.
\bjtitle{Advanced Intelligent Systems}
\bvolume{6}(\bissue{3}),
\bfpage{2300579}
(\byear{2024})
\end{barticle}
\endbibitem

\bibitem[\protect\citeauthoryear{Chen et~al.}{2022}]{chen2022side}
\begin{barticle}
\bauthor{\bsnm{Chen}, \binits{L.-W.}},
\bauthor{\bsnm{Chen}, \binits{Z.}},
\bauthor{\bsnm{Schindler}, \binits{W.}},
\bauthor{\bsnm{Zhao}, \binits{X.}},
\bauthor{\bsnm{Schmidt}, \binits{H.}},
\bauthor{\bsnm{Du}, \binits{N.}},
\bauthor{\bsnm{Polian}, \binits{I.}}:
\batitle{On side-channel analysis of memristive cryptographic circuits}.
\bjtitle{IEEE Transactions on Information Forensics and Security}
\bvolume{18},
\bfpage{463}--\blpage{476}
(\byear{2022})
\end{barticle}
\endbibitem

\end{thebibliography}

\section*{Acknowledgements}
This work is supported by the German Research Foundation (DFG) Projects MemDPU (Grant Nrs. DU 1896/3-1 and ME 4612/1-1). I.P., N.D., L.C., Z.C. and X.Z. acknowledge the funding support by DFG project MemCrypto (GrantNr. DU 1896/2-2 and Grant Nr. PO 1220/15–2). S.M., F.L. and C.B., acknowledge the funding support in part by the Federal Ministry of Education and Research (BMBF, Germany) through the project NEUROTEC II with the grant numbers 16ME0398K and 16ME0399. M. D. acknowledges support from the National Science Foundation under Grant No. ECCS-2229880.

\section*{Ethics Declarations}
The authors declare no competing interests. The data that support the plots within this paper and other findings of this study are available from the corresponding author upon reasonable request.

\subsection*{Author Contributions}
N.D. and S.M. conceived the main conceptual idea and computational framework of mixed-mode computing. I. P. designed the crossbar-oriented automation tool M$^3$S, and optimized control sequence for \(N\)-bit carry-ripple adder and 4-input Sbox. N.D. and I.P. wrote the manuscript with the support from S.M.. C.B. provides the energy evaluation of \(N\)-bit carry-ripple adder implementations. K. L., Z.C. and X.Z carried out the experimental work and analysed the data. L. C., and F.L. aided in interpreting the results. U.H. fabricated the physical crossbar based on BiFeO$_{\text{3}}$ memristors and performed SEM analysis. H.K. initiates and supervised the work. M.D. supported data interpretation and provided critical feedback, especially in computing reliability in memory-oriented logic design considered in this work. All coauthors discussed the results and implications at all states and contributed to the improvement of the manuscript text.

 \begin{appendices}

 \end{appendices}

\end{document}


\title[Article Title]{Mixed-Mode In-Memory Computing: Towards High-Performance Logic Processing in Memristive Crossbar Array}


\author*[1,2]{\fnm{Nan} \sur{Du}}
\email{nan.du@leibniz-ipht.de}

\author[3]{\fnm{Ilia} \sur{Polian}}

\author[4, †]{\fnm{Christopher} \sur{Bengel} \footnote[0]{† Christopher Bengel works now at HELLA GmbH \& Co. KGaA | Rixbecker Strasse 75, 59552 Lippstadt, Germany. }}

\author[1,2]{\fnm{Kefeng} \sur{Li}}

\author[1,2]{\fnm{Ziang} \sur{Chen}}

\author[1,2]{\fnm{Xianyue} \sur{Zhao}}

\author[1]{\fnm{Uwe} \sur{Hübner}}

\author[3]{\fnm{Li-Wei} \sur{Chen}}

\author[5]{\fnm{Feng} \sur{Liu}}


\author[6]{\fnm{Massimiliano} \sur{Di Ventra}}

\author[5]{\fnm{Stephan} \sur{Menzel}}

\author[1,2]{\fnm{Heidemarie} \sur{Krüger}}


\affil[1]{\orgname{Leibniz Institute of Photonic Technology (IPHT)}, \orgaddress{\street{Albert-Einstein-Str. 9}, \city{Jena}, \postcode{07745}, \country{Germany}}}

\affil[2]{\orgdiv{Institute for Solid State Physics}, \orgname{Friedrich Schiller University Jena}, \orgaddress{\street{Helmholtzweg 3}, \city{Jena}, \postcode{07743}, \country{Germany}}}

\affil[3]{\orgdiv{Institute of Computer Engineering and Computer Architecture}, \orgname{University of Stuttgart}, \orgaddress{\street{Pfaffenwaldring 47}, \city{Stuttgart}, \postcode{70569}, \country{Germany}}}

\affil[4]{\orgdiv{Institute of Materials in Electrical Engineering and Information Technology}, \orgname{RWTH Aachen University}, \orgaddress{\street{Sommerfeldstraße 18}, \city{Aachen}, \postcode{52074}, \country{Germany}}}

\affil[5]{\orgdiv{Peter Grünberg Institut (PGI-7)}, \orgname{Forschungszentrum Jülich}, \orgaddress{\street{Wilhelm-Johnen-Straße}, \city{Jülich}, \postcode{52428}, \country{Germany}}}

\affil[6]{\orgdiv{Department of Physics}, \orgname{University of California, San Diego}, \orgaddress{\street{9500 Gilman Drive}, \city{La Jolla}, \postcode{CA 92093-0319}, \country{USA}}}


\maketitle
\renewcommand{\thefigure}{S\arabic{figure}}
\setcounter{figure}{0}
\renewcommand{\thetable}{S\arabic{table}}
\setcounter{table}{0}
\section*{\centering SUPPLEMENTARY INFORMATION }
\renewcommand\thesection{\Alph{section}}

\section{Classification of representative memristive logic designs}\label{secA}

\begin{figure}[!h]
\centering\includegraphics[width=\linewidth]{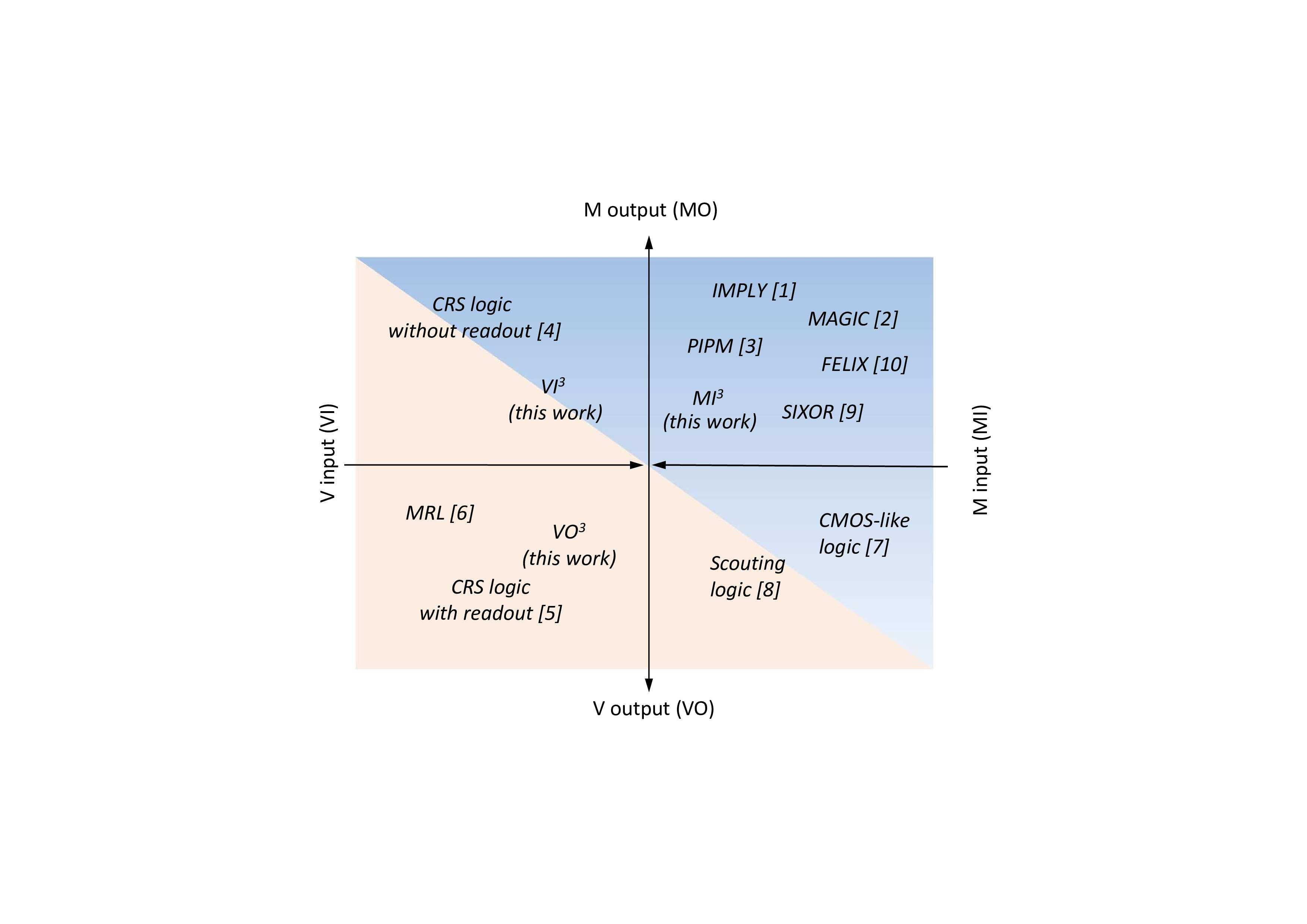}
\caption{Landmap of representative memristive logic designs distributed in logic kernels MI, MO, VI and VO. 
}
\label{fig_S1}
\end{figure}

The intrinsic functional behavior of memristors, characterized by voltage-triggered latching effects and memristance-based memorization, sets them apart from their CMOS counterparts. The stored memristance (M) and the applied voltages (V) at their terminals act as both logical inputs and outputs, facilitating their use in logic processing. Nanoscale memristive devices harness these unique properties to perform logic operations, offering a promising approach for in-memory computing paradigms where logic processing and computation occur directly within the memory block. As illustrated in Fig. \ref{fig_S1}, the memristive logic designs presented in the literature can be systematically classified based on distinct logic kernels: MI, MO, VI, and VO. These designs are structured as combinations of one input kernel (MI or VI), representing the logic input variable, and one output kernel (MO or VO), representing the output logic variable. Through this categorization, we analyze the strengths and limitations of these designs, offering valuable insights into their compatibility with specific logic kernels.

 \begin{table}[h]

 \tiny
 \caption{
Summary of representative memristive logic designs in comparison to 3-input $\mathrm{MI}^{3}$, $\mathrm{VI}^{3}$ and $\mathrm{VO}^{3}$ logic operations.
}\label{tab_logic}
 \begin{tabular}{@{}cccccccccllc@{}}
\toprule

  \begin{tabular}[c]{@{}c@{}} Logic \\design  \\ \end{tabular} &
  \begin{tabular}[c]{@{}c@{}}Logic \\inputs\\ \end{tabular} &
  \begin{tabular}[c]{@{}c@{}}Logic \\outputs\\  \end{tabular} &
  \begin{tabular}[c]{@{}c@{}}Logic \\functions\\  \end{tabular} &
  \begin{tabular}[c]{@{}c@{}}Computation\\ style\\  \end{tabular} &
  \begin{tabular}[c]{@{}c@{}}\# \\ devices †\\ \end{tabular} &
  \begin{tabular}[c]{@{}c@{}}\# \\ cycles \\ \end{tabular} &
  \begin{tabular}[c]{@{}c@{}}Gate\\ error rate\\ \end{tabular} 

  \\ \midrule

  \begin{tabular}[c]{@{}c@{}} IMPLY \\\cite{borghetti2010memristive} \\ \end{tabular} &
  \begin{tabular}[c]{@{}c@{}}MI \\2-input\\  \end{tabular} &
  \begin{tabular}[c]{@{}c@{}} MO \\ \\  \end{tabular} &
  \begin{tabular}[c]{@{}c@{}} IMPLY \\ \\   \end{tabular} &
  \begin{tabular}[c]{@{}c@{}} Sequential \\ \\   \end{tabular} &
  \begin{tabular}[c]{@{}c@{}}2M+1R \\ \\  \end{tabular} &
  \begin{tabular}[c]{@{}c@{}}1\\  \\ \end{tabular} &
  \begin{tabular}[c]{@{}c@{}} High \\  \\  \end{tabular} 

  \\ \\

  \begin{tabular}[c]{@{}c@{}} MAGIC \\\cite{kvatinsky2014magic} \\  \\ \end{tabular} &
  \begin{tabular}[c]{@{}c@{}}MI \\2-input\\   \\ \end{tabular} &
  \begin{tabular}[c]{@{}c@{}} MO \\ \\  \\  \end{tabular} &
  \begin{tabular}[c]{@{}c@{}} e.g.NOR, \\ NAND \\  \\   \end{tabular} &
  \begin{tabular}[c]{@{}c@{}} Sequential \\ \\  \\   \end{tabular} &
  \begin{tabular}[c]{@{}c@{}}3M\\ \\   \\ \end{tabular} &
  \begin{tabular}[c]{@{}c@{}}1\\  \\ \\  \end{tabular} &
  \begin{tabular}[c]{@{}c@{}} High \\  \\  \\ \end{tabular} 
  
  \\ 

    \begin{tabular}[c]{@{}c@{}} PIPM \\\cite{papandroulidakis2014boolean} \\  \\ \end{tabular} &
  \begin{tabular}[c]{@{}c@{}}MI \\2-input\\   \\ \end{tabular} &
  \begin{tabular}[c]{@{}c@{}} MO \\ \\  \\  \end{tabular} &
  \begin{tabular}[c]{@{}c@{}} e.g.NOR, \\ XNOR \\   \\  \end{tabular} &
  \begin{tabular}[c]{@{}c@{}} Parallel \\ \\  \\   \end{tabular} &
  \begin{tabular}[c]{@{}c@{}}2M\\ \\  \\  \end{tabular} &
  \begin{tabular}[c]{@{}c@{}}1\\  \\  \\ \end{tabular} &
  \begin{tabular}[c]{@{}c@{}} Low \\  \\   \\ \end{tabular}

  \\ 

  \begin{tabular}[c]{@{}c@{}} CRS logic \\ without readout \\ \cite{linn2010complementary}  \\ \end{tabular} &
  \begin{tabular}[c]{@{}c@{}}VI \\2-input\\   \\ \end{tabular} &
  \begin{tabular}[c]{@{}c@{}} MO \\ \\ \\   \end{tabular} &
  \begin{tabular}[c]{@{}c@{}} 14 gates \\ except\\ XOR/XNOR \\   \end{tabular} &
  \begin{tabular}[c]{@{}c@{}} Sequential \\ \\    \\ \end{tabular} &
  \begin{tabular}[c]{@{}c@{}}1M\\ \\  \\  \end{tabular} &
  \begin{tabular}[c]{@{}c@{}}1\\  \\  \\ \end{tabular} &
  \begin{tabular}[c]{@{}c@{}} Very low \\  \\  \\  \end{tabular} 
 
  \\ \\  

  \begin{tabular}[c]{@{}c@{}} CRS logic  \\with readout\\\cite{you2014exploiting} \\ \end{tabular} &
  \begin{tabular}[c]{@{}c@{}}VI \\2-input\\   \\ \end{tabular} &
  \begin{tabular}[c]{@{}c@{}} VO \\ (2-input\\ VI) \\  \end{tabular} &
  \begin{tabular}[c]{@{}c@{}} 16 gates \\ possible\\  \\   \end{tabular} &
  \begin{tabular}[c]{@{}c@{}} Sequential \\ \\   \\  \end{tabular} &
  \begin{tabular}[c]{@{}c@{}}1M\\ \\ \\   \end{tabular} &
  \begin{tabular}[c]{@{}c@{}}2\\  \\  \\ \end{tabular} &
  \begin{tabular}[c]{@{}c@{}} Very low \\  \\  \\  \end{tabular} 

  \\  \\ 

  \begin{tabular}[c]{@{}c@{}} MRL \\ \cite{kvatinsky2012mrl} \\ \\ \end{tabular} &
  \begin{tabular}[c]{@{}c@{}}VI \\2-input\\   \\ \end{tabular} &
  \begin{tabular}[c]{@{}c@{}} VO \\  \\   \\ \end{tabular} &
  \begin{tabular}[c]{@{}c@{}} AND, OR \\ \\  \\   \end{tabular} &
  \begin{tabular}[c]{@{}c@{}} Parallel \\ \\   \\  \end{tabular} &
  \begin{tabular}[c]{@{}c@{}}2M\\ \\  \\  \end{tabular} &
  \begin{tabular}[c]{@{}c@{}}1\\  \\  \\ \end{tabular} &
  \begin{tabular}[c]{@{}c@{}} Low \\  \\  \\  \end{tabular} 

  \\  
  
    \begin{tabular}[c]{@{}c@{}} CMOS-like\\  logic \\\cite{vourkas2012novel} \\ \end{tabular} &
  \begin{tabular}[c]{@{}c@{}}MI \\2-input\\  \\  \end{tabular} &
  \begin{tabular}[c]{@{}c@{}} VO \\  \\  \\  \end{tabular} &
  \begin{tabular}[c]{@{}c@{}} NOR, NAND \\ \\  \\   \end{tabular} &
  \begin{tabular}[c]{@{}c@{}} Sequential \\ \\   \\  \end{tabular} &
  \begin{tabular}[c]{@{}c@{}}4M\\ \\  \\  \end{tabular} &
  \begin{tabular}[c]{@{}c@{}}1\\  \\  \\ \end{tabular} &
  \begin{tabular}[c]{@{}c@{}} Low \\  \\   \\ \end{tabular} 

  \\ \\ 
  
    \begin{tabular}[c]{@{}c@{}} Scouting\\  logic \\\cite{xie2017scouting} \\ \end{tabular} &
  \begin{tabular}[c]{@{}c@{}}MI \\2-input\\  \\  \end{tabular} &
  \begin{tabular}[c]{@{}c@{}} VO \\  \\  \\  \end{tabular} &
  \begin{tabular}[c]{@{}c@{}} AND, OR, \\XOR \\  \\   \end{tabular} &
  \begin{tabular}[c]{@{}c@{}} Sequential \\ \\   \\  \end{tabular} &
  \begin{tabular}[c]{@{}c@{}}2M+2T\\ \\  \\  \end{tabular} &
  \begin{tabular}[c]{@{}c@{}}1\\  \\  \\ \end{tabular} &
  \begin{tabular}[c]{@{}c@{}} Very low \\  \\   \\ \end{tabular} 

  \\ \\ 

    \begin{tabular}[c]{@{}c@{}} SIXOR\\  \cite{taherinejad2021sixor} \\ \\ \end{tabular} &
  \begin{tabular}[c]{@{}c@{}}MI \\2-input\\  \\  \end{tabular} &
  \begin{tabular}[c]{@{}c@{}} MO \\  \\  \\  \end{tabular} &
  \begin{tabular}[c]{@{}c@{}} XOR \\ \\  \\   \end{tabular} &
  \begin{tabular}[c]{@{}c@{}} Sequential \\ \\   \\  \end{tabular} &
  \begin{tabular}[c]{@{}c@{}}5M+1T\\ \\  \\  \end{tabular} &
  \begin{tabular}[c]{@{}c@{}}1\\  \\  \\ \end{tabular} &
  \begin{tabular}[c]{@{}c@{}} Very high \\  \\   \\ \end{tabular} 

  \\ 

    \begin{tabular}[c]{@{}c@{}} FELIX\\  \cite{gupta2018felix} \\ \\ \end{tabular} &
  \begin{tabular}[c]{@{}c@{}}MI \\n-input\\  \\  \end{tabular} &
  \begin{tabular}[c]{@{}c@{}} MO \\  \\  \\  \end{tabular} &
  \begin{tabular}[c]{@{}c@{}} NOR,OR \\ \\  \\   \end{tabular} &
  \begin{tabular}[c]{@{}c@{}} Sequential \\ \\   \\  \end{tabular} &
  \begin{tabular}[c]{@{}c@{}}nM\\ \\  \\  \end{tabular} &
  \begin{tabular}[c]{@{}c@{}}1\\  \\  \\ \end{tabular} &
  \begin{tabular}[c]{@{}c@{}} Very high \\  \\   \\ \end{tabular} 
 
  \\ 

    \begin{tabular}[c]{@{}c@{}} MI$^{3}$ \\ (this work) \\ \\ \end{tabular} &
  \begin{tabular}[c]{@{}c@{}}MI \\3-input\\  \\  \end{tabular} &
  \begin{tabular}[c]{@{}c@{}} MO \\  \\  \\  \end{tabular} &
  \begin{tabular}[c]{@{}c@{}} - \\ \\  \\   \end{tabular} &
  \begin{tabular}[c]{@{}c@{}} Sequential \\ \\   \\  \end{tabular} &
  \begin{tabular}[c]{@{}c@{}}3M\\ \\  \\  \end{tabular} &
  \begin{tabular}[c]{@{}c@{}}1\\  \\  \\ \end{tabular} &
  \begin{tabular}[c]{@{}c@{}} High \\  \\   \\ \end{tabular} 

  \\ 

    \begin{tabular}[c]{@{}c@{}} VI$^{3}$ \\ (this work) \\ \\ \end{tabular} &
  \begin{tabular}[c]{@{}c@{}} MI(/VI) \\3-input\\  \\  \end{tabular} &
  \begin{tabular}[c]{@{}c@{}} MO \\  \\  \\  \end{tabular} &
  \begin{tabular}[c]{@{}c@{}} - \\ \\  \\   \end{tabular} &
  \begin{tabular}[c]{@{}c@{}} Sequential \\ \\   \\  \end{tabular} &
  \begin{tabular}[c]{@{}c@{}}1M\\ \\  \\  \end{tabular} &
  \begin{tabular}[c]{@{}c@{}}1\\  \\  \\ \end{tabular} &
  \begin{tabular}[c]{@{}c@{}} Very low \\  \\   \\ \end{tabular} 

  \\ 

    \begin{tabular}[c]{@{}c@{}} VO$^{3}$ \\ (this work) \\ \\ \end{tabular} &
  \begin{tabular}[c]{@{}c@{}}VO \\3-input\\  \\  \end{tabular} &
  \begin{tabular}[c]{@{}c@{}} MO \\  \\  \\  \end{tabular} &
  \begin{tabular}[c]{@{}c@{}} - \\ \\  \\   \end{tabular} &
  \begin{tabular}[c]{@{}c@{}} Sequential \\ \\   \\  \end{tabular} &
  \begin{tabular}[c]{@{}c@{}}3M\\ \\  \\  \end{tabular} &
  \begin{tabular}[c]{@{}c@{}}1\\  \\  \\ \end{tabular} &
  \begin{tabular}[c]{@{}c@{}} Very low \\  \\   \\ \end{tabular} 

  \\ 
\midrule  
\end{tabular}

\footnotetext{ †  In the column labeled “\# of devices”, M represents memristor cells, and R represents resistor cells.}

\end{table}

\begin{figure}[!h]
\centering\includegraphics[width=\linewidth]{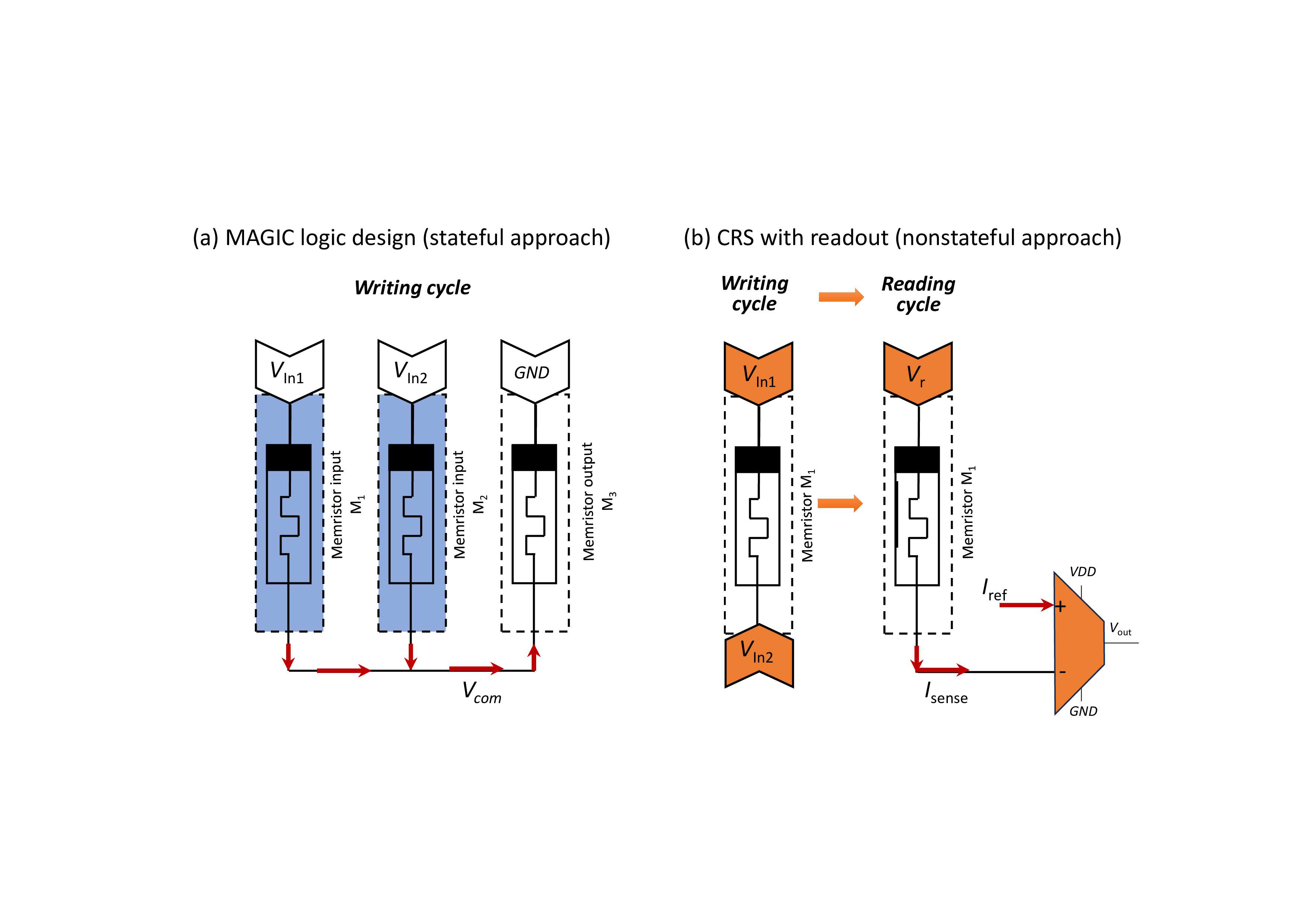}
\caption{Comparative illustration between (a) stateful MAGIC \cite{kvatinsky2014magic} and (b) nonstateful CRS \cite{you2014exploiting} logic designs, where the logic input variables are marked in blue and orange, respectively. }
\label{fig_VM}
\end{figure}

For instance, the representative memristive logic designs such as Material Implication Logic (IMPLY) \cite{borghetti2010memristive}, Memristor-Aided Logic (MAGIC) \cite{kvatinsky2014magic}, and Fast and Energy-Efficient Logic in Memory (FELIX) \cite{gupta2018felix} exemplify the fusion of MI and MO kernels, wherein both inputs and outputs are represented by nonvolatile memristive states M, also known as the stateful logic concept. For instance, the MAGIC logic design illustrated in Fig. \ref{fig_VM}a, comprises at least two input memristive cells ($\text{M}_{\text{1}}$/$\text{M}_{\text{2}}$) and one output cell ($\text{M}_{\text{3}}$).  When voltages ($V_{\text{in1}}$ and $V_{\text{in2}}$) are applied, an input-dependent resistive voltage divider forms between the inputs and the output. Under specific input configurations, the output cell selectively switches its state if the voltage drop ($V_{\text{com}}$) exceeds a certain threshold. Such logic designs often achieve universal gate types, storing the output permanently within the crossbar array upon generation. Note that, in this work, ``universality'' refers to the capability of one logic kernel to realize arbitrary functions using the corresponding logic designs in a cascading manner by using the logic inputs $\{\text{const-0}, \text{const-1}, x_1, \overline x_1, x_2, \overline x_2, \ldots \}$. For instance, considering one logic design within the VI kernel—specifically the CRS logic design \cite{linn2010complementary} without readout allowed—to realize 3-input logic functions through cascading operations (sequential writing without current sensing) can achieve only 104 out of the 256 possible functions ($2^8$), demonstrating that the VI kernel is not universal. This minimizes the need for additional memory operations and facilitates the straightforward reuse of the logic output for subsequent operations. Nonetheless, these designs face following 
limitations: 

\begin{itemize}
  \item Prolonged operation sequences during logic processing are common due to the restricted number of gate types applicable exploiting 
specific memristive technologies. For instance, in Ref. \cite{kim2021memristive}, the RESET process (LRS $\rightarrow$ HRS) is implemented in the output cell of the MAGIC logic design to achieve two-input Boolean NOR/NAND gates. In comparison, Ref. \cite{hoffer2020experimental} demonstrates the implementation of the SET process (HRS $\rightarrow$ LRS) in the output cell to realize NOT-material Implication (NIMP), OR, and NOT gate types. 
  \item Practical logic cascading may encounter an elevated error rate of up to 70\% \cite{in2020universal,kim2020stateful}, attributed to stochastic variability inherent in nanoscale memory devices, as experimentally analyzed in published works. This high error rate renders large-scale logic processing impractical.
  \item Voltage-divider circuit designs introduce additional uncertainty in resulted memristive states of input and output cells (see details in descriptions in Supplementary Information \ref{secC}).
  \item Recent investigations underscore the necessity for more than three input memristive cells within a single stateful logic gate to achieve higher computing efficiency. This requirement is evidenced by designs such as the single-cycle in-memristor XOR (SIXOR) \cite{taherinejad2021sixor} and FELIX logic \cite{gupta2018felix}. However, we contend that a stateful approach incorporating more than three cells is not practically feasible due to the heightened error rates observed experimentally even with just three cells. Including multiple devices within a single logic operation would significantly elevate the error rates.
\end{itemize}

These issues exacerbate the error rate during cascading of such stateful logic gates. In our study, we introduce the universal MI$^{3}$ logic operation, which represents an expanded iteration of the MAGIC logic operation. This advancement not only broadens the input variables from 2 to 3, facilitating the cascading process between MI and VI kernels as well as within the MI kernel (see detailed discussions at the end of this section), but also addresses significant challenges such as input drift and partial switching issues within input and output cells (see detailed discussions in Supplementary Information \ref{secC}), by flexibly combining logic operations in VI kernel before and after MI$^{3}$. It is noteworthy that the goal of our work is not to resolve all limitations of individual kernel but to leverage the complementary strengths of MI and VI kernels to address the shortcomings of each. By combining the universal capabilities of MI$^{3}$ with the robustness and bidirectional transition flexibility of VI$^{3}$ logic designs, the mixed-mode approach achieves enhanced resilience, reduced processing cycles, and the elimination of VO readout, enabling universal and more reliable logic processing.

The representative logic designs such as complementary resistive switching (CRS) logic with readout \cite{you2014exploiting} exemplifies the fusion of VI and VO kernels, where two subsequent operational steps are required (also known as one type of non-stateful logic concepts). 
As illustrated in Table \ref{tab_logic}, CRS logic with readout employing a single memristive cell ($\text{M}_{\text{1}}$) as input and output cell. Input-dependent writing and reading voltage values are applied to both cell terminals in both writing and reading cycles. Notably, the logic inputs are delineated by voltage applied to the device terminals (marked in orange in Fig. \ref{fig_VM}). In the writing cycle, upon applying specific input combinations, which generate a voltage difference between the terminals, the cell undergoes a deterministic switch to either LRS or HRS through the SET/RESET process. Otherwise, it remains in its initial stored state. The output is permanently stored within the crossbar array directly as it is computed. To acquire the output, expressed either as current or voltage sensed from the memristive cell, a reading cycle with readout operation in VO kernel is mandated. 
This logic design facilitates deterministic switching processes on a single device and demonstrates resilience to device variations. 
Additionally, it entails significantly reduced area and latency costs compared to its stateful counterparts, as a diverse range of logic gate types 
can be achieved using just one cell. 
Nonetheless, these designs face following limitations: 
\begin{itemize}
  \item 
  Such a logic design without a readout process can not realize arbitrary functions in a cascading manner by using the input variables of the function to be computed (literals) $\{\text{const-0}, \text{const-1}, x_1, \overline x_1, x_2, \overline x_2, \ldots \}$. 
  For example, the CRS logic \cite{linn2010complementary}, exemplifies the integration of VI and MO kernels without readout VO kernel. In CRS logic, voltage applied to device terminals serves as inputs, while output is represented through nonvolatile memristive states M (also known as one type of nonstateful logic concepts). Notably, in the work in Ref. \cite{linn2010complementary}, it demonstrated that CRS logic without readout can implement 14 types of logic gates (excluding XOR and XNOR) using a single cell. However, using such a logic design to implement 3-input logic functions in a cascading manner (without additional initialization) can realize only 104 out of the 256 possible arbitrary functions. 
  \item During logic processing, in both writing and reading cycles, data for logic operations or cascading must be accessed by peripherals to be applied as voltage values to the memristor's two terminals. This necessitates repeated usage of the readout VO kernel, requiring peripheral circuitry and introducing additional latency and power cost. Further insights into the readout VO kernel are elaborated in Supplementary Information \ref{secB}.
\end{itemize}

In this work, we propose a 3-input VI$^{3}$ logic operation, representing an extended iteration of CRS logic (without readout). Unlike conventional CRS logic design, this approach expands the scope of input variables from 2 to 3 and establishes a complete gate set through integration with MI$^{3}$ logic operation, thereby eliminating the need for a readout operation. This elimination of the readout operation during logic cascading is a significant advantage of our work, which will be detailed further in the next section (in Supplementary Information \ref{secC}). Besides that, the 3-input logic operations proposed in this work offer several distinct advantages: Firstly, they embody a synergistic design by integrating two types of 2-input gates, such as the $\overline{p} \cdot q \quad \text{and} \quad \overline{p} + q$ gates ($p$ and $q$ are logic input variables in 2-input logic gates) from the CRS logic family, into a single 3-input operation. This integration reduces circuit-level design complexity and enables faster processing per cycle. Additionally, the 3-input configuration fully exploits the dual physical properties of memristors—stored resistance and applied electrode voltage—by using both resistance states and operating voltages as input variables. This approach achieves efficient and accurate logic operations without requiring additional readout logic, significantly improving overall performance. Secondly, the 3-input design provides substantial efficiency gains in logic cascading. Unlike traditional 2-input gates that require memristive devices to be re-initialized before each operation, the 3-input design allows the output from one operation to be immediately reused as input in subsequent operations, bypassing re-initialization steps. This direct integration enhances throughput, reduces the number of sequential stages, and supports a streamlined and efficient design flow, ensuring smooth operation between MI and VI kernels. Furthermore, the 3-input configuration simplifies the implementation of complex logic functions within automation tools by reducing the need for additional stages and design complexity. This makes it a practical solution for managing memristive systems in crossbar arrays, improving both automation tool development and overall circuit design. The last but not the least, our work adopts a holistic co-design methodology that integrates device, circuit, and system-level considerations to maximize performance. By addressing these levels in a coordinated manner, the approach leverages interdependencies that enhance system capabilities. The 3-input logic design optimizes automation tool development, which in turn facilitates the effective implementation of the design within crossbar architectures, achieving a high degree of efficiency, flexibility, and functionality.

The Boolean expressions that represent the operations of both the MI$^{3}$ and VI$^{3}$ logic operations are described as follows: the Boolean function for the MI$^{3}$ operation is

$m_{\text{y}_{1}} =  (m_{\text{x}_{1}} \cdot \overline{m}_{\text{x}_{2}} \cdot \overline{m}_{\text{x}_{3}})$, 

while the Boolean function for the VI$^{3}$ operation is as 

$m_{\text{y}_{1}} = (\overline{m}_{\text{x}_{1}} \cdot v_{\text{x}_{2}} \cdot \overline{v}_{\text{x}_{3}}) + (m_{\text{x}_{1}} \cdot \overline{v}_{\text{x}_{2}} \cdot v_{\text{x}_{3}}) + (m_{\text{x}_{1}} \cdot v_{\text{x}_{2}} \cdot \overline{v}_{\text{x}_{3}}) + (m_{\text{x}_{1}} \cdot v_{\text{x}_{2}} \cdot v_{\text{x}_{3}})$. 

While simpler logic functions like XOR or XNOR can also be implemented, such examples do not fully demonstrate the strengths of our mixed-mode approach in terms of cycle efficiency and reduced device usage. In this work, we experimentally demonstrate the capabilities of our system by implementing not only a full adder but also computationally intensive functions such as the 4-bit Sbox. The 4-bit Sbox, with its high nonlinearity and resistance to cryptographic attacks, serves as a challenging benchmark to highlight the robust capabilities and optimized performance of our system in managing real-world, complex computational tasks.

\section{Readout VO kernel and its limitations}\label{secB}

The readout VO kernel is implemented through one readout logic step, necessitating the sensing of memristance state stored in the cell by applying input-dependent reading voltage values across the two terminals of the memristor. If VO kernel is allowed, the current or voltage values sensed from memristive cells are used as logic output or input variables in the cascaded logic gates.

\begin{figure}[!h]
\centering\includegraphics[width=\linewidth]{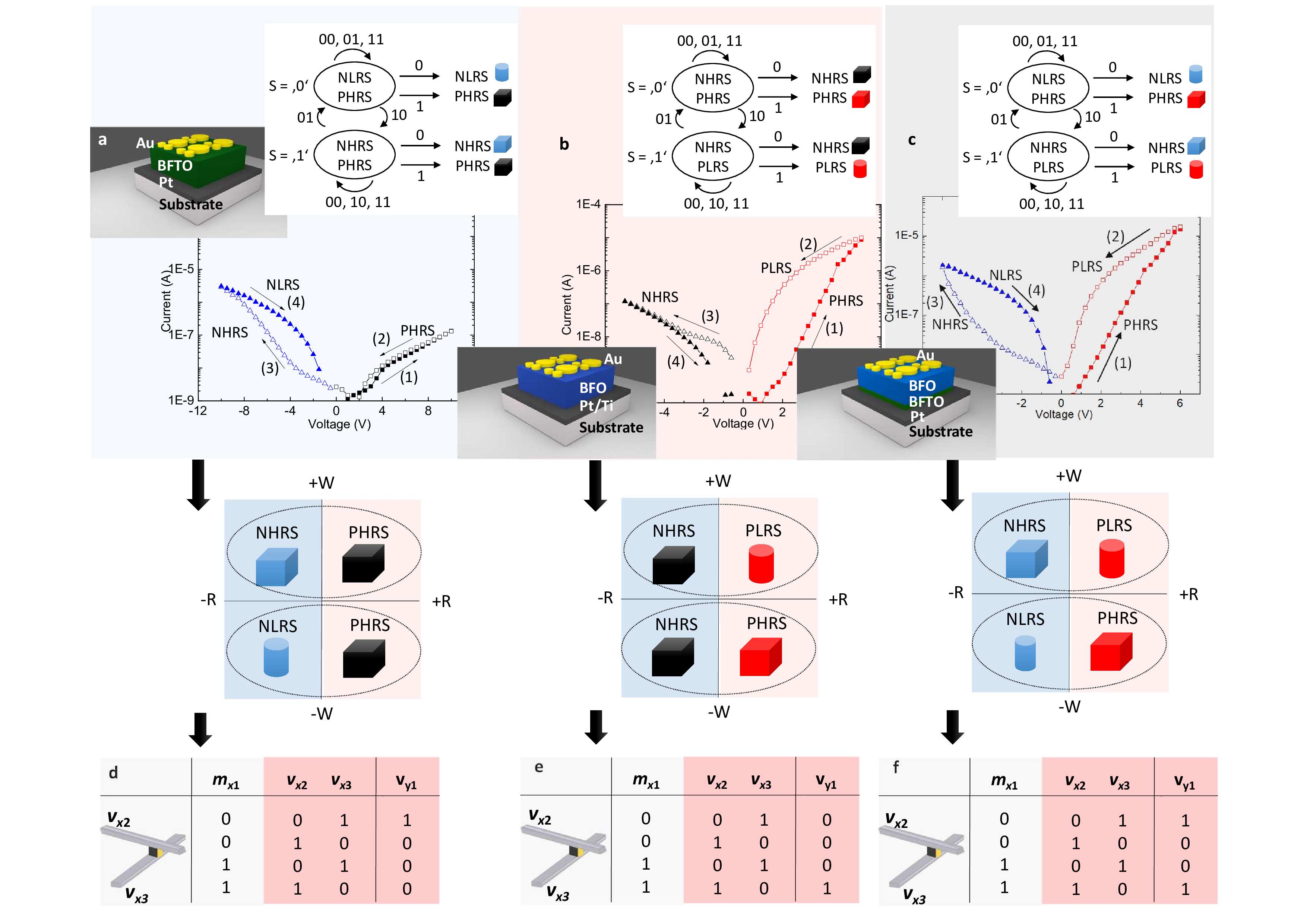}
\caption{Demonstration of readout VO$^{3}$ logic operations in BiFeO$_{\text{3}}$ memristor series. For (a) BiFeTiO$_{\text{3}}$, (b) BiFeO$_{\text{3}}$ and (c) BiFeTiO$_{\text{3}}$/BiFeO$_{\text{3}}$ memristive devices, the device schematics, $I-V$ characteristics, logic related switching dynamics, and the state definitions according to the polarity of writing and reading biases are demonstrated (Cylinders: LRS, Cubes: HRS. Red colored shapes represent reconfigurable states in positive bias range, while blue colored ones in negative bias range. Non-reconfigurable states are presented by black colored shapes). The truth tables of technology dependent readout logic VO$^{3}$ are determined for (d) BiFeTiO$_{\text{3}}$, (e) BiFeO$_{\text{3}}$ and (f) BiFeTiO$_{\text{3}}$/BiFeO$_{\text{3}}$ memristive devices. Note that the state S in (a) corresponds to $m_{\text{x}_{1}}$ in (d-f), indicating the physical configuration of the memristor after a write pulse is applied. }
\label{fig_S2}
\end{figure}

Some devices support a rich readout behavior.  Fig. \ref{fig_S2} demonstrates the readout logic determined in $\mathrm{BiFeO}_{3}$ memristor series, including $\mathrm{BiFeTiO}_{3}$, $\mathrm{BiFeO}_{3}$ and $\mathrm{BiFeTiO}_{3}/ \mathrm{BiFeO}_{3}$ memristive devices. The fabrication process of experimental BiFeO$_{\text{3}}$ memristor series is described in Experimental Section. The experimentally tested $I-V$ characteristics of BiFeTiO$_{\text{3}}$, BiFeO$_{\text{3}}$ and BiFeTiO$_{\text{3}}$/BiFeO$_{\text{3}}$ memristive devices, recorded under ramping pulses with voltage step of $0.1 \mathrm{~V}$ and step time of $0.1 \mathrm{~s}$ are shown in Fig. \ref{fig_S2}a, Fig. \ref{fig_S2}b, and Fig. \ref{fig_S2}c, respectively. The arrows indicate the sweeping direction of the applied ramping voltage values, which are applied to the respective top electrodes. The BiFeO$_{\text{3}}$ memristor series are demonstrating bipolar resistive switching behaviors, while possessing different hysteretic switching dynamics. For example, the hysteretic behavior can be found in negative or positive bias range in BiFeTiO$_{\text{3}}$ (Fig. \ref{fig_S2}a) or BiFeO$_{\text{3}}$ (Fig. \ref{fig_S2}b) memristive devices, respectively, due to the constructed rectifying/nonrectifying contact with flexible Schottky-like barrier height near to top electrode or bottom electrode interfaces \cite{du2018field}. The deposition of a BiFeTiO$_{\text{3}}$ film followed by a BiFeO$_{\text{3}}$ film on Pt/Ti substrate defines the BiFeTiO$_{\text{3}}$/BiFeO$_{\text{3}}$ bilayer structure as illustrated schematically in the inset of Fig. \ref{fig_S2}c. Flexible barriers formed at both top and bottom interfaces for the bilayer structure results in hysteretic behaviors both in the negative and positive bias range in BiFeTiO$_{\text{3}}$/BiFeO$_{\text{3}}$ memristive device \cite{you2014exploiting}.
In BiFeO$_{\text{3}}$ memristor series, the positive low resistance state (PLRS) and negative high resistance state (NHRS) are attainable upon an application of a writing pulse with positive amplitude in BiFeO$_{\text{3}}$ and BiFeTiO$_{\text{3}}$/BiFeO$_{\text{3}}$ devices, while the negative low resistance state (NLRS) and positive high resistance state (PHRS) can be recorded after applying a negative writing pulse in BiFeTiO$_{\text{3}}$ and BiFeTiO$_{\text{3}}$/BiFeO$_{\text{3}}$ devices. Therefore, in BiFeO$_{\text{3}}$ memristor series, it is possible to readout the PLRS and PHRS in a non-destructive manner by a positive reading bias of $2 \mathrm{~V}$ to top electrode and grounding bottom electrode, while attaining the NLRS and NHRS by a positive reading bias of $2 \mathrm{~V}$ to bottom electrode and grounding top electrode. The performance metrics, including retention, endurance, switching ratio, and variations of the BiFeO$_{\text{3}}$ memristor and its variants, have been comprehensively analyzed and published in our previous works \cite{Shuai2011Zhou, Shuai2013Du, You2014Shuai, Du2015Kiani, zhao2024understanding}.

We determine the 3-input VO$^{3}$ logic operation, which can be in principle integrated with VI$^{3}$ and MI$^{3}$ logic operations in mixed-mode computing paradigm. The readout VO kernel is strongly technology dependent, and the 3-input VO$^{3}$ logic functions enabled by BiFeTiO$_{\text{3}}$, BiFeO$_{\text{3}}$ and BiFeTiO$_{\text{3}}$/BiFeO$_{\text{3}}$ memristors are shown in Fig. \ref{fig_S2}d, Fig. \ref{fig_S2}e and Fig. \ref{fig_S2}f, respectively. For instance, each type of BiFeO$_{\text{3}}$ memristor
supports “Reading 0 ($v_{\text{x}_{2}} $ = $\bar{v}_{\text{x}_{3}} $= 0)” and “Reading 1 ($v_{\text{x}_{2}}$ = $\bar{v}_{\text{x}_{3}} $ = 1)” operations, which return the memristance states in the negative and in the positive bias ranges, respectively.  
Combining VI and VO kernel for realizing logic functions presents advantages compared to combining VI and MI, including further reduction of cycle count with the use of readout VO kernel, owing to its ability to implement various logic gate types. However, several critical considerations must be addressed when exploiting the VO kernel, particularly in comparison to MI: 

\begin{itemize}
  \item Needs of Sense Amplifiers: Readout VO kernel often involves sensing the small changes in resistance in the memristive cells. Sense amplifiers, positioned at each BL within the crossbar structure of peripheral circuitry, are commonly employed for this purpose. Tailored to the specific sensing resolutions demanded by various applications, diverse designs of sensing amplifiers are formulated featuring huge difference in power consumption. For instance, to achieve lower resolution sensing, a reference current in the $\mu$A range is employed in the sensing amplifier, discerning between `1' or `0' states by comparing it with the current from the memristor, typically operating between the nA in HRS and mA in LRS range. Conversely, for a required high resolution sensing, these amplifiers need to provide enough gain to accurately detect the resistance states in both states, which typically requires much higher power consumption for high accuracy. 
  \item Requirement for Output-Dependent Control Logic: The logic output, derived from the current sensed by the sense amplifier, is utilized to trigger the sourcing block. For instance, a logic `1' triggers the application of the writing voltage, while a logic `0' triggers grounding to the electrode. This output-dependent control logic block sources the input required in VI kernel based on the sensed output from VO kernel during logic cascading, thereby incurring additional power consumption and latency. 
  \item Read-Write Operations: In memory devices featuring destructive readout, performing read operations in memristive cells may disturb the stored data, necessitating subsequent write operations to restore the original state. This additional read-write cycle leads to heightened power consumption. However, harnessing such destructive reading in logic design has the potential to decrease the initialization cycles needed. 
\end{itemize}

As discussed, readout VO kernel can indeed consume considerable power and latency, especially in scenarios requiring high precision, speed, and reliability. Efforts are ongoing to optimize circuit designs, materials, and algorithms to mitigate these power challenges.  
It is essential to highlight that within our mixed-mode computing approach, we strategically utilize the readout operations in VO kernel at the beginning of logical processing. This facilitates the retrieval of input data from memory to peripherals, thereby enabling the VI kernel. Simultaneously, this approach eliminates the need for data copying or transmission operations, ensuring that computations can be performed efficiently at any memory location without necessitating subsequent storage adjustments. 
Despite in this work we intentionally avoid utilizing the VO kernel during logical cascading to streamline computational processes, it is worthy to mention that VO kernel can play a significant role in intermediate stages, particularly when MI$^{3}$ operations are cascaded, and the input memristors for subsequent stages are not physically co-located with the output memristors of prior stages. In such scenarios, the VO kernel can facilitate efficient data transfer between disjoint locations, reducing the need for additional copying and transmission cycles. For instance, the use of the VO kernel during logic processing is demonstrated in Supporting Information \ref{secG}. We recognize the potential of the VO kernel to relay memristance values effectively across crossbars in these situations, and further exploration of this capability is a key area planned for future work.

\section{Unreliable logic processing in MI kernel}\label{secC}

\begin{figure}[!h]
\centering\includegraphics[width=\linewidth]{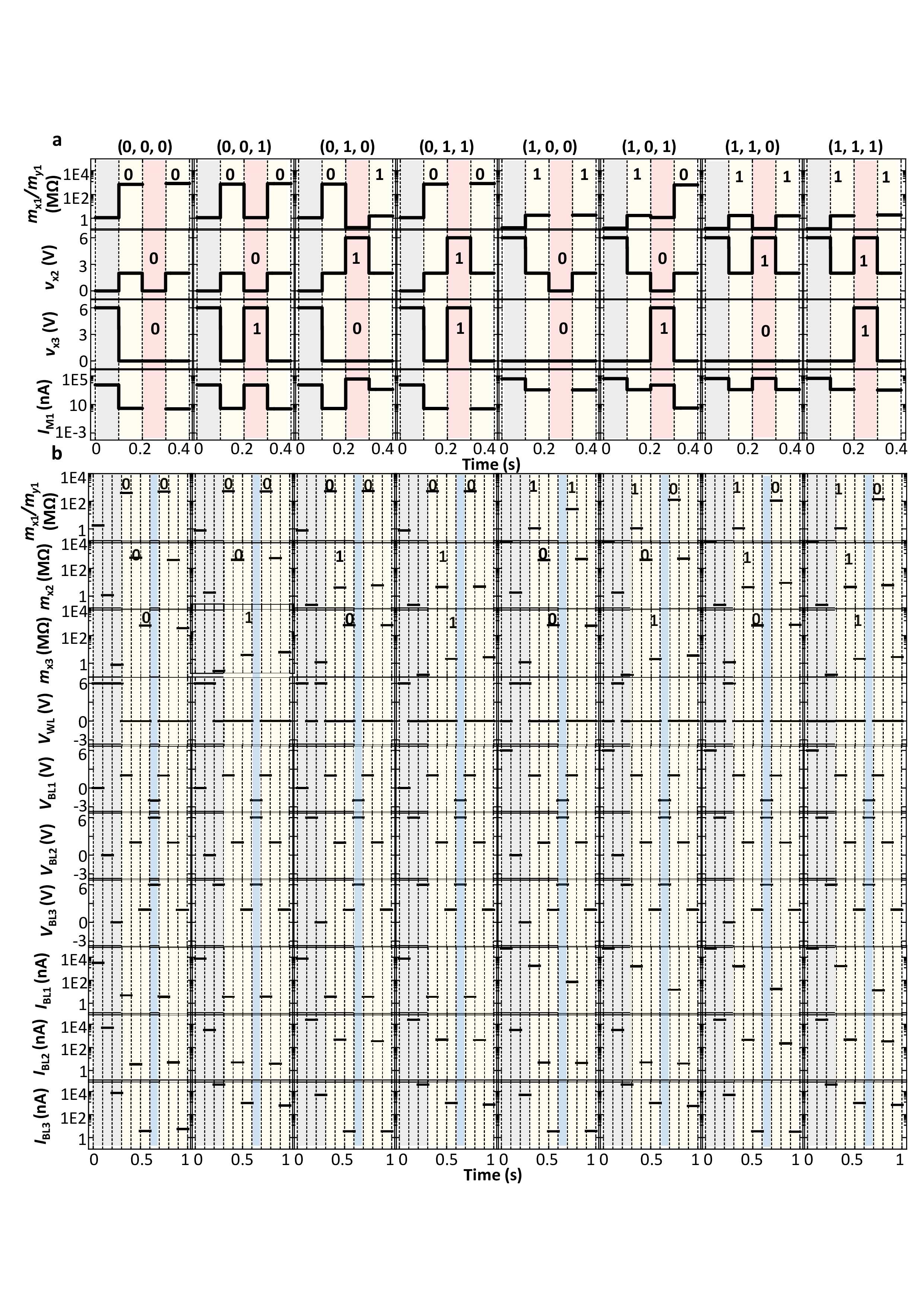}
\caption{Extended experimental demonstration of (a) VI$^{3}$ and (b) MI$^{3}$ logic operations by using BiFeO$_{\text{3}}$ memristive crossbar.}
\label{fig_S3}
\end{figure}

Notwithstanding their fascinating functionalities, the nanoscale memory devices are inherently subject to  stochastic variability \cite{kim2021memristive,bengel2021utilizing}. Originated from their stochastic nature of the switching processes, such variabilities are emerging as major issues in practical operation, and become problematic especially in practical logic cascading in large-scale logic processing. It becomes nearly fatal in stateful MI logic kernel, when multiple cells have to be accessed simultaneously while the voltage divider effect is determined by the common node voltage among the cells. The typical error rates in different cascaded stateful logic operations are evaluated, which are ranging from $54 \%$ to $92.8 \%$ as reported in Ref. \cite{in2020universal}. Most publications about state-of-the-art logic gates and their cascading are assuming an ideal operation and do not account for such errors \cite{talati2016logic,kvatinsky2013memristor,rohani2017improved,cheng2019functional}, which is highly unrealistic and impractical. Only few efforts are made so far to perform error correction \cite{in2020universal,kim2020stateful} by adopting the auxiliary peripheral CMOS-based circuits beside the memory array, which unfortunately deteriorate the computing efficiency and significantly sacrifice the advantages in stateful operations.

In this study, we select the self-rectifying BiFeO$_{\text{3}}$ memristor as our case study. Apart from their high density in passive array, which eliminates the necessity for a transistor beneath each memristor, BiFeO$_{\text{3}}$ memristors offer rich dynamical behavior within BiFeO$_{\text{3}}$ series, making them a valuable point of comparison across various devices and logic kernels. A primary motivation for choosing the self-rectifying BiFeO$_{\text{3}}$ memristor lies in its minimal device variations, in terms of cell-to-cell (C2C) and device-to-device (D2D) variations, in contrast to filamentary switching memristive cells. These low variations inherently mitigate the elevated error rates associated with device discrepancies. 

In this work, we propose MI$^{3}$ logic operation in MI kernel utilizing BiFeO$_{\text{3}}$ memristors enables the execution of stateful logic concept to realize logic functions by using 3 parallel connected cells in one WL/BL within single one cycle, and directly storing logic variables as memristance in memristive cell, enabling cascading without the need for resistance state readout. In comparison to that, VI$^{3}$ operations in VI kernel can accomplish the logic function by switching the memristive state directly by applying writing bias on cell terminals in single one cell within single one cycle. The experimental demonstration of VI$^{3}$ and MI$^{3}$ logic operations by using BiFeO$_{\text{3}}$ memristors are demonstrated in Fig. \ref{fig_S3}. 
It includes the memristance of each cell $M_\text{i}$ before and after logic operations, the voltage applied to the top electrode $V_\text{TE}/V_\text{BLi}$, the voltage applied to the bottom electrode biases $V_\text{BE}/V_\text{WL}$, and the absolute value of current from top electrode to bottom electrode through the cell $|I_\text{Mi}|/|I_\text{BLi}|$. 

As depicted in Fig. \ref{fig_S3}a for VI$^{3}$ operation, the resistance state of the cell $M_\text{1}$ is initialized according to $m_{\text{x}_{1}}$ (highlighted in gray) and subsequently verified through a sequential readout operation. This operation involves applying a reading bias of 2 V across each memristor from top electrode to bottom electrode (highlighted in yellow, consistent across all state checking steps in this study). By employing the voltage combinations $v_{\text{x}_{2}}$ and $v_{\text{x}_{3}}$ to the top electrode and bottom electrode of the memristive cell (indicated in red), the logic output state $m_{\text{y}_{1}}$ is further validated sequentially through readout biases of 2 V applied to the top electrode at the conclusion of the test (highlighted in yellow). As anticipated, the logic output $m_{\text{y}_{1}}$ demonstrates ideal LRS and HRS memristive values according to the `1' or `0' in truth table, attributed to the deterministic switching achieved by applying logic-dependent biases to electrodes on a single cell. 

As depicted in Fig. \ref{fig_S3}b for MI$^{3}$ operation, three memristive cells in a WL are sequentially initialized into the resistance state according to $m_{\text{x}_{1-3}}$ (marked in gray), and pre-checked by readout operation sequentially by applying a reading bias of 2 V across each memristor from top electrode to bottom electrode (marked in yellow). By applying $V_\text{in}$ = 5.3 V on $M_\text{2}$ and $M_\text{3}$ while grounding $M_\text{1}$, the output of MI$^{3}$ operation $m_{\text{y}_{1}}$, which is stored as memristance state in the cell $M_\text{1}$, is further verified sequentially by readout biases of 2 V applied to the top electrode at the end of the test (marked in yellow). It is noteworthy that MI$^{3}$ (or VI$^{3}$) could also be applied to cells along a BL (rather than WL) by applying reversed biases to bottom electrodes through WL.  
The switching process in output cell $M_\text{1}$ is observed in logic input combinations `101', `110', and `111', wherein the cell $M_\text{1}$ transitions from its initial LRS to HRS, resulting in a logic output of `0'. For all other logic input combinations, the memristance state remained unchanged before and after the MI$^{3}$ operation. 

The non-ideal effects of the device significantly influence the performance of the proposed 3-input logic computation method, particularly in MI$^{3}$ and VI$^{3}$ operations. As discussed in the main article and Supplementary Information \ref{secA}, the greatest impact arises from the switching dynamics and stochastic variability inherent in nanoscale memory devices. Switching dynamics dictate the types of logic gates that can be implemented, with some gates requiring additional processes such as RESET or SET operations, leading to prolonged operation sequences and reduced efficiency. Stochastic variability further exacerbates the issue by introducing high error rates—up to 70\% in experimental studies—particularly in cascaded operations. These effects collectively reduce the scalability and accuracy of the computation method. 
Next, using BiFeO$_{\text{3}}$-based MI$^3$ operation as an example (with experimental data shown in Fig. \ref{fig_S3}), we discuss two major issues in logic designs within the MI kernel, i.e. ``Input drift'' and ``Partial switching'' issues. Both issues are inevitable in stateful logic operations. Notably, with much more significant device variability issue in abrupt switching devices compared to analog ones, it further reduces the programming window and leads to an even higher error rate in logic processing. 

\begin{itemize}
  \item ``Input drift'' issue: Sufficient positive programming bias $V_\text{in}$ is required for switching output cell to desired memristance state according to truth table. However, too high positive programming bias $V_\text{in}$ applied to the top electrodes of cells $M_{2}$ and $M_{3}$ may induce drift in the memristive states towards LRS when the logic input is HRS `0', a phenomenon known as the ``input drift'' issue. This challenge constitutes one of the primary hurdles in the realization of a stateful logic kernel. 
 \item ``Partial switching'' issue: Due to the voltage divider effect, with logic input combinations `101', `110', and `111', the insufficient positive bias $V_\text{in}$ applied to $M_\text{2}$ and $M_\text{3}$ during MI$^3$ operation, i.e. insufficient positive bias $V_\text{com}$ to the bottom electrode of $M_\text{1}$ can result in compromised HRS (compromised logic `0') compared to the initial HRS value in BiFeO$_{\text{3}}$ memristor.  This indicates a potential "partial switching" issue in the output cell inherent to MI$^3$ logic operations. One approach to mitigathe partial switching issue from `1' to `0' in cell $M_{1}$ is to amplify the positive writing bias applied to their top electrodes, thereby elevating the voltage $V_\text{com}$. However, this amplification simultaneously increases the resistance value in cell $M_{1}$ for the output logic `1' in the logic combination case `100' due to the heightened positive bias applied to its bottom electrode. 
\end{itemize}

Both aforementioned ``Input drift'' and ``Partial switching'' issues are inherent in the stateful logic process (cannot be alleviated). Under mixed-mode computing paradigm, our aim is to mitigate these issues through a co-design strategy that enhances computing accuracy, particularly by leveraging the unique self-rectifying and analog switching characteristics of BiFeO$_{\text{3}}$ memristors. This emphasizes the critical importance of technology-specific optimization in adapting memristor devices for logic designs. 
Firstly, to ensure stable and accurate outputs, we carefully analyzed the switching kinetics of BiFeO$_{\text{3}}$ memristors. These kinetics favor easier transitions from LRS to HRS compared to HRS to LRS. Based on this, we structured the MI$^3$ operation to apply positive bias to the top electrodes of two parallel-connected memristors while grounding the top electrode of the third memristor. This configuration allows us to exploit cascading for ``re-initializing'' the logic `1', minimizes the influence of partial switching in the input cells, and enhances stability within the cascading MI kernel. Consequently, we designed the MI$^3$ operation to produce a more reliable `0' output by optimizing the bias voltage $V_\text{in}$ to 5.3 V.  For instance, as demonstrated in experimental results in Fig. \ref{fig_S3}, the MI$^3$ operation demonstrates correct transitions from LRS to HRS with input combinations `101', `110', and `111' (HRS of $M_\text{1}$ after MI$^3$: 111.4 M$\si{\ohm}$ compared to initialized HRS of $M_\text{1}$ at 402.4 M$\si{\ohm}$) as shown in Fig. \ref{fig_S3}. As expected, with $V_\text{in}$ = 5.3 V, the MI$^3$ operation with input combination `100' results in a compromised LRS in M1 after MI$^3$ (LRS of $M_\text{1}$ after MI$^3$: 27.2 M$\si{\ohm}$ compared to initialized LRS of $M_\text{1}$ at 1.1 M$\si{\ohm}$). Thus this bias $V_\text{in}$ = 5.3 V ensures stable `0' outputs across various input combinations, reducing errors in cascaded operations (see experimental results in Fig. \ref{fig_S3}). As next step, in the definition of M3S automation tool is explicitly designed to allow the output cell of MI$^3$ operation can only be reassigned as an output cell in subsequent MI$^3$ operations if it has been used at least once as an input cell in an MI$^3$ or VI$^3$ operation. This intentional reassignment avoids scenarios where the output `1' from a prior MI$^3$ gate is recomputed into another `1' in the next MI$^3$ cycle, ensures that a more precisely defined input or output value is applied in logic cascading, thereby minimizing error propagation commonly observed in traditional stateful approaches.

\section{Technology dependency in VI$^{3}$ logic operations}
\label{secD}

As aforementioned, the VO and MI logic kernels are inherently reliant on the underlying technology. However, it has been observed that the 3-input VI$^{3}$ operation demonstrates a generic feature: the same logic function can be realized across multiple memristive technologies.

\begin{table}[h]
\caption{Definition of logic input variables in different nanoscale memory devices for realizing 3-input VI$^{3}$ logic function. }\label{tabS1}%
\begin{tabular}{@{}cccccc@{}}
\hline
\begin{tabular}{l}   \end{tabular}&
\begin{tabular}{l} $m_{\text{x}{1}}$: $^{\prime}1^{\prime}/ ^{\prime}0^{\prime}$  \end{tabular}   &
\begin{tabular}{l} $v_{\text{x}{2}}$:$^{\prime}1^{\prime}/ ^{\prime}0^{\prime}$ \end{tabular} &
\begin{tabular}{l} $v_{\text{x}{3}}$: $^{\prime}1^{\prime}/ ^{\prime}0^{\prime}$ \end{tabular} \\
\hline

\begin{tabular}[c]{@{}c@{}} Bipolar \\ReRAM \cite{jin2015transport} \\\\\end{tabular}  &
\begin{tabular}[c]{@{}c@{}} LRS/HRS\\\\\end{tabular}   &
\begin{tabular}[c]{@{}c@{}} $\mathrm{Vw} / \mathrm{GND}$ \\\\\end{tabular} &
\begin{tabular}[c]{@{}c@{}} $\mathrm{Vw} / \mathrm{GND}$ \\\\ \end{tabular} \\ 

\begin{tabular}[c]{@{}c@{}} Unipolar \\ReRAM\cite{yin2020two}\\ \\ \end{tabular} &
\begin{tabular}[c]{@{}c@{}} LRS/HRS \\\\\end{tabular}   &
\begin{tabular}[c]{@{}c@{}} $V_{\mathrm{SET}} / \mathrm{GND}$ \\\\ \end{tabular} &
\begin{tabular}[c]{@{}c@{}} $V_{\mathrm{RESET}} / \mathrm{GND}$ \\\\\end{tabular} \\
 
\begin{tabular}[c]{@{}c@{}} Complementary \\ReRAM $\cite{you2014exploiting}$\\ \\ \end{tabular} & 
\begin{tabular}[c]{@{}c@{}} PLRS(NHRS)/ \\ PHRS(NLRS) \\ \end{tabular} & 
\begin{tabular}[c]{@{}c@{}} $\mathrm{Vw} / \mathrm{GND}$ \\ \\\end{tabular} & 
\begin{tabular}[c]{@{}c@{}} $\mathrm{Vw} / \mathrm{GND}$ \\ \\ \end{tabular} \\

\begin{tabular}[c]{@{}c@{}} PCM \cite{burr2010phase}\\\\ \end{tabular} & 
\begin{tabular}[c]{@{}c@{}} LRS/HRS \\\\\end{tabular} & 
\begin{tabular}[c]{@{}c@{}} Short high voltage \\ $\mathrm{Vw} / \mathrm{GND}$\\ \end{tabular} & 
\begin{tabular}[c]{@{}c@{}} Long medium \\ $\mathrm{Vw} / \mathrm{GND}$  \\\end{tabular} \\

\begin{tabular}[c]{@{}c@{}} STT-MRAM \cite{kultursay2013evaluating} \\\\ \end{tabular} & 
\begin{tabular}[c]{@{}c@{}} LRS/HRS \\\\ \end{tabular} & 
\begin{tabular}[c]{@{}c@{}} $\mathrm{Vw} / \mathrm{GND}$ \\\\ \end{tabular} & 
\begin{tabular}[c]{@{}c@{}} $\mathrm{Vw} / \mathrm{GND}$\\\\  \end{tabular} \\
\hline

\end{tabular}
\end{table}

This generic feature is because that the design of 3-input VI$^{3}$ logic function is rooted from the intrinsic switching behavior of memory devices: the memory device switches from original input state from $m_{\text{x}_\text{1}}$ to output state $m_{\text{y}_\text{1}}$ only if the applied voltage/current across the cell (difference between $v_{\text{x}_\text{2}}$ and $v_{\text{x}_\text{3}}$) initiate the underlying switching process.

It is worthy to mention that the performance of the 3-input logic method can vary depending on the underlying device technology. This variation is influenced by factors such as the switching dynamics of the technology, which determine the types of logic gates that can be realized, and device variability, which affects logic accuracy, especially for resistance-controlled logic design. 
Table \ref{tabS1} demonstrates that the same 3-input VI$^{3}$ logic function can be implemented using various series of BiFeO$_{\text{3}}$ memristors, as well as more general types of redox-based random access memory (ReRAM), phase change memory (PCM) \cite{burr2010phase}, and spin transfer torque magneto-resistive RAM (STT-MRAM) \cite{kultursay2013evaluating}. This implementation involves customizing the definitions of $v_{\text{x}_\text{2}}$ and $v_{\text{x}_\text{3}}$ to initiate switching based on the specific characteristics of each underlying technology.
To adopt the proposed mixed-mode computing architecture, the chosen technology must satisfy two essential criteria: (1) it should enable resistance-controlled logic to support mixed-mode computing, and (2) it should exhibit low device variability to achieve high logic accuracy. Additionally, good endurance is also an important consideration for practical applications.
While the selection of the optimal device largely depends on the target application, in general, technologies that offer higher density, lower energy consumption, and faster operation would be most desirable.

\section{High parallel logic processing in mixed-mode computing}\label{secE}

Memristive crossbar configurations offer significant parallel computing capabilities by leveraging the inherent parallelism in their architecture, offering high speed and improved energy efficiency. This section provides a summary of the parallelism computing capabilities inherent in the two-dimensional crossbar architecture, integrated into the design of the M$^3$S tool.

We exemplify the parallel computing capabilities facilitated by memristive crossbar configurations (Fig. \ref{fig_S10}), utilizing the proof-of-principle demonstrator $N$-bit carry-ripple adder with $\mathrm{VI}^{3}$ and $\mathrm{MI}^{3}$ as a case study. 

\begin{figure}[!h]
\centering\includegraphics[width=\linewidth]{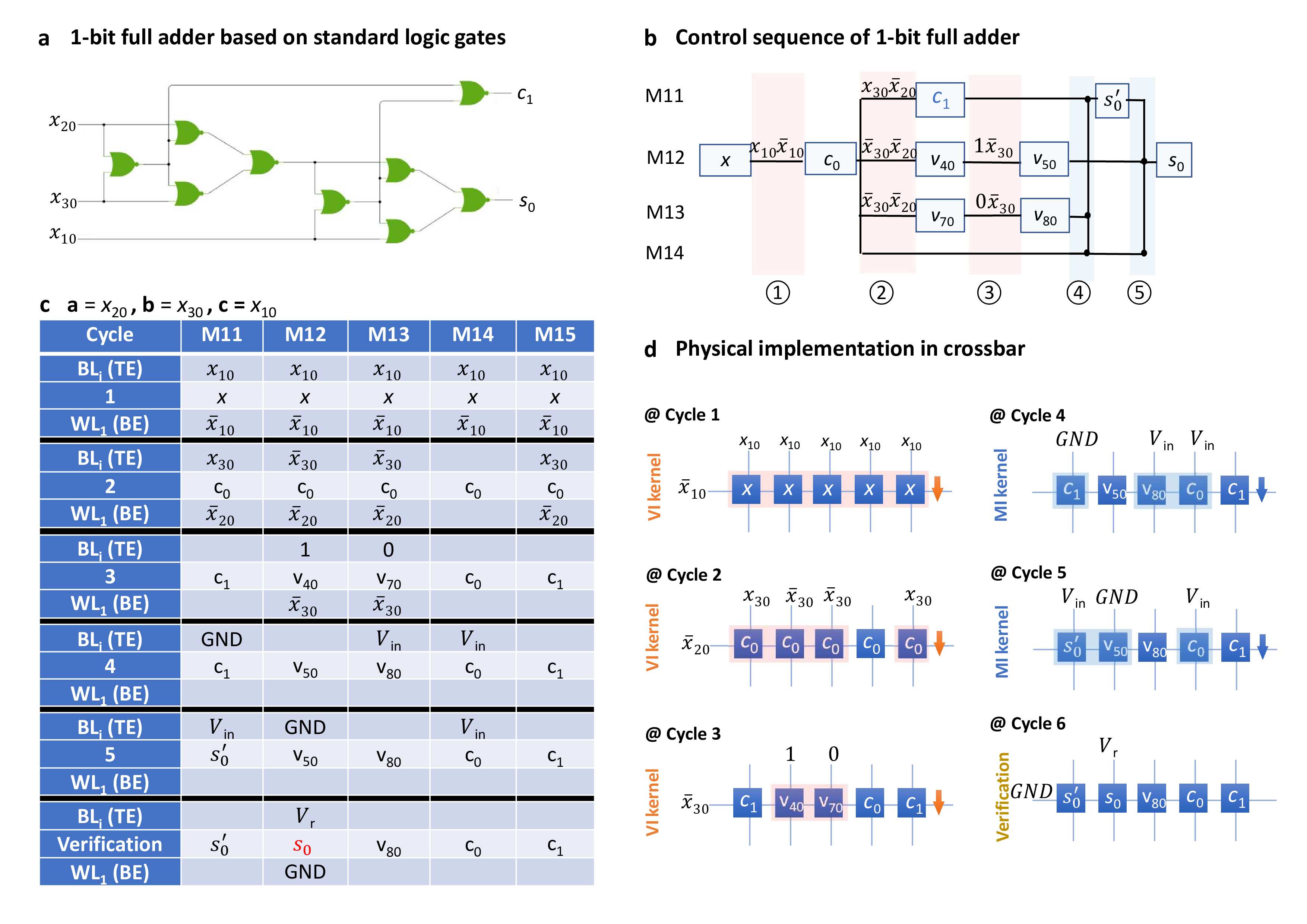}
\caption{Implementation of $1$-bit full adder using (a) standard logic gates (NOR gates) in comparison to the design using (b) mixed-mode computing according to the control sequence shown in Fig. \ref{fig_3}b. (c) Step-by-step control cycles for applying logic inputs through BLs and WLs in the crossbar configuration. (d) Physical implementation in memristor-based crossbar for each cycle. }
\label{fig_S11_1}
\end{figure}

For example, as demonstrated in Fig. \ref{fig_S11_1}, in the case of the 1-bit carry-ripple adder, each column representing a cell (M$_{11}$–M$_{14}$) needed for computation, while M15 stores the output carry bit c1. The five columns represent the five cycles required for completing the 1-bit carry-ripple adder, broken down into three cycles for $\mathrm{VI}^{3}$ operations and two for $\mathrm{MI}^{3}$ operations. The $\mathrm{VI}^{3}$ operations are executed across all cells simultaneously, enabling efficient processing. For each cycle, we specify the WL (Word Line) and BL (Bit Line) inputs used.  
For implementing $1$-bit carry-ripple adder, our mixed-mode approach requires 5 cells and 5 cycles, while a purely in-memory computing approach without reading would require 10 cells and 13 cycles (as referenced in Ref. \cite{talati2016logic} in Tab.\ref{tab_adder} in Supplementary Information \ref{secI}). However, the efficiency gains of the mixed-mode paradigm become more apparent as the bit number increases. For instance, in Fig. \ref{fig_S11}, the $4$-bit carry-ripple adder can be realized in a $4 \times 5$ passive crossbar array with 1R configuration, where no readout is allowed in logic cascading. Implementing an 8-bit full adder in our mixed-mode approach requires only 33 cells and 12 cycles, compared to a purely in-memory computing approach, which would require 87 cells and 97 cycles (also as referenced in Ref. \cite{talati2016logic} in Tab.\ref{tab_adder} in Supplementary Information \ref{secI}). This comparison highlights the efficiency of mixed-mode computing in terms of cell and cycle optimization, particularly for multi-bit and complex operations.

Due to the distinct nature of each operation during logic processing, different parallel strategies are possible. Assuming to compute one arbitrary function with logic input variables $x_{\text{1}_\text{i}}$, $x_{\text{2}_\text{i}}$, and $x_{\text{3}_\text{i}}$, with $i$ representing $i$-th bit in each logic input, the parallel strategies can be described as follows: 

\begin{figure}[!h]
\centering\includegraphics[width=\linewidth]{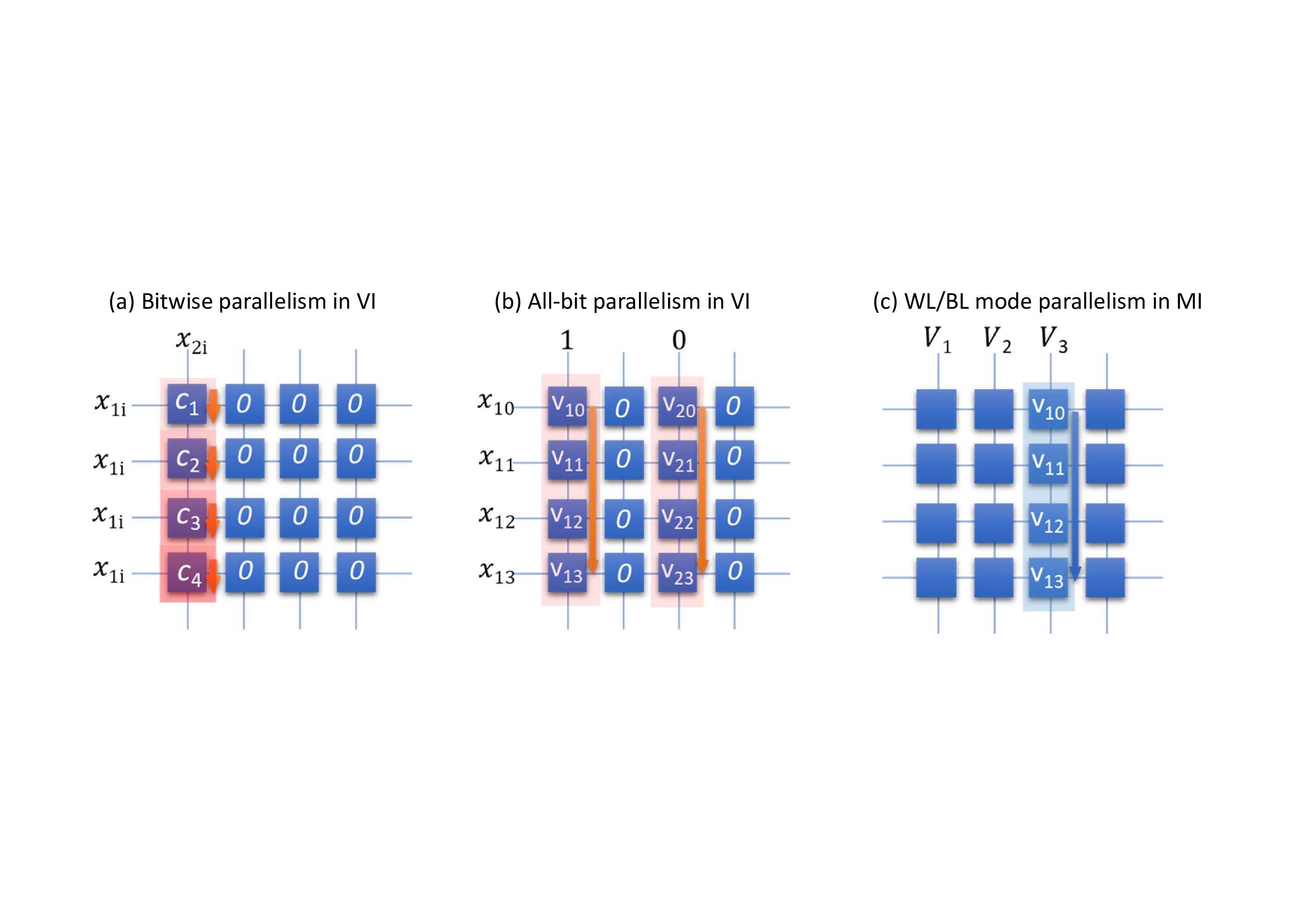}
\caption{Illustration of parallelism computing facilitated by memristive crossbar configurations.}
\label{fig_S10}
\end{figure}

\begin{figure}[!h]
\centering\includegraphics[width=\linewidth]{Table S3.jpg}
\caption{Implementation flow in memristive crossbar of 4-bit carry-ripple adder according to control sequence by M$^3$S demonstrated in Fig. \ref{fig_3}b.}
\label{fig_S11}
\end{figure}

\begin{itemize}
  \item  Bitwise parallelism in cascading VI (Fig. \ref{fig_S10}a): the bit-wise parallelism is facilitated by applying multiple $i$-th bit of logic inputs to WLs (or BLs) for performing multiple VI operations in a single operational cycle. For example, as demonstrated in Fig. \ref{fig_S11}, the carry bit $c_\text{i+1}$ in a modular adder can be iteratively computed based on $c_\text{i}$ in cycles 2-5. For computing bit $c_\text{1}$ in BL1, all cells in BL1 are applied by $x_{\text{1}_\text{1}}$ to all WLs and $x_{\text{2}_\text{1}}$ to BL1. $c_\text{2}$ can be further computed by applying $x_{\text{1}_\text{2}}$ to WL2-4 and $x_{\text{2}_\text{2}}$ to BL1, then $c_\text{2}$ is stored in cells M21, M31 and M41 in BL1. Such bitwise parallelism in cascading VI is particularly suitable in modular computing, where the $(i+1)$-th bit is iteratively computed based on the value of $i$-th bit.
	  \item  All-bit parallelism in cascading VI (Fig. \ref{fig_S10}b): All-bit parallelism is realized by applying constant input ‘1’ and ‘0’ to BLs (or WLs), and the corresponding WLs (or BLs) are applied by each bit in logic inputs  $x_{\text{1}_\text{i}}$ in the same cycle. All-bit parallelism is preferred in modular design as it highly reduces the cycle number in $n$-bit implementations (counted as one cycle). To enable all-bit parallelism, the constant logic input ‘1’/‘0’ are preferred as WL or BL operands. As demonstrated in Fig. \ref{fig_S11},  the all-bit parallelism in cascading VI is explored in cycle 6.
	  \item  WL/BL mode parallelism in cascaded MI (Fig. \ref{fig_S10}c): In the BL mode of cascaded MI operations, parallelism can be achieved by applying logic input independent pulses, such as , $V_\text{1-3}$ through three BLs. Typically $V_\text{1}$, $V_\text{2}$ $\textgreater$ $V_\text{3}$, and the BL cells applied by $V_\text{3}$ store the logic outputs as memristance state after applying MI. The bias differences between $V_\text{1}$,$V_\text{2}$ and $V_\text{3}$ are usually comparable with $V_\text{w}$, which represents the amplitude of the writing bias of the cell. All WLs are kept floating during this process. This BL mode operation enables parallel MI operations in three BLs across all floated WLs, including both desired and undesired cells. As demonstrated in Fig. \ref{fig_S11},  the bitwise parallel operation in the MI kernel in WL mode is also explored in cycle 7 and cycle 8. It is verified that the $\mathrm{MI}^{3}$ operations demonstrated in the adder design, can apply flexibly any outputs of $\mathrm{VI}^{3}$ as input cells or output cells, and requires no additional initialization step of the output cells in comparison to conventional MAGIC gate designs.
\end{itemize}

Note that the parallel computing shown in cascading VI kernel can be utilized also in cascading VO kernel (not shown here).

\section{Algorithms used in crossbar-oriented mapping and synthesis tool M$^3$S}\label{secF} 

\begin{figure}[!h]
    \centering \includegraphics[width=1\linewidth]{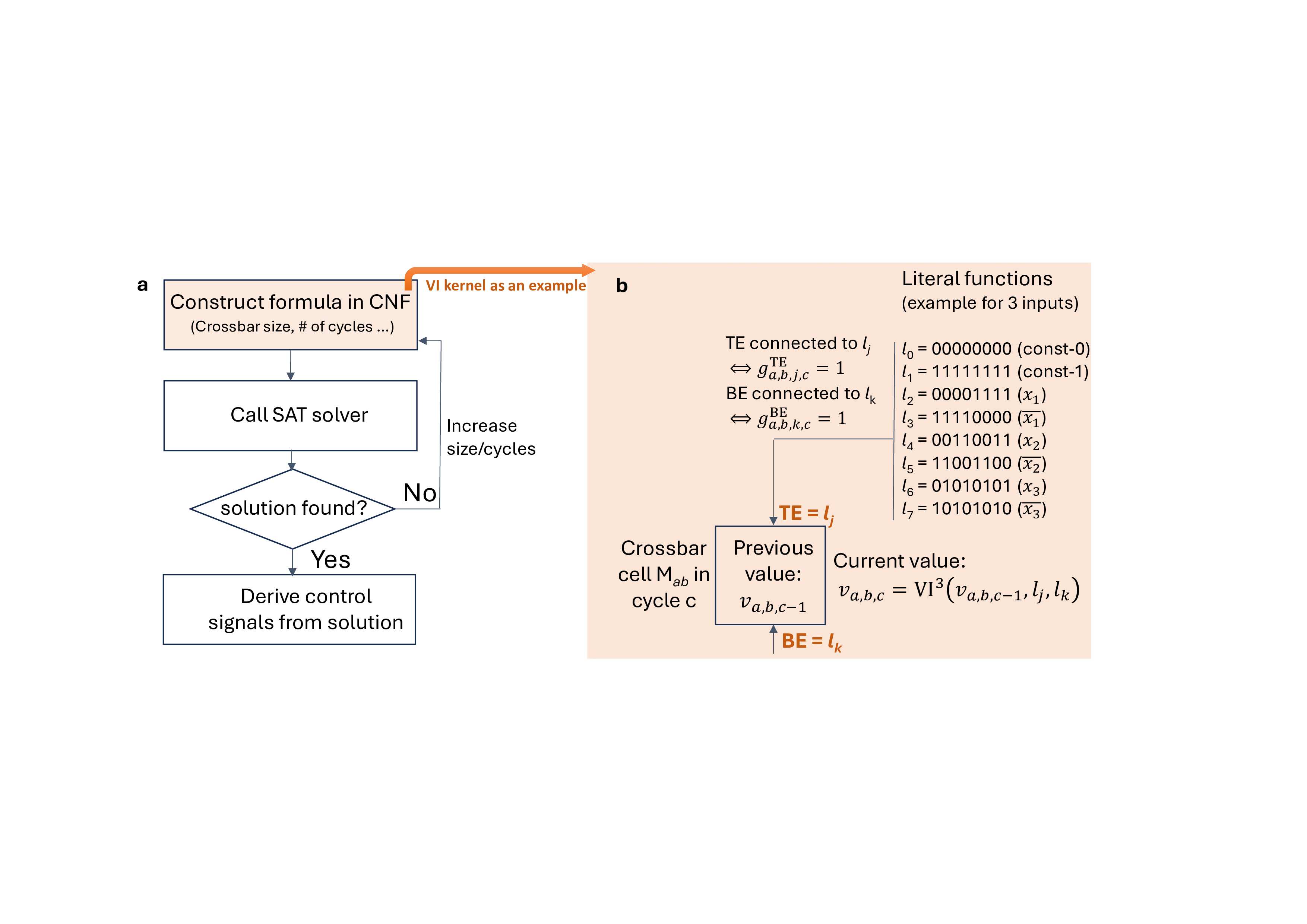}
    \caption{(a) Flowchart illustrating the working principle of the M$^3$S automation tool for synthesizing and mapping logic circuits on memristive crossbars. The constructed Boolean formula, expressed in conjunctive normal form (CNF), serves as input to a Boolean satisfiability (SAT) solver to determine a solution. (b) Illustration of the V-mode operation in Eq. \ref{eq:cnf}, providing an example of literal functions for $n$ = 3 inputs. The table outlines the function table entries for each variable $l_{j,k}$ in the formula, demonstrating how the SAT solver assigns one variable for the top electrode (TE) and one for the bottom electrode (BE) to control which of the (2$n$+2) literal functions is applied during each cycle. For each crossbar cell M$_{ab}$ in cycle $c$, the Boolean satisfiability solver ensures that exactly one variable $g^\text{TE}_{a,b,j,c}$  is set to $1$ for the top electrode and exactly one variable $g^\text{BE}_{a,b,j,c}$  is set to $1$ for the bottom electrode, while all other variables are set to $0$. This precise selection determines which of the (2$n$+2) literal functions is applied during that cycle, enabling accurate control of the memristive logic operation.}
    \label{fig:solver}
\end{figure}

As has been described in the main Article, M$^3$S takes as inputs the truth tables of Boolean functions $f(x_1, x_2, \ldots)$ to be synthesized and the number of cycles when V-mode and M-mode operations are performed. It then constructs a Boolean satsifiability formula in conjunctive normal form, and solving it using an automatic Boolean satisfiability solver.
A conjunctive normal form is comprised of disjunctive clauses over propositional (binary) variables, and its solution entails finding an assignment to the variables that satisfies all clauses \cite{BHMW:21}. The flowchart of the M$^3$S automation tool is demonstrated in Fig. \ref{fig:solver}, using an example of the VI kernel and the construction of Eq. \ref{eq:cnf}. As shown in the flowchart, the process begins with the construction of a Boolean formula in conjunctive normal form format, which is input into a Boolean satisfiability solver. The solver determines whether a solution exists and, if successful, the automation tool interprets the solution and implements it on the memristive cells within the crossbar.

Recall that V-mode operations VI$^3$ have three inputs, one of which is the prior state (resistance) of the same device in the crossbar and two are voltages that come from peripherals. 
Upon inputting the (single- or multiple-output) Boolean function $f$ into the M$^3$S tool,  the constants 0 and 1, and the function $f$'s literals $x_1, \overline{x_1}, \ldots x_n, \overline{x_n}$ (with $n$ being the number of $f$'s inputs), are serving as logic inputs in VI$^3$ operations, applied to the cell's top electrode and bottom electrode. For simplicity, we assign an index to each available literal function, with constant-0 having index 1. The tool automatically produces a conjunctive normal form defined over variables that represent the minterms of the realized function $f$ and the outputs of each V-mode and M-mode operations, and a number of additional auxiliary variables. To obtain the optimal solution in terms of the number of required crossbar cells, the designer can start with a small crossbar that has no valid solution and gradually increase its size until the conjunctive normal form is solved. 
Therefore, the Boolean satisfiability formula contains variables $g^\text{TE}_{a,b,j,c}$ and $g^\text{BE}_{a,b,j,c}$ for all crossbar locations $(a, b)$ and all cycles $c$ when V-mode operations are executed. $g^\text{TE}_{a,b,j,c}$ is set to 1 when the top electrode of the memristor at location $(a, b)$ is driven by literal function $l_j$; the bottom electrode is analogously described by $g^\text{BE}_{a,b,k,c}$. The Boolean satisfiability formula will include, for all M-mode cycles, expressions
\begin{equation}
    \bigwedge_{\begin{array}{c}\scriptscriptstyle 1 \leq j,k \leq 2n + 2\\[-4pt]\scriptscriptstyle 1 \leq q \leq 2^n\end{array}} \left( (g_{a,b,j,c}^\text{TE} \wedge g_{a,b,k,c}^\text{BE}) \to
\left(v_{a,b,c,q} \equiv \text{VI}^3(v_{a,b,c-1,q}, l_{j,q}, l_{k,q}) \right)\right).\label{eq:cnf}
\end{equation}
Here, the VI$^3$ operation in Figure \ref{fig_2}c is transformed into conjunctive normal form, as are implication (``$\to$'') and equivalence (``$\equiv$'') operators. The variables $g_{a,b,j,c}^\text{TE}$ and $g_{a,b,k,c}^\text{BE}$ are set to 1 when the top electrode (bottom electrode) of cell M$_{ab}$ is connected to literal function with index $j$ ($k$). According to the memristance input variable $v_{a,b,c-1,q}$ in M$_{ab}$ in cycle $c-1$, the output variable $v_{a,b,c,q}$ in cycle $c$ is set to 1 when the $q$-th entry of the VI$^3$ truth table is 1 and 0 otherwise. For instance, if the solution provides $g_{a,b,j,c}^\text{TE} = 1$ while all other $g_{a,b,1,c}^\text{TE} = 0$, it signifies that during cycle c, the top electrode of M$_{ab}$ (in BL-b) is driven by $l_1 = \text{const-0}$. Similarly, if $g_{a,b,6,c}^\text{BE} = 1$, then the bottom electrode of M$_{ab}$ (in WL-a) is driven by $l_6 = \overline{x_2}$. One more example of Boolean satisfiability formula expressions for this functionality is given in the main Article.

The M-mode operations have three input as well, where each of the input can be either a literal, or the state of any memristor in the crossbar. Let, for simplicity, the list of all allowed input functions be $\alpha_1, \alpha_2, \ldots$, and let the $q$-th truth table entry of function $\alpha_i$ be $\alpha_{i, q}$. The Boolean satisfiability formula will include, for each crossbar location $(a, b)$ and each M-mode cycle $c$, three Boolean variables $g^\text{In1}_{a,b,j,c},g^\text{In2}_{a,b,j,c},g^\text{In3}_{a,b,k,c}$, where $g^\text{In$k$}_{a,b,j,c}$ is 1 if and only if input $k$ of the memristor at location $(a, b)$ is connected to function $\alpha_j$ during cycle $c$. It will also include variables $r_{a,b,c,q}$ to represent the $q$-th entry of the truth table of the function realized on $(a, b)$ during cycle $c$. Then, the Boolean satisfiability formula will include, for all M-mode cycles, expressions
\begin{equation}
    \bigwedge_{a,b,c,q,i,j,k} \left(g^\text{In1}_{a,b,i,c} \wedge g^\text{In2}_{a,b,j,c} \wedge g^\text{In3}_{a,b,j,c}\right) \to \left(r_{a, b, c, q} \equiv \text{MI}^3(\alpha_{i, q}, \alpha_{j, q}, \alpha_{k, q})\right)
\end{equation}
In addition to expressions describing the V-mode and the M-mode operations, the Boolean satisfiability formula includes clauses to, e.g., enforce that precisely one of the $g$ variables assumes the logical value of 1, i.e., that each operation's inputs are well-defined. In addition, the crossbar structure is enforced by requiring that the inputs of the top electrodes of different memristors on the same column are identical, same with the bottom electrodes of memristors on the same row. (In principle, any particular connectivity beyond crossbars could be expressed as well; e.g., a structure where each memristor is addressable individually would correspond to just removing this last restriction altogether.) Moreover, the functions at the outputs of the circuit are locked (by means of unit clauses) to the values from the truth table of function $f$, i.e., the user's specification.

A solution of the constructed Boolean satisfiability formula contains the information which operation input is connected to which literal or output. It also includes all values assumed by the memristor outputs during all considered cycles. If the formula is unsatisfiable, this constitutes a formal proof that no valid circuit is realizable within the given crossbar size and number of V-mode and M-mode cycles. This gives the designer a handle to compute an optimal circuit with a provably minimal size or latency by finding a solution for (possibly too large) crossbar dimensions and number of cycles and then gradually reducing them until the formula becomes unsatisfiable. 
By incorporating parallel cascading computing in both V- and M-modes within a two-dimensional crossbar setup, M$^3$S can offer an optimal solution for executing function $f$ while minimizing crossbar size and cycle count. Additionally, it is possible to minimize the number of potentially unreliable M-mode operations, designers can adjust the number of cycles in the second phase, starting from 0, until the satisfiability of the conjunctive normal form produced by M$^3$S is achieved. 
Its downside is its limited scalability due to long run times of Boolean satisfiability solving software. This is in line with optimal synthesis methods for classical CMOS technology, which work in practice (except for restricted classes of functions) only for functions with up to 5-6 inputs. Even in the CMOS technology that is capable of realizing circuits with billions of logic gates, optimal synthesis is considered useful because it can be applied hierarchically; we expect the same to hold for the memristive case.

\section{Proof-of-principle demonstration of $N$-bit carry-ripple adder exploiting $\mathrm{VI}^{3}$ and $\mathrm{VO}^{3}$ Operations}\label{secG}

This section showcases the implementation of an $N$-bit carry-ripple adder utilizing VI$^{3}$ and VO$^{3}$ operations. Despite the significant power and latency consumption in peripherals associated with readout operations (Supplementary Information \ref{secB}), the VI$^{3}$ and VO$^{3}$-based $N$-bit carry-ripple adder solution exhibits remarkably low latency and cell count, surpassing state-of-the-art designs.

\begin{figure}[!h]
\centering\includegraphics[width=\linewidth]{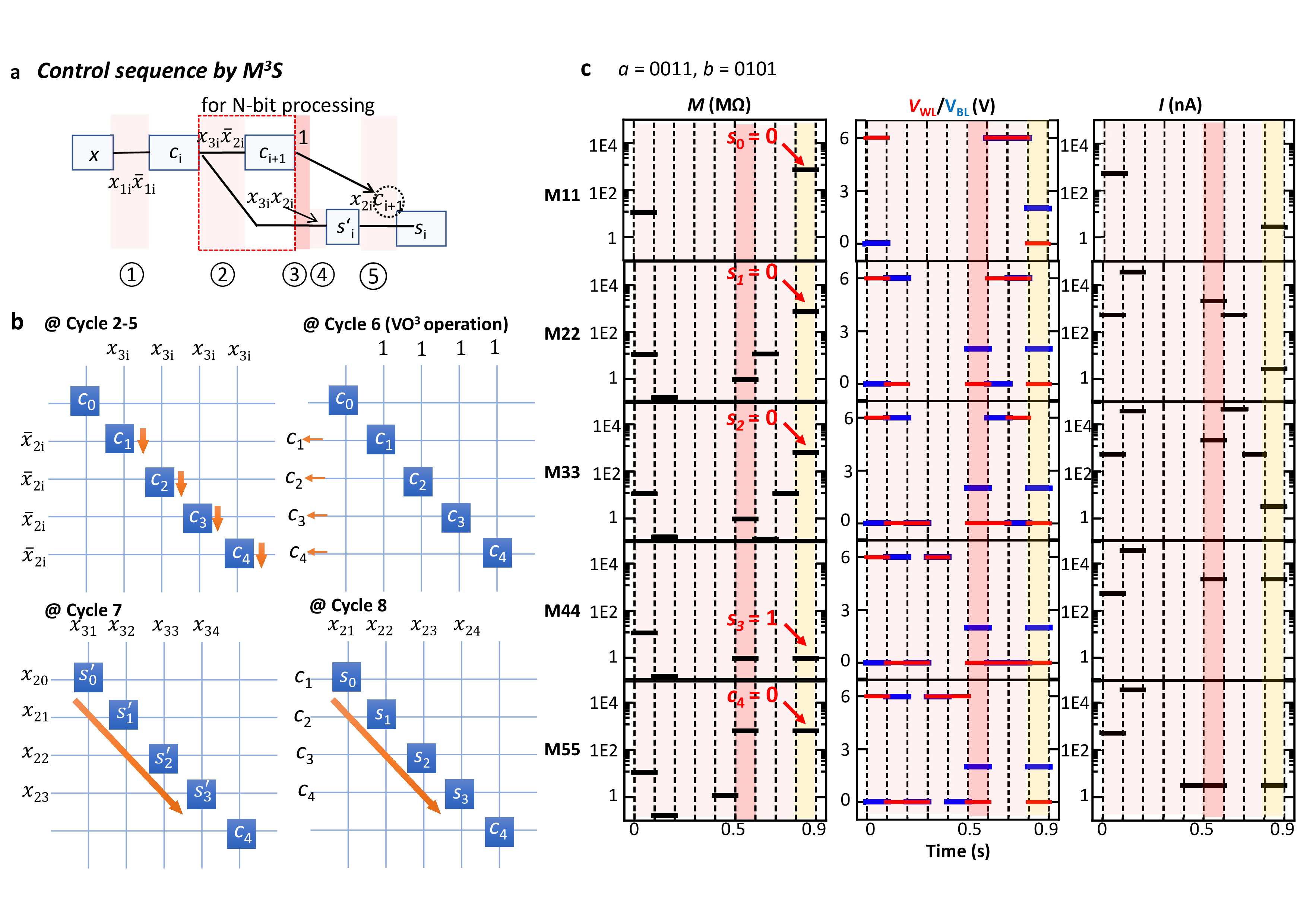}
\caption{Experimental implementation of $N$-bit carry-ripple adder by exploiting M$^{3}$S based on $\mathrm{VI}^{3}$ and $\mathrm{VO}^{3}$ . (a) Control sequence for $N$-bit carry-ripple adder by using BiFeO$_{\text{3}}$ memristive crossbar structure. (b) Demonstration of sequential operations for 4-bit carry-ripple adder with arbitrary inputs, which requires in total 13 cycles including four readout logic cycles (rLIV = 1, marked in dark red color). (c) Experimental results of 4-bit carry-ripple adder, i.e. memristance of each cell $M$, voltage on WL to shared bottom electrode $V_{\mathrm{WL}}$, voltage on $\mathrm{BL}$ to individual top electrode 
 $V_{\mathrm{BL}}$, and current $I$ from top electrode to bottom electrode across each cell. }
\label{FigS_AdderVO}
\end{figure}

 \begin{table}[h]
 \tiny
 \caption{Comparison of the implemented $N$-bit carry-ripple adder based on VI$^{3}$ and  VO$^{3}$ with readout logic and the representative designs by exploiting memory-centric computing with readout logic during cascading.}\label{tab_readout}
 \begin{tabular}{@{}cccccccccllc@{}}
\toprule

  \begin{tabular}[c]{@{}c@{}}Logic \\  primitives\\ \\ \end{tabular} &
    \begin{tabular}[c]{@{}c@{}}\# Mem-cells \\of $N$-bit adder \\ (N = 8) \end{tabular} &
  \begin{tabular}[c]{@{}c@{}}\# Cycles\\ of $N$-bit adder\\ (N = 8) \\\end{tabular} &
  \begin{tabular}[c]{@{}c@{}}Area-Delay-\\Product\\ with N = 8 \\ \end{tabular} &
  \begin{tabular}[c]{@{}c@{}}Synthesis\\ methodology\\ \\ \end{tabular} &
  \begin{tabular}[c]{@{}c@{}}Full adder\\ types \\ \\ \end{tabular} &
  \begin{tabular}[c]{@{}c@{}}References,\\ year\\ \\\end{tabular} \\ \midrule

  \begin{tabular}[c]{@{}c@{}}RIMP, NIMP, XOR\\ (2-input logic)\\ \\\end{tabular} &
    \begin{tabular}[c]{@{}c@{}} 2(N+1)\\ 18\\ \\  \end{tabular} &
  \begin{tabular}[c]{@{}c@{}}2N+4 \\20\\ \\ \end{tabular} &
  \begin{tabular}[c]{@{}c@{}}360\\  \\ \\ \end{tabular} &
  \begin{tabular}[c]{@{}c@{}}Handcraft\\  \\ \\ \end{tabular} &
  \begin{tabular}[c]{@{}c@{}}Carry ripple \\adder \\ \\  \end{tabular} &
  \begin{tabular}[c]{@{}c@{}} \cite{ siemon2015complementary}, 2015 \\ \\ \\ \end{tabular} \\

  \begin{tabular}[c]{@{}c@{}}XOR, NAND \\ (2-input logic)\\ \\\end{tabular} &
    \begin{tabular}[c]{@{}c@{}} 2N+1\\ 17\\ \\  \end{tabular} &
  \begin{tabular}[c]{@{}c@{}}2N+1 \\ 17\\ \\ \end{tabular} &
  \begin{tabular}[c]{@{}c@{}}289\\  \\ \\ \end{tabular} &
  \begin{tabular}[c]{@{}c@{}}Handcraft\\  \\ \\ \end{tabular} &
  \begin{tabular}[c]{@{}c@{}}Carry ripple \\adder \\ \\  \end{tabular} &
  \begin{tabular}[c]{@{}c@{}} \cite{wang2018efficient}, 2018 \\ \\ \\ \end{tabular} \\

    \begin{tabular}[c]{@{}c@{}}RIMP/NIMP \\ (2-input logic)\\ \\\end{tabular} &
    \begin{tabular}[c]{@{}c@{}} $(5 + \log_2(N)) \cdot N$\\ 64\\ \\  \end{tabular} &
  \begin{tabular}[c]{@{}c@{}}$(21 + \log_2(N)-1) \cdot 8$ \\ 37\\ \\ \end{tabular} &
  \begin{tabular}[c]{@{}c@{}}2368\\  \\ \\ \end{tabular} &
  \begin{tabular}[c]{@{}c@{}}Handcraft\\  \\ \\ \end{tabular} &
  \begin{tabular}[c]{@{}c@{}}Sklansky Tree  \\adder \\ \\  \end{tabular} &
  \begin{tabular}[c]{@{}c@{}} \cite{siemon2019sklansky}, 2019 \\ \\ \\ \end{tabular} \\

  \begin{tabular}[c]{@{}c@{}}XOR+Majority \\ (2-input logic)\\ \\\end{tabular} &
    \begin{tabular}[c]{@{}c@{}} 3N\\ 24\\ \\  \end{tabular} &
  \begin{tabular}[c]{@{}c@{}}2N+2 \\ 18\\ \\ \end{tabular} &
  \begin{tabular}[c]{@{}c@{}}432\\  \\ \\ \end{tabular} &
  \begin{tabular}[c]{@{}c@{}}Standard\\tool  \\ \\ \end{tabular} &
  \begin{tabular}[c]{@{}c@{}}Carry ripple \\adder \\ \\  \end{tabular} &
  \begin{tabular}[c]{@{}c@{}} \cite{pinto2021robust}, 2021 \\ \\ \\ \end{tabular} \\

    \begin{tabular}[c]{@{}c@{}}Majority, NOT \\ (2-input logic)\\ \\\end{tabular} &
    \begin{tabular}[c]{@{}c@{}} $(8N+16)\cdot6$*\\ 480*\\ \\  \end{tabular} &
  \begin{tabular}[c]{@{}c@{}}$4\log_2(N) + 6$ \\ 18\\ \\ \end{tabular} &
  \begin{tabular}[c]{@{}c@{}}8640\\  \\ \\ \end{tabular} &
  \begin{tabular}[c]{@{}c@{}}Standard\\tool  \\ \\ \end{tabular} &
  \begin{tabular}[c]{@{}c@{}}Parallel-prefix \\adder \\ \\  \end{tabular} &
  \begin{tabular}[c]{@{}c@{}} \cite{reuben2021accelerated}, 2021 \\ \\ \\ \end{tabular} \\

    \begin{tabular}[c]{@{}c@{}}XOR+Majority \\ (2-input logic)\\ \\\end{tabular} &
    \begin{tabular}[c]{@{}c@{}} 10N+1\\ 81\\ \\  \end{tabular} &
  \begin{tabular}[c]{@{}c@{}}8+2N \\ 24\\ \\ \end{tabular} &
  \begin{tabular}[c]{@{}c@{}}1944\\  \\ \\ \end{tabular} &
  \begin{tabular}[c]{@{}c@{}}Handcraft\\   \\ \\ \end{tabular} &
  \begin{tabular}[c]{@{}c@{}}Carry ripple \\adder \\ \\  \end{tabular} &
  \begin{tabular}[c]{@{}c@{}} \cite{brackmann2024improved}, 2023 \\ \\ \\ \end{tabular} \\

    \begin{tabular}[c]{@{}c@{}} VI$^{3}$,VO$^{3}$ \\ (3-input logic)\\ \\\end{tabular} &
    \begin{tabular}[c]{@{}c@{}} N+1\\ 9\\ \\  \end{tabular} &
  \begin{tabular}[c]{@{}c@{}}N+4 \\ 12\\ \\ \end{tabular} &
  \begin{tabular}[c]{@{}c@{}}108\\  \\ \\ \end{tabular} &
  \begin{tabular}[c]{@{}c@{}}M$^{3}$S\\tool  \\ \\ \end{tabular} &
  \begin{tabular}[c]{@{}c@{}}Carry ripple \\adder \\ \\  \end{tabular} &
  \begin{tabular}[c]{@{}c@{}} This work \\ \\ \\ \end{tabular} \\
  \hline
\end{tabular}

  \footnotetext[*]{ The area cost of sense amplifier has been considered and included by the authors. }
\end{table}

Fig. \ref{FigS_AdderVO} demonstrates the experimental implementation of $N$-bit carry-ripple adder based on VI$^{3}$ and VO$^{3}$ by using memristive crossbar based on BiFeO$_{\text{3}}$ cells. The control sequence by M$^{3}$S is demonstrated in Fig. \ref{FigS_AdderVO}a. The VI$^{3}$ operations are marked with background of red color, including the required readout VO$^{3}$ cycle highlighted in dark red, and there is no $\mathrm{MI}^{3}$ operations needed. One of the eight functions $\left\{x_{1}, \bar{x}_{1}, x_{2}, \bar{x}_{2}, x_{3}, \bar{x}_{3}\right.$, const-0, const- 1$\}$ is applicable as logic inputs in VI$^{3}$ or VO$^{3}$ operations at either of the memristive cell's two electrodes with top electrodes controlled by BLs, while bottom electrodes are controlled by WLs. As shown in Fig. \ref{FigS_AdderVO}a, a full carry-ripple adder design by using VI$^{3}$ and VO$^{3}$ requires 2 rows, indicating 2 cells for implementing $1$-bit full adder, including one for computing and storing the output-carry bit $c_{i+1}$, and the other one for computing and storing output-sum bit $s_{i}$. The 5 numbered columns indicate the 5 cycles required for implementing $1$-bit full adder, 4 of VI$^{3}$ operations (marked in light red) and 1 of VO$^{3}$ (marked in dark red).

Starting with unknown state $x$, the $c_\text{i+1}$, computed through VI$^{3}$ operations during cycles 1 and 2, are stored as memristive states within both cells. Subsequently, $c_\text{i+1}$ is retrieved via a non-destructive VO$^{3}$ operation, with logic inputs $v_\text{x2}$ = $\bar{v}_\text{x3}$ = 1, during cycle 3. This readout value, $c_\text{i+1}$, further serves as $v_\text{x3}$ in the subsequent VI$^{3}$ operation during cycle 5 for the computation of $s_\text{i}$ on cell 2, on which the value $s^{\prime}_\text{i}$ is computed and stored by VI$^{3}$ during cycle 4. 

As an example, the $4$-bit carry-ripple adder with input $ a= x_\text{2} = 0011$ and $b = x_\text{3} = 0101$ is experimentally tested on a passive crossbar based on BiFeO$_{\text{3}}$ memristive cells. In order to maximize the parallelism, we implement the automation flow in Fig. \ref{FigS_AdderVO}a in the diagonal cells of a crossbar array. 
Starting with $c_\text{0}$, all carry bits $c_\text{i+1}$ are computed by bitwise parallel $\mathrm{VI}^{3}$ operations (cycles 2-5), and readout in cycle 6 in parallel from WLs. The sum bits of $s^{\prime}_\text{i}$ (cycle 7) and $s_\text{i}$ (cycle 8) are computed by the parallel VI$^{3}$ operations on all bits. Fig. \ref{FigS_AdderVO}c shows the memristance, voltages on WLs and BLs, and absolute values of current across the required 5 cells during 8 cycles. 
The carry bit $c_\text{4}$ is computed and stored in cell M55. The correctness of output states is verified by applying reading biases as $2 \mathrm{~V}$ to top electrode and $0 \mathrm{~V}$ to bottom electrode in each output cell; the observed states $s_\text{0-3}$ and $c_\text{4}$ are indicated in the diagram. 

$\mathrm{N}$-bit carry-ripple adder implementation based on $\mathrm{VI}^{3}$ and $\mathrm{VO}^{3}$ can be considered as an area-optimized solution, which requires the minimum possible $N+1$ cells for processing a $N$-bit adder.
Only the iterative computation of $c_\text{i+1}$ has to be repeated for each input bit of $a$ and $b$. The cell number required by $N$-bit carry-ripple adder based on VI$^{3}$ and VO$^{3}$ is limited by the number of computing bits $N$ in addition processing. The readout of $c_\text{i+1}$ in cycle 3, the computation of $s^{\prime}_\text{i}$ (cycle 4) and $s_\text{i}$ (cycle 5) can be realized in a parallel manner (counted as 1 cycle for $N$ bit). Computing $N$-bit carry-ripple adder requires $N+1$ cells (one for saving the output-carry bit $c_\text{i+1}$ ) and $N+4$ cycles, while maximizing the parallelism during $\mathrm{VI}^{3}$ $\mathrm{VO}^{3}$ operations.  

However, it is noteworthy that to leverage diagonal cells for $N$-bit carry-ripple adder will block all the cells in the applied WLs and BLs, even there is no operation required on them. As an alternative, the $N$-bit adder in Fig. \ref{FigS_AdderVO}a can also be implemented by applying memristive cells in one WL or BL, reaching the minimum possible $N+1$ cells with slightly increased cycle number of $3N+1$ cycles (not shown here). The comparison of the proposed $N$-bit carry-ripple adder based on VI$^{3}$ and VO$^{3}$ with the previously published adders is summarized in Tab. \ref{tab_readout}. The proposed $\mathrm{N}$-bit carry-ripple adder based on VI$^{3}$ and VO$^{3}$ demonstrates the best computing performance in terms of cell number and cycle number in comparison to state-of-the-art approaches.

\section{Energy evaluation in mixed-mode computing}\label{secH}

A primary advantage of memristive devices over CMOS is their ability to combine both logic and memory functions within a single device, which is particularly relevant for logic-in-memory computing schemes. As data processing requirements grow, the inefficiencies of moving data back and forth between memory and the CPU in traditional von Neumann architectures become increasingly problematic. Data movement between the memory and processor in CMOS-based systems can consume up to 1000 times more energy than the logic operations themselves. In fact, this energy limitation is one of the most pressing issues in the computing industry as the energy required to compute grows exponentially while the global energy production grows only linearly, year after year. As noted in Supplementary Information \ref{secD}, our approach is transferable to a variety of memristive devices, and the performance of our approach is closely tied to the properties of these devices. Substantial advancements in memristor technology are actively being made. Recent progress includes write speeds reaching 20 ps \cite{Csontos2023}, feature sizes down to 2x2 nm$^2$ \cite{Pi2019}, multilevel switching capabilities with over 2048 discrete levels in a single device \cite{Rao2023}, and stackability of up to 8 layers for 3D integration \cite{Lin2020}. These advancements hint at future improvements in speed, energy efficiency, and scalability for memristor-based computing. In literature it was shown that hybrid computing concepts offloading some of the processing in the memory, in logic in memory applications \cite{LeGallo2018Sebastian, LeGallo2018} and in neuromorphic computing applications \cite{LeGallo2024, Ankit2019, Wan2022, Huang2024, Shafiee2016}, leads to performance improvements in particular by reducing the amount of data to be moved between memory and CPU. 
\textcolor{blue}{Building on these advancements, we further emphasize that our approach integrates both memristance (M) and voltage (V) as active logic variables within a single computational process. This mixed-mode operation introduces computing capabilities that were previously unexplored. In contrast to existing in-memory computing methods, which utilize either memristance or voltage for logic computation (e.g., Refs. \cite{LeGallo2018Sebastian, LeGallo2018, LeGallo2024, Ankit2019, Wan2022, Huang2024, Shafiee2016}), our approach incorporates both, providing an additional degree of freedom to computing architectures. Rather than replacing conventional in-memory logic, our method augments existing approaches, potentially unlocking additional computational possibilities while carefully managing switching overhead.} 
\textcolor{blue}{For instance, our approach can be particularly beneficial for hyperdimensional computing (HDC) \cite{Karunaratne2020}, which has shown great promise in machine learning tasks such as text classification, biomedical signal processing, and sensor fusion. HDC operations like binding and bundling can be efficiently performed in-memory, with writing-based operations limited to preparing the associative memory array, thereby reducing endurance concerns. As reported in Ref. \cite{Karunaratne2020}, memristors are utilized for in-memory logic operations for binding. Unlike prior approaches that rely on memristor-based XOR lookup tables or digital logic gates, our mixed-mode approach leverages both voltage and memristance as logic variables, enhancing computational efficiency.} 
\textcolor{blue}{To make memory technology competitive with conventional technologies, several critical factors must be addressed: 1) Endurance Considerations: The endurance of memristors is a critical factor in memory technology. While conventional non-volatile memory technologies require high endurance ($\geq 10^{16}$ cycles) for practical development, our approach can mitigate endurance concerns through distributed and parallel switching strategies. For instance, if we assume sequential additions in a memory block, individual devices would only switch a limited number of times. The endurance requirement can be approximated as: }
\begin{equation}
\textit{Required endurance} = \frac{\# \textit{switching cycles per device}}{\# \textit{operations performed}}
\end{equation}
\textcolor{blue}{By efficiently distributing operations across memory cells and utilizing techniques such as write-sharing, the required endurance per device can be significantly reduced, making the approach more practical. 2) Energy Efficiency Considerations: A key metric for competitiveness is that the energy cost of performing operations must be lower than the energy required for moving data to the CPU. We estimate energy consumption based on the number of switching events: }
\begin{equation}
E_{\textit{op}} \leq \frac{E_{\textit{data movement}}}{\# \textit{switching events}}
\end{equation}
\textcolor{blue}{Additionally, charging and discharging the crossbar array introduce additional energy overheads, as described in prior studies (e.g., \cite{5597913}). In such cases, optimizing the frequency of charge/discharge cycles is crucial to maintaining energy efficiency. Future designs incorporating smarter charge management strategies and optimizing peripheral circuitry will further enhance efficiency.}

In the context of the mixed-mode computing paradigm, the energy cost associated with logic computation is heavily influenced by the characteristics of the underlying memristive technology, particularly its switching kinetics. For a specific device technology, this energy cost is further dictated by the number of write and readout operations required for a given implementation. As memristive technology continues to advance, these developments will likely enable even greater efficiency and scalability for mixed-mode and logic-in-memory computing paradigms. Given the adaptability of mixed-mode computing, facilitated by crossbar-oriented automatic flow tools that seamlessly accommodate diverse technologies and facilitate transferability, we assess the energy consumption of logic computation within this paradigm by quantifying the number of write and readout operations. The $N$-bit carry-ripple adder based on VI$^{3}$ and MI$^{3}$ demonstrated in Fig. \ref{fig_3} requires $\left(3 N^{2}+21 N+2\right) / 2$ programming operations (write operations) and no readout operations during logic cascading, whereas the $\mathrm{N}$-bit carry-ripple adder based on $\mathrm{VI}^{3}$ and $\mathrm{VO}^{3}$ in Fig. \ref{FigS_AdderVO} requires in total $\left(N^{2}+7 N+2\right) / 2$ programming operations as well as $N$ read operations. 
For example, if the energy cost of readout VO kernel 
is the same as programming exploiting MI kernel, 
the adder with readout will always be more efficient. 
However, as discussed in Supplementary Information \ref{secB}, in practical cases, the readout VO kernel requires additional power and latency from peripherals, e.g. the output dependent control logics, makes it more expensive than programming operations in MI kernel.
Thus, while assuming a constant average energy cost of each programming operation, the overall energy cost of $N$-bit carry-ripple adder is not only depending on the ratio of energy required for reading and programming on $N$ in the memory cells, but also strongly influenced by the complexity in the individual peripheral design surrounding the memory blocks.

 \begin{sidewaystable}
 \tiny
 \caption{Comparison of the implementation for $N$-bit adders between the mixed-mode approaches of this work and the representative ones by exploiting memory-centric computing (no readout during logic cascading).}
 \label{tab_adder}
 \begin{tabular}{@{}cccccccccllc@{}}
\toprule

  \begin{tabular}[c]{@{}c@{}}Logic operations\\ (x-input logic$^1$)\\ \\ \end{tabular} &
    \begin{tabular}[c]{@{}c@{}} Logic kernel \\classification \\  \end{tabular} &
  \begin{tabular}[c]{@{}c@{}}\# Mem-cells\\ of $N$-bit adder\\ (N = 8) \\\end{tabular} &
  \begin{tabular}[c]{@{}c@{}}\# cycles\\ of $N$-bit adder\\ (N = 8) \\\end{tabular} &
  \begin{tabular}[c]{@{}c@{}}\# Other devices \\of $N$-bit adder\\ (N = 8) \\ \end{tabular} &
  \begin{tabular}[c]{@{}c@{}}Area-Delay-\\Product\\ (8-bit adder) \\ \end{tabular} &
  \begin{tabular}[c]{@{}c@{}} Full adder\\ types \\ \\ \end{tabular} &
  \begin{tabular}[c]{@{}c@{}}Synthesis\\ methodology\\ \\ \end{tabular} &
  \begin{tabular}[c]{@{}c@{}}Array\\compatible?\\ \\ \end{tabular} &
  \begin{tabular}[c]{@{}c@{}}References,\\ published year\\ \end{tabular} \\ \midrule
 
  \begin{tabular}[c]{@{}c@{}}IMPLY,FALSE\\ (2-input logic) \\ \\ \end{tabular} &
      \begin{tabular}[c]{@{}c@{}}MI\\ \\ \\  \end{tabular} &
  \begin{tabular}[c]{@{}c@{}}3N+5\\ (29)\\  \\ \end{tabular} &
  \begin{tabular}[c]{@{}c@{}}88N+48\\ (252)\\ \\ \end{tabular} &
  \begin{tabular}[c]{@{}c@{}}Resistor: 1\\ (1)\\  \\  \end{tabular} &
    \begin{tabular}[c]{@{}c@{}}22560\\ \\ \\ \end{tabular} &
    \begin{tabular}[c]{@{}c@{}}Carry ripple\\adder\\ \end{tabular} &
    \begin{tabular}[c]{@{}c@{}}Standard tool\\ \\ \\ \end{tabular} &
   \begin{tabular}[c]{@{}c@{}} Y \\ \\ \\ \end{tabular} &
    \begin{tabular}[c]{@{}c@{}} \cite{lehtonen2009stateful}, 2009\\ \\ \\ \end{tabular} \\
    
  \begin{tabular}[c]{@{}c@{}}IMPLY,FALSE\\ (2-input logic)\\ \\ \end{tabular} &
     \begin{tabular}[c]{@{}c@{}} MI \\ \\ \\ \end{tabular} &
  \begin{tabular}[c]{@{}c@{}}3N+3\\ (27)\\ \\ \end{tabular} &
  \begin{tabular}[c]{@{}c@{}}29N\\ (232)\\ \\ \end{tabular} &
  \begin{tabular}[c]{@{}c@{}}Resistor: 1\\ (1)\\ \\ \end{tabular} &
   \begin{tabular}[c]{@{}c@{}} 6496\\ \\ \\  \end{tabular} &
    \begin{tabular}[c]{@{}c@{}}Carry ripple\\  adder\\ \\  \end{tabular} &
    \begin{tabular}[c]{@{}c@{}}Standard tool, \\ optimized   for\\ series   operation\\  \end{tabular} &
    \begin{tabular}[c]{@{}c@{}}Y \\ \\ \\ \end{tabular} &
    \begin{tabular}[c]{@{}c@{}} \cite{kvatinsky2013memristor},   2013\\ \\ \\  \end{tabular} \\
    
  \begin{tabular}[c]{@{}c@{}}IMPLY,   FALSE\\  (2-input  logic)\\ \\ \end{tabular} &
    \begin{tabular}[c]{@{}c@{}}MI \\ \\ \\ \end{tabular} &
  \begin{tabular}[c]{@{}c@{}}9N\\(72)\\ \\ \end{tabular} &
  \begin{tabular}[c]{@{}c@{}}5N+18\\ (58)\\ \\ \end{tabular} &
  \begin{tabular}[c]{@{}c@{}}Resistor: 1\\ (1)\\ \\ \end{tabular} &
  \begin{tabular}[c]{@{}c@{}}4234 \\ \\ \\ \end{tabular} &
  \begin{tabular}[c]{@{}c@{}}Carry ripple\\  adder\\ \\  \end{tabular} &
  \begin{tabular}[c]{@{}c@{}}Standard tool,\\  optimized for\\  parallel operation\\  \end{tabular} &
  \begin{tabular}[c]{@{}c@{}}Y$^*$ \\  \\ \\    \end{tabular} &
 \begin{tabular}[c]{@{}c@{}}  \cite{kvatinsky2013memristor},  2013\\ \\ \\ \end{tabular}  \\

  \begin{tabular}[c]{@{}c@{}}AND, OR, NOT\\ (2-input   logic)\\ \\  \end{tabular} &
    \begin{tabular}[c]{@{}c@{}}  VI \\ \\ \\ \end{tabular} &
  \begin{tabular}[c]{@{}c@{}}-\\(128)\\ \\ \end{tabular} &
  \begin{tabular}[c]{@{}c@{}}-\\ (1)\\ \\ \end{tabular} &
  \begin{tabular}[c]{@{}c@{}}Transistors: -\\ (112)\\ \\ \end{tabular} &
   \begin{tabular}[c]{@{}c@{}} 240 \\ \\ \\ \end{tabular} &
   \begin{tabular}[c]{@{}c@{}} Carry ripple\\ adder\\ \\  \end{tabular} &
  \begin{tabular}[c]{@{}c@{}}Standard tool \\ with optimized algorithm \\ for cell number\\ \end{tabular} &
  \begin{tabular}[c]{@{}c@{}}  N \\ \\ \\ \end{tabular} &
  \begin{tabular}[c]{@{}c@{}}  \cite{liu2015signal}, 2015\\ \\ \\ \end{tabular} \\

  \begin{tabular}[c]{@{}c@{}}NOR  \\ (2-input   logic)\\ \\ \end{tabular} &
    \begin{tabular}[c]{@{}c@{}}MI\\ \\ \\  \end{tabular}&
   \begin{tabular}[c]{@{}c@{}}  5 \\ \\ \\  \end{tabular}&
  \begin{tabular}[c]{@{}c@{}}15N\\ (120)\\ \\ \end{tabular} &
  \begin{tabular}[c]{@{}c@{}}- \\ \\ \\ \end{tabular}&
  \begin{tabular}[c]{@{}c@{}}600 \\ \\ \\ \end{tabular}&
  \begin{tabular}[c]{@{}c@{}}Carry ripple\\  adder\\ \\  \end{tabular}&
  \begin{tabular}[c]{@{}c@{}}Standard   tool \\  with   area optimized\\ \\  \end{tabular}&
 \begin{tabular}[c]{@{}c@{}} Y\\ \\ \\  \end{tabular}&
  \begin{tabular}[c]{@{}c@{}} \cite{talati2016logic}, 2016\\ \\ \\ \end{tabular} \\

  \begin{tabular}[c]{@{}c@{}}NOR\\  (2-input   logic)\\ \\ \end{tabular} &
    \begin{tabular}[c]{@{}c@{}}MI\\ \\ \\  \end{tabular} &
  \begin{tabular}[c]{@{}c@{}}11N-1\\ (87)\\ \\ \end{tabular} &
  \begin{tabular}[c]{@{}c@{}}12N+1\\ (97)\\ \\ \end{tabular} &
  \begin{tabular}[c]{@{}c@{}}-\\ \\ \\  \end{tabular} &
  \begin{tabular}[c]{@{}c@{}}8439 \\ \\ \\ \end{tabular} &
  \begin{tabular}[c]{@{}c@{}}Carry ripple\\  adder\\ \\  \end{tabular} &
  \begin{tabular}[c]{@{}c@{}}Standard tool \\ with latency optimized \\ \\ \end{tabular} &
  \begin{tabular}[c]{@{}c@{}}Y \\ \\ \\ \end{tabular} &
  \begin{tabular}[c]{@{}c@{}} \cite{talati2016logic}, 2016\\ \\ \\ \end{tabular}  \\

  \begin{tabular}[c]{@{}c@{}}IMPLY, FALSE\\ (2-input logic)\\ \\ \\ \end{tabular} &
    \begin{tabular}[c]{@{}c@{}}MI \\ \\ \\ \end{tabular} &
  \begin{tabular}[c]{@{}c@{}}2N+3\\ (19)\\ \\ \\ \end{tabular} &
  \begin{tabular}[c]{@{}c@{}}22N\\ (176)\\ \\ \\ \end{tabular} &
  \begin{tabular}[c]{@{}c@{}}Resistor: 1\\ (1)\\ \\ \\ \end{tabular} &
  \begin{tabular}[c]{@{}c@{}}3520\\ \\ \\  \end{tabular} &
  \begin{tabular}[c]{@{}c@{}}Carry ripple\\ adder \\ \\ \end{tabular} &
  \begin{tabular}[c]{@{}c@{}}optimized algorithm \\   specific for CRA \\  \end{tabular} &
  \begin{tabular}[c]{@{}c@{}}Y \\ \\ \\ \end{tabular} &
  \begin{tabular}[c]{@{}c@{}} \cite{rohani2017improved}, 2017 \\ \\ \\ \end{tabular} \\

  \begin{tabular}[c]{@{}c@{}}NOR, NOT \\ (2-input logic)\\ \\ \end{tabular} &
    \begin{tabular}[c]{@{}c@{}}MI\\ \\ \\  \end{tabular} &
  \begin{tabular}[c]{@{}c@{}}-\\ (N = 4: 48)\\   \end{tabular} &
  \begin{tabular}[c]{@{}c@{}}-\\ (N = 4: 101) \\  \end{tabular} &
  \begin{tabular}[c]{@{}c@{}}-\\ \\ \\  \end{tabular} &
  \begin{tabular}[c]{@{}c@{}} 4848 \\ (N = 4)\\ \\  \end{tabular} &
  \begin{tabular}[c]{@{}c@{}}Carry look-\\ ahead adder\\ \\  \end{tabular} &
  \begin{tabular}[c]{@{}c@{}}Standard   tool \\ \\ \\ \end{tabular} &
  \begin{tabular}[c]{@{}c@{}}Y \\ \\ \\ \end{tabular} &
  \begin{tabular}[c]{@{}c@{}} \cite{thangkhiew2017efficient}, 2017\\ \\ \\ \end{tabular}  \\

  \begin{tabular}[c]{@{}c@{}}NOR, NOT\\ (2-input logic)\\ \\ \end{tabular} &
    \begin{tabular}[c]{@{}c@{}}MI \\ \\ \\ \end{tabular} &
  \begin{tabular}[c]{@{}c@{}}14N+1\\ (113)\\ \\ \end{tabular} &
  \begin{tabular}[c]{@{}c@{}}12N+1\\ (97)\\ \\ \end{tabular} &
  \begin{tabular}[c]{@{}c@{}}-\\ \\ \\  \end{tabular} &
  \begin{tabular}[c]{@{}c@{}}2793\\ \\ \\  \end{tabular} &
 \begin{tabular}[c]{@{}c@{}} Carry ripple\\ adder\\ \\  \end{tabular} &
  \begin{tabular}[c]{@{}c@{}}Standard tool\\ \\ \\   \end{tabular} &
  \begin{tabular}[c]{@{}c@{}}Y\\ \\ \\  \end{tabular} &
  \begin{tabular}[c]{@{}c@{}} \cite{thangkhiew2017efficient}, 2017\\ \\ \\  \end{tabular} \\

  \begin{tabular}[c]{@{}c@{}}AND, OR, NOT\\ (2-input logic)\\ \\ \\ \end{tabular} &
    \begin{tabular}[c]{@{}c@{}}VI \\ \\ \\ \end{tabular} &
  \begin{tabular}[c]{@{}c@{}}-\\  (228)\\ \\ \end{tabular} &
  \begin{tabular}[c]{@{}c@{}}1\\ (1)\\ \\ \end{tabular} &
  \begin{tabular}[c]{@{}c@{}}Transistors: -\\ (64)\\ \\ \end{tabular} &
  \begin{tabular}[c]{@{}c@{}}292 \\ \\ \\ \end{tabular} &
  \begin{tabular}[c]{@{}c@{}}Carry save \\ adder\\ \\  \end{tabular} &
  \begin{tabular}[c]{@{}c@{}}Standard tool \\ with area optimized \\  \end{tabular} &
  \begin{tabular}[c]{@{}c@{}}N \\ \\ \\ \end{tabular} &
  \begin{tabular}[c]{@{}c@{}} \cite{mandal2019design}, 2019\\ \\ \\ \end{tabular}  \\ 

  \begin{tabular}[c]{@{}c@{}}IMPLY,   OR, NOR,\\  COPY operation\\ (2-input   logic)\\ \end{tabular} &
    \begin{tabular}[c]{@{}c@{}}MI\\ \\ \\  \end{tabular} &
  \begin{tabular}[c]{@{}c@{}}11N\\ (88)\\ \\ \end{tabular} &
  \begin{tabular}[c]{@{}c@{}}6N+6\\  (54)\\ \\ \end{tabular} &
  \begin{tabular}[c]{@{}c@{}}Resistor: 1\\ (1)\\ \\ \end{tabular} &
  \begin{tabular}[c]{@{}c@{}}4806 \\ \\ \\ \end{tabular} &
  \begin{tabular}[c]{@{}c@{}}Carry ripple\\ adder\\ \\  \end{tabular} &
  \begin{tabular}[c]{@{}c@{}}Handcraft\\ \\ \\  \end{tabular} &
  \begin{tabular}[c]{@{}c@{}}Y\\ \\ \\  \end{tabular} &
  \begin{tabular}[c]{@{}c@{}} \cite{cheng2019functional}, 2019\\ \\ \\  \end{tabular} \\ 

  \begin{tabular}[c]{@{}c@{}}IMPLY, FALSE\\ (2-input   logic)\\ \\ \end{tabular} &
    \begin{tabular}[c]{@{}c@{}}MI\\ \\ \\  \end{tabular} &
  \begin{tabular}[c]{@{}c@{}}2N+3\\ (19)\\ \\ \end{tabular} &
  \begin{tabular}[c]{@{}c@{}}17N\\ (136)\\ \\ \end{tabular} &
  \begin{tabular}[c]{@{}c@{}}Resistor: 1\\ (1)\\ \\ \end{tabular} &
  \begin{tabular}[c]{@{}c@{}}2720\\ \\ \\ \end{tabular}  &
  \begin{tabular}[c]{@{}c@{}}Carry ripple\\  adder \\ \\ \end{tabular} &
  \begin{tabular}[c]{@{}c@{}}Standard tool \\ optimized for \\ semiparallel processing \\ \end{tabular} &
  \begin{tabular}[c]{@{}c@{}}Y$^*$ \\ \\ \\  \end{tabular} &
  \begin{tabular}[c]{@{}c@{}} \cite{rohani2019semiparallel}, 2019 \\ \\ \\ \end{tabular} \\

  \begin{tabular}[c]{@{}c@{}}IMPLY,   FALSE\\ COPY operation\\ (3-input   logic)\\ \end{tabular} &
    \begin{tabular}[c]{@{}c@{}}MI\\ \\ \\  \end{tabular} &
  \begin{tabular}[c]{@{}c@{}}6N+6\\ (54)\\ \\ \end{tabular} &
  \begin{tabular}[c]{@{}c@{}}2N+15\\ (31)\\ \\ \end{tabular} &
  \begin{tabular}[c]{@{}c@{}}Resistor: 1\\ (1)\\ \\ \end{tabular} &
  \begin{tabular}[c]{@{}c@{}}1705\\ \\ \\  \end{tabular} &
  \begin{tabular}[c]{@{}c@{}}Carry ripple\\ adder \\ \\ \end{tabular} &
 \begin{tabular}[c]{@{}c@{}} Handcraft\\ \\ \\  \end{tabular} &
  \begin{tabular}[c]{@{}c@{}}Y \\ \\ \\ \end{tabular} &
  \begin{tabular}[c]{@{}c@{}} \cite{siemon2019stateful}, 2019 \\ \\ \\ \end{tabular} \\

  \begin{tabular}[c]{@{}c@{}}IMPLY,   AND\\ (2-input   logic)\\ \\ \end{tabular} &
    \begin{tabular}[c]{@{}c@{}}MI\\ \\ \\  \end{tabular} &
  \begin{tabular}[c]{@{}c@{}}5N+1\\ (41)\\ \\ \end{tabular} &
  \begin{tabular}[c]{@{}c@{}}6N+8\\ (56)\\ \\ \end{tabular} &
  \begin{tabular}[c]{@{}c@{}}Resistor: 1\\ (1)\\ \\ \end{tabular} &
  \begin{tabular}[c]{@{}c@{}}2352\\ \\ \\  \end{tabular} &
  \begin{tabular}[c]{@{}c@{}}Carry ripple\\ adder\\ \\  \end{tabular} &
  \begin{tabular}[c]{@{}c@{}}Handcraft \\ \\ \\ \end{tabular} &
  \begin{tabular}[c]{@{}c@{}}Y$^*$ \\  \\   \end{tabular} &
  \begin{tabular}[c]{@{}c@{}} \cite{xu2020stateful}, 2020\\ \\ \\  \end{tabular} \\

  \begin{tabular}[c]{@{}c@{}}AND, OR, NOT\\ (2-input logic)\\ \\ \end{tabular} &
    \begin{tabular}[c]{@{}c@{}}VI \\ \\ \\ \end{tabular} &
  \begin{tabular}[c]{@{}c@{}}18N\\  (144)\\ \\ \end{tabular} &
  \begin{tabular}[c]{@{}c@{}}1\\ (1)\\ \\ \\ \end{tabular} &
  \begin{tabular}[c]{@{}c@{}}Transistors: \\ 8N \\ (64)\\ \end{tabular} &
  \begin{tabular}[c]{@{}c@{}}208\\ \\ \\  \end{tabular} &
  \begin{tabular}[c]{@{}c@{}}Carry ripple\\ adder\\ \\  \end{tabular} &
  \begin{tabular}[c]{@{}c@{}}Standard   tool, \\   optimized   for  \\  energy   and \\ step delay \end{tabular} &
 \begin{tabular}[c]{@{}c@{}} Y$^*$\\   \\ \\    \end{tabular} &
 \begin{tabular}[c]{@{}c@{}}  \cite{ali2021hybrid}, 2021 \\ \\ \\ \end{tabular} \\ 

  \begin{tabular}[c]{@{}c@{}}XOR\\  (2-input logic**)\\ \\ \end{tabular} &
    \begin{tabular}[c]{@{}c@{}}MI\\ \\ \\  \end{tabular} &
  \begin{tabular}[c]{@{}c@{}}6N+3\\  (51)\\ \\ \end{tabular} &
  \begin{tabular}[c]{@{}c@{}}2N+2\\  (18)\\ \\ \end{tabular} &
  \begin{tabular}[c]{@{}c@{}}-\\ \\ \\  \end{tabular} &
  \begin{tabular}[c]{@{}c@{}}918 \\ \\ \\ \end{tabular} &
 \begin{tabular}[c]{@{}c@{}} Carry ripple\\ adder\\ \\  \end{tabular} &
  \begin{tabular}[c]{@{}c@{}}Standard   tool\\ \\ \\  \end{tabular} &
  \begin{tabular}[c]{@{}c@{}}Y$^*$\\  \\    \end{tabular} &
  \begin{tabular}[c]{@{}c@{}} \cite{taherinejad2021sixor}, 2021 \\ \\ \\ \end{tabular} \\

  \begin{tabular}[c]{@{}c@{}}IMPLY,   FALSE\\ (2-input logic)\\ \\ \end{tabular} &
    \begin{tabular}[c]{@{}c@{}}MI\\ \\ \\  \end{tabular} &
  \begin{tabular}[c]{@{}c@{}}5N+1\\ (41)\\ \\ \end{tabular} &
  \begin{tabular}[c]{@{}c@{}}N+4\\ (12)\\ \\ \end{tabular} &
  \begin{tabular}[c]{@{}c@{}}MOSFETs: \\ 8N+1\\ (65)\\ \end{tabular} &
  \begin{tabular}[c]{@{}c@{}}1272\\ \\ \\  \end{tabular} &
  \begin{tabular}[c]{@{}c@{}}Carry ripple\\  adder\\ \\  \end{tabular} &
 \begin{tabular}[c]{@{}c@{}} Handcraft\\ \\   \end{tabular} &
  \begin{tabular}[c]{@{}c@{}}Y$^*$ \\  \\ \end{tabular} &

  \begin{tabular}[c]{@{}c@{}} \cite{fu2022high}, 2022\\ \\ \\  \end{tabular} \\
  \begin{tabular}[c]{@{}c@{}}IMPLY, COPY \\ operation\\ (2-input logic)\\ \end{tabular} &
   \begin{tabular}[c]{@{}c@{}} MI \\ \\ \\ \end{tabular} &
  \begin{tabular}[c]{@{}c@{}}19(N/2)+6\\  (82)\\ \\ \end{tabular} &
  \begin{tabular}[c]{@{}c@{}}3N+27\\ (51)\\ \\ \end{tabular} &
  \begin{tabular}[c]{@{}c@{}}Resistor: 1 \\ (1)\\ \\ \end{tabular} &
  \begin{tabular}[c]{@{}c@{}}4233 \\ \\ \\ \end{tabular} &
  \begin{tabular}[c]{@{}c@{}}Carry select \\ adder \\ \\ \end{tabular} &
  \begin{tabular}[c]{@{}c@{}}Standard   tool \\ \\ \\ \end{tabular} &
 \begin{tabular}[c]{@{}c@{}} Y \\ \\ \\ \end{tabular} &
  \begin{tabular}[c]{@{}c@{}} \cite{kaushik2023imply}, 2023 \\ \\ \\ \end{tabular} \\ 

  \begin{tabular}[c]{@{}c@{}}IMPLY, COPY\\ operation\\ (2-input logic)\\ \end{tabular} &
    \begin{tabular}[c]{@{}c@{}} MI\\ \\ \\  \end{tabular} &
  \begin{tabular}[c]{@{}c@{}}-\\ (136)\\ \\ \end{tabular} &
  \begin{tabular}[c]{@{}c@{}}-\\  (54)\\ \\ \end{tabular} &
  \begin{tabular}[c]{@{}c@{}}Resistor: 1 \\ (1)\\ \\ \end{tabular} &
  \begin{tabular}[c]{@{}c@{}}7398\\ \\ \\  \end{tabular} &
  \begin{tabular}[c]{@{}c@{}}Conditional \\ carry adder\\ \\  \end{tabular} &
  \begin{tabular}[c]{@{}c@{}}Standard   tool\\ \\ \\  \end{tabular} &
 \begin{tabular}[c]{@{}c@{}}Y \\ \\ \\ \end{tabular} &
  \begin{tabular}[c]{@{}c@{}} \cite{kaushik2023imply}, 2023 \\ \\ \\ \end{tabular} \\
 
  \textbf{\begin{tabular}[c]{@{}c@{}}MI$^3$,   VI$^3$ \\operations\\ (3-input logic) \end{tabular} }&
     \textbf{\begin{tabular}[c]{@{}c@{}}MI, VI\\ \\ \\  \end{tabular} }&
   \textbf{\begin{tabular}[c]{@{}c@{}}4N+1\\  (33)\\ \\ \end{tabular} }&
  \textbf{ \begin{tabular}[c]{@{}c@{}}N+4\\  (12)\\ \\ \end{tabular} }&
   \textbf{\begin{tabular}[c]{@{}c@{}}- \\ \\ \\ \end{tabular} }&
   \textbf{\begin{tabular}[c]{@{}c@{}}396 \\ \\ \\ \end{tabular} }&
   \textbf{\begin{tabular}[c]{@{}c@{}}Carry ripple\\  adder\\ \\  \end{tabular} }&
  \textbf{\begin{tabular}[c]{@{}c@{}} M$^3$S tool  \\ \end{tabular} }&
   \textbf{\begin{tabular}[c]{@{}c@{}}Y \\ \\ \\ \end{tabular} }&
  \textbf{ \begin{tabular}[c]{@{}c@{}}This work\\ \\ \\  \end{tabular}}\\ \bottomrule
  
\end{tabular}
 \footnotetext[1]{x determines the number of logic input variables in the gate design.}
  \footnotetext[*]{  The works in \cite{kvatinsky2013memristor}, \cite{rohani2019semiparallel}, \cite{ali2021hybrid} and \cite{taherinejad2021sixor} require significant structural modification. The work in \cite{xu2020stateful} require 3D array with antiparallel memristors. The work in \cite{fu2022high} require combined arrays. }
 \end{sidewaystable}

\section{Comparison with state-of-the-art implementation of full adders}\label{secI}

Table \ref{tab_adder} presents a comprehensive comparison of representative $N$-bit full adder designs alongside adder design proposed in this study. Notably, all existing and representative works featured in the comparison table leverage memristive logic designs from either the MI or VI kernel. In contrast, our work distinguishes itself as the first to exploit both kernels for arithmetic logic processing. 
As promised, our approach, employing mixed-mode logic processing through VI$^{3}$ and MI$^{3}$ operations, showcases an optimal balance between cell and cycle numbers, evident from the Area-Delay-Product column for 8-bit adders. 

It is crucial to emphasize that, unlike the other entries in the list which are simulation-based studies on addition processing without readout, our work stands out as experimental, as indicated by the bold entry in the final row. 
\textcolor{blue}{Comparing Ref. \cite{talati2016logic} with our approach, it is true that the area-delay product of our approach is more advantageous for small- to medium-bit adders but becomes less favorable beyond 14-bit implementations. However, there are important trade-offs that must be considered: the approach in Ref. \cite{talati2016logic} does not inherently store the result of the computation, which means that additional storage would be required if result storage is necessary, leading to an area overhead that also scales with $N$. }
Furthermore, it is noteworthy that only few studies demonstrating experimental implementations of addition functions in the existing literature, due to the challenges posed by reliability issues in memristor cells (Supplementary Information \ref{secC}). 
For instance, Ref. \cite{kim2019family} presents the experimental implementation of individual logic gate designs, with the full adder demonstrated via simulation by leveraging these experimental results. Ref. \cite{adam2016optimized} presents an experimental study on an 8-bit full adder employing a three-dimensional stack of monolithically integrated bipolar memristors. Such robust and optimized IMPLY logic designs is achieved by substituting the load resistor with a current source, though at the expense of compatibility with the conventional crossbar structure. 

It is further noteworthy that all full adder designs using the VI kernel \cite{mandal2019design, ali2021hybrid, liu2015signal} exhibit superior Area-Delay-Product for 8-bit adders, due to the single cycle processing. However, they all necessitate significant structural modifications or require CMOS gates for practical implementation, thus lack crossbar compatibility. 
Moreover, most studies exploiting MI kernel (including the Refs. \cite{lehtonen2009stateful, kvatinsky2013memristor, talati2016logic, rohani2017improved, thangkhiew2017efficient, rohani2019semiparallel, xu2020stateful, fu2022high} listed in the Table \ref{tab_adder}) have often assumed precise alignment of inputs with designed logic gates. However, data distribution within a memristor crossbar can span multiple locations, necessitating in-memory data relocation such as copying or transmission within the crossbar, leading to inevitable additional power costs. In contrast, we employ the VO kernel to transmit the existing logic input variables stored in the crossbar to peripherals before cascading operations commence. This approach eliminates the need for data copying and transmission, and further enables the VI kernel.

\bibliography{sn-supplementary-bibliography}